\documentclass[preprint,amsmath,amssymb,aps,prx,superscriptaddress,onecolumn]{revtex4-2}
\usepackage{orcidlink}

\definecolor{NavyBlue}{rgb}{0.0, 0.0, 128}  

\usepackage{graphicx}
\usepackage{hyperref}
\hypersetup{colorlinks=true,allcolors=NavyBlue}

\usepackage{array}
\usepackage{tabularx}

\usepackage{bm}

\usepackage{afterpage}
\usepackage{capt-of}
\usepackage{newfloat}

\DeclareFloatingEnvironment[fileext=loc,name=Caption]{captionfloat}

\newcommand{\h}{ {\bf h} }
\newcommand{\Qbf}{ {\bf Q} }
\newcommand{\E}{ {\bf E} }
\newcommand{\qbf}{ {\bf q} }
\usepackage{float}

\usepackage{amsthm}

\usepackage[singlelinecheck=false,justification=raggedright]{caption} 
\newcommand{\rhogstarS}{ \rhog^{\star \mathrm{S} }}
\newcommand{\rhogstarH}{ \rhog^{\star \mathrm{H} }}
\newcommand{\astar}{ a^{\star }}

\usepackage{textcomp} 
\usepackage{wasysym}
\newcommand{\GBAij}{ {\bf G}_{\mathrm{BA},ij } }
\newcommand{\GBAik}{ {\bf G}_{\mathrm{BA},ik } }

\newcommand{\GBAH}{ \GBA^{\mathrm{H}} }
\newcommand{\Gdagger}{ \Gbf^\dagger }

\newcommand{ \dgHT }{\Delta \g^{\mathrm{H}}_T}
\newcommand{ \dgST }{\Delta \g^\mathrm{S}_T}

\newcommand{\GBAS}{ \GBA^{\mathrm{S}} }

\newcommand{\rhogstar}{ \rhog^{\star }}

\let\var\relax
\DeclareMathOperator{\var}{var}

\newcommand{\Tr}{\mathrm{Tr}}

\newcommand{\x}{{\bf x}}

\newcommand{\p}{{\bf p}}

\newcommand{\covbf}{{\bf cov}}

\newcommand{\rbf}{{\bf r}}
\newcommand{\rbfA}{{\bf r}_{\mathrm{A}}}
\newcommand{\rbfB}{{\bf r}_{\mathrm{B}}}

\newcommand{\rbfBct}{\rbf_{\mathrm{B}}^{\mathrm{ct}}}
\newcommand{\rbfBip}{\rbf_{\mathrm{B}}^{\mathrm{ip}}}

\newcommand{\g}{{\bf g}}

\newcommand{\gBA}{ {\bf g}_\mathrm{BA} }
\newcommand{\gdagger}{ \g^{\dagger} }

\newcommand{\Gbf}{{\bf G}}

\newcommand{\Gbfhat}{ { \hat{\bf G} } }
\newcommand{\GAB}{ {\bf G}_\mathrm{AB} }
\newcommand{\GBA}{ {\bf G}_\mathrm{BA} }
\newcommand{\GABhat}{ \hat{\bf G}_\mathrm{AB} }
\newcommand{\GBAhat}{ \hat{\bf G}_\mathrm{BA} }
\newcommand{\Ghat}{ \hat{G} }
\newcommand{\ahat}{ a }

\newcommand{\GBAhatij}{ \hat{ \mathrm{G} }_{\mathrm{BA}, ij} }

\newcommand{\W}{{\bf W}}

\newcommand{\WB}{ {\bf W}_\mathrm{B} }
\newcommand{\WA}{ {\bf W}_\mathrm{A} }

\newcommand{\B}{{\bf B}}

\newcommand{\Dbf}{{\bf D}}
\newcommand{\dbf}{{\bf d}}

\newcommand{\Sigmabf}{{\bf \Sigma}}

\newcommand{\Hbf}{{\bf H}}
\newcommand{\Pbf}{{\bf P}}
\newcommand{\Hdagger}{\Hbf^\dagger}
\newcommand{\adagger}{a^\dagger}

\newcommand{\rhog}{ \rho_{\mathrm{G}} }
\newcommand{\rhow}{ \rho_{\mathrm{W}} }

\newcommand{\R}{\mathbb{R}}

\newcommand{\vectorform}{\operatorname{vec}}

\newcommand{\appropto}{\mathrel{\vcenter{
\offinterlineskip\halign{\hfil$##$\cr
\propto\cr\noalign{\kern2pt}\sim\cr\noalign{\kern-2pt}}}}}

\usepackage{bbold}
\newcommand{\one}{ {\bf I} }
\newcommand{\onevec}{ {\bf 1} }

\newcommand{\zero}{{\bf 0}}
\newcommand{\cl}{\mathcal{L}}

\newcommand{\cn}{\mathcal{N}}

\DeclareMathOperator{\tr}{Tr}

\begin{document}

\title{One nose but two nostrils: Learn to align with sparse connections between two olfactory cortices}

\author{Bo Liu\orcidlink{0000-0002-2819-608X}}
\affiliation{Center for Brain Science and Department of Molecular and Cellular Biology, Harvard University, Cambridge, Massachusetts, USA.}
\affiliation{Kempner Institute for the Study of Natural and Artificial Intelligence, Harvard University, Cambridge, Massachusetts, USA.}

\author{Shanshan Qin\orcidlink{0000-0001-8012-5687}}
\affiliation{John A. Paulson School of Engineering and Applied Sciences, Harvard University, Cambridge, Massachusetts, USA.}
\affiliation{Present address: Center for Computational Neuroscience, Flatiron Institute, New York, NY 10010, USA}

\author{Venkatesh Murthy\orcidlink{0000-0003-2443-4252}}
\email{vnmurthy@fas.harvard.edu}
\affiliation{Center for Brain Science and Department of Molecular and Cellular Biology, Harvard University, Cambridge, Massachusetts, USA.}
\affiliation{Kempner Institute for the Study of Natural and Artificial Intelligence, Harvard University, Cambridge, Massachusetts, USA.}

\author{Yuhai Tu\orcidlink{0000-0002-4589-981X}}
\email{yuhai@us.ibm.com}
\affiliation{IBM Thomas J. Watson Research Center, Yorktown Heights, NY, USA.}
\date{\today}

\begin{abstract}

The integration of neural representations in the two hemispheres is an important problem in neuroscience.  Recent experiments revealed that odor responses in cortical neurons driven by separate stimulation of the two nostrils are highly correlated. This bilateral alignment points to structured inter-hemispheric connections, but detailed mechanism remains unclear. Here, we hypothesized that continuous exposure to environmental odors shapes these projections and modeled it as online learning with local Hebbian rule. We found that Hebbian learning with sparse connections achieves bilateral alignment, exhibiting a linear trade-off between speed and accuracy. We identified an inverse scaling relationship between the number of cortical neurons and the inter-hemispheric projection density required for desired alignment accuracy, i.e., more cortical neurons allow sparser inter-hemispheric projections. We next compared the alignment performance of local Hebbian rule and the global stochastic-gradient-descent (SGD) learning for artificial neural networks. We found that although SGD leads to the same alignment accuracy with modestly sparser connectivity, the same inverse scaling relation holds. We showed that their similar performance originates from the fact that the update vectors of the two learning rules align significantly throughout the learning process. This insight may inspire efficient sparse local learning algorithms for more complex problems.

\end{abstract}

\maketitle

\section{Introduction}

Animals with bilateral symmetry must integrate information from the two sides of the brain to form a coherent picture of the environment. This problem has been studied in vision and audition, where similarities and asymmetries in information sampled through the two sides can offer interpretable computational clues \cite{grothe2014natural, McAlpine2009,blake1973psychophysical,cumming2001physiology}. But it is less clear how information from two sensors is integrated in olfaction \cite{rajan2006rats,porter2005brain,mainland2002one,rabell2017spontaneous,dalal2020bilateral}.
 
In many species including rodents, the two nares are separated by a septum that prevents inhaled odors from transferring between nostrils. Since the two nostrils may sample independent pockets of air \cite{kikuta2008compensatory,wilson1999respiratory,eccles2000nasal}, bilateral information can be used, in principle, to discover features of the odor stimulus not readily available to a single nostril. By comparing odor concentration across the two nostrils, foraging animals may locate odor sources under conditions where smooth gradients exist \cite{catania2013stereo,rabell2017spontaneous,gire2016mice,bhattacharyya2015robust,findley2021sniff}. Such asymmetric information processing from the two nostrils has also been observed in humans, who can sense the direction of an odor source presented close to their nose \cite{porter2005brain,wu2020humans}. In contrast to computations emphasizing the differences between the two sides, many behaviors require making common inferences that are independent of the stimulated nostril. For instance, animals must be able to identify relevant smells of food or a predator, regardless of the odor source's location relative to their nose and independent of any asymmetries in air flow through the nostrils \cite{kucharski1987new,mainland2002one}.
 
These behaviors using two nostrils require neural computations that rely on inter-hemispheric communication, which occurs via axons crossing through the anterior commissure \cite{kucharski1990anterior,dalal2020bilateral,rabell2017spontaneous}. Two brain regions, anterior olfactory nucleus (AON) and anterior piriform cortex (APC), are strong candidates to integrate information from the two nostrils \cite{brunjes2005field,kikuta2008compensatory,kikuta2010neurons,wilson1997binaral,dalal2020bilateral,yan2008precise,hagiwara2012optophysiological} – we will refer to these regions collectively as olfactory cortex (OC). The olfactory bulb (OB), the first stage of circuit processing in the olfactory system,  sends projections exclusively to the ipsilateral OC. Neurons in the OC, however, send projections to the opposite hemisphere through the anterior commissure, and make functional synapses  \cite{haberly1978association,brunjes2005field,hagiwara2012optophysiological,russo2020synaptic,martin2018development}.

Single OC neurons can respond to odors presented separately to ipsilateral nostril or contralateral nostril. Although direct projections from the OB account for ipsilateral responses, inter-hemispheric projections are the likely substrate for the responses of OC neurons to odors presented to the contralateral nostril \cite{kikuta2008compensatory,kikuta2010neurons,wilson1997binaral,dalal2020bilateral,dikeccligil2023odor} [Fig.~\ref{fig Background and model setup}(a)]. It is unclear whether, and how similar, the responses to ipsi- and contralateral stimuli are in the OC. Systematically different responses can be used to compute the direction from which an odor stimulus arises in natural situations \cite{kikuta2010neurons}. However, useful computations may not be possible if the responses are entirely uncorrelated.

Recently, we used a panel of diverse odors to probe the structure of odor responses in individual mouse olfactory cortical neurons to ipsilaterally and contralaterally presented stimuli \cite{Grimaud2021}. We found that individual neurons responded to odors presented to the contralateral nostril in a similar way to the same odors presented ipsilaterally \cite{Grimaud2021}. This similar tuning, which we refer to as bilateral alignment, was significantly better than chance and is not easily explained by the apparently random and disperse projections to the OC \cite{haberly2001parallel,sosulski2011distinct,miyamichi2011cortical,ghosh2011sensory,schaffer2018odor}. Matching responses to odors presented separately to the two nostrils must therefore arise from coordinated connectivity in the inter-hemispheric projections, which must somehow be aligned with the ipsilateral projections. To achieve such coordinated connectivity, thus bilateral alignment, an attractive and plausible hypothesis is that Hebbian or correlation-based synaptic plasticity shapes synapses formed by inter-hemispheric axons [Fig.~\ref{fig Background and model setup}(a)].

In the present work, we hypothesized that the continuous exposure of both nostrils to environmental odors shapes the functional properties of inter-hemispheric projections, and tested whether online learning with local Hebb’s rule is sufficient to account for matched responses found experimentally. Furthermore, mammals have millions of olfactory cortical neurons \cite{srinivasan2019scaling}, while the typical number of synapses onto one neuron is $10^3-10^4$ \cite{cragg1967density,schuz1989density}; thus, a cortical neuron in one hemisphere is unlikely to connect with all the neurons in the other hemisphere \cite{hagiwara2012optophysiological,wang2020cell}, leading to the questions of whether sparse inter-hemispheric connectivity could achieve bilateral alignment, and if so, how sparse/dense the connections need to be. Finally, within the context of bilateral alignment, we studied the differences and similarities between the local Hebb's rule and gradient-based global learning rules in particular the stochastic gradient descent (SGD) rule (algorithm) that is prevalent in training artificial neural networks.

The paper is organized as follows: First, we define the problem by describing the online Hebbian learning process in a simplified network model of the olfactory cortex and the testing criterion for bilateral alignment. We then present three main findings from our study: 1) Online Hebbian learning can successfully achieve bilateral alignment with highly sparse inter-hemispheric projections, and the learning rate modulates the trade-off between alignment accuracy and convergence speed. 2) For a given degree of alignment, the required density of inter-hemispheric projections scales inversely with the number of cortical neurons, meaning that more cortical neurons allow sparser connections. 3) The SGD learning rule achieves the same alignment performance with modestly sparser connectivity but exhibits the same scaling relation. Furthermore, we showed that the similar performance for local Hebbian rule and global SGD rule is due to the fact that the update vectors of the two learning rules are partially aligned during the learning process.

\section{Model setup and training process}
To study whether online Hebbian learning can lead to bilateral alignment, we use a minimum learning model as illustrated in Figs.~\ref{fig Background and model setup}(a)-(c), where (a) and (b) illustrate the circuits for contral- and ipsi-lateral responses respectively. As shown by the neural network in Fig.~\ref{fig Background and model setup}(c), odor elicited the same OB responses in two sides, denoted as $\x\in \R^{m}$ (anatomically, $m$ stands for the number of OB glumeruli \cite{murthy2011olfactory,uchida2014coding}, and for simplicity, we refer to the responses of output neurons of the glumeruli as ``OB responses".). The projections from OBs to OCs ($\WA$ and $\WB$) are modeled to be largely random. This lead to different cortical odor representations in two hemispheres $\rbfA, \rbfB\in \R^{n}$, which are also modulated by the inter-hemispheric interactions ($\GAB$ and $\GBA$).

For simplicity, the OB response $\x$ to an odor is drawn from Gaussian distribution with $\zero$ mean: $\x\sim\mathcal{N}\left(\zero, {\bf \Sigma}\right)$ (negative firing rates indicate suppressed activities of OB output neurons, compared to their baseline firing rates).  Here ${\bf \Sigma} $ is the input covariance matrix, and for analytical tractability, we take ${\bf \Sigma} =  \gamma^2{\bf I}_m$ (independent and identically distributed or IID Gaussian distribution) with $\gamma $ the input strength. To characterize the sparse OB-to-OC projections \cite{babadi2014sparseness,schaffer2018odor}, we denote the density of $\WA$ and $\WB$ as $\rhow$ (the fraction of their non-zero elements). For analytical tractability, we assume that the nonzero elements of $\WA$ and $\WB$ are also IID Gaussian variables, i.e., $\W_{ij}\sim \mathcal{N}\left(0, 1\right), \forall  \W_{ij}\neq0 $, which are sampled at the beginning and fixed after that. In contrast, the inter-hemispheric projection matrix $\GBA$ will be updated during learning [the blue arrows in Fig.~\ref{fig Background and model setup}(a)-(c)], whose connectivity density is $\rhog$. We initialize the non-zero entries of $\GBA$ with IID Gaussian distribution $\GBAij\sim \mathcal{N}\left(0, \sigma_0^2\right),\forall \GBAij\neq0$ [the same for $\GAB$, which is not shown in Fig.~\ref{fig Background and model setup}(c) to avoid clutter], where the variance $\sigma_0^2$ is chosen to be similar to that of the final learned weights.

During learning, each step contains three parts [Fig.~\ref{figS Model setup and training process}(b)-(c)]: firstly, one training odor is ``delivered" to two nostrils [Fig.~\ref{fig Background and model setup}(c) and Fig.~\ref{figS Model setup and training process}(a)]. The two cortical odor representations, i.e., the firing rate vectors $\rbfA$ and $ \rbfB $, are then determined by the steady state solution of the following neural dynamics:
\begin{equation}
\begin{aligned}
\tau\dfrac{d\rbfA}{dt} &= - \rbfA +\tanh( \WA\x + \GAB\rbfB), \\
\tau\dfrac{d\rbfB}{dt} &= - \rbfB + \tanh(\WB\x + \GBA\rbfA ), \\
\end{aligned}\label{Eq neural dynamics}
\end{equation}
where $\tau$ is the single-neuron integration time constant ($\tau=10\, ms$ is used in this study) and we adopted the hyperbolic tangent function $\tanh()$ as the activation function. 

Secondly, assuming a time scale separation between the neural dynamics and learning (synaptic update), we update $\GBA$ by a discrete Hebb's rule (similar for $\GAB$):
\begin{equation}\label{Eq the sparse Hebbian update rule}
    \begin{aligned}
    \Delta\GBA &= \eta ( \rbfB\rbfA^\top - \beta \GBA )\odot \GBAhat,\\
    \end{aligned}
\end{equation}
with $\eta$ the learning rate and $\beta$ the weight decay constant. Here, $\odot$ represents element-wise product; $\GBAhat$ is a fixed sparsity mask matrix with binary entries (0 or 1, the probability of $\GBAhatij=1$ is $\rhog$), to enforce the sparsity of inter-hemispheric connections (note that $\GBAhat$ is also applied to the initial weights). This Hebb's rule is local because the right hand side of Eq.~\eqref{Eq the sparse Hebbian update rule} only involves the pre-synaptic and the post-synaptic neurons.

Finally, with the Hebbian-updated $\GBA$ and $\GAB$, following experiments, we simulate the contralateral response to this single training odor [$\rbfBct$, Fig.~\ref{fig Background and model setup}(a)], as well as the ipsilateral response $\rbfBip$ [Fig.~\ref{fig Background and model setup}(b)]. Equations for $\rbfBct$ and $\rbfBip$ can be seen in Appendix \ref{Equations of ipsi- and contra-lateral responses}. To characterize the performance of online Hebbian learning at step $T$, we randomly generate $K$ test odors from the same distribution, then simulate their ipsi- and contra-lateral responses. We define the test Bilateral Alignment Level (BAL) as the mean cosine similarity $ a(T):= \langle\cos(\rbfBip,\rbfBct)\rangle_K$, which characterizes the alignment performance at step $T$. %

\section{Results}

\subsection{Dynamics of learning for bilateral alignment: learning rate modulates the speed-accuracy tradeoff}
\label{Hebb’s rule achieves bilateral alignment with sparse connections, despite some generalization gap controlled by learning rate}

To study whether sparse inter-hemispheric projections are sufficient for bilateral alignment, we started with $(m,n, \rhow,\rhog)= (20,500, 0.1,0.05)$, $N=1000 $ steps, and a learning rate of $ \eta = 0.01\  \mathrm{s}^2$. Other parameters are listed in Appendix Table \ref{Table Parameter values and the interpretations}. As shown by the blue curve in Fig.~\ref{fig_two_eta_speed_accuracy_tradeoff}(a), the test BAL converged to 0.51 within 200 steps, meaning that Hebb’s rule can achieve bilateral alignment using sparse connections. 

We then adopted a smaller learning rate $\eta = 0.001\  \mathrm{s}^2$, which led to higher test BAL but slower convergence [the red curve in Fig.~\ref{fig_two_eta_speed_accuracy_tradeoff}(a)]. Furthermore, we varied $\eta$ with other parameters fixed, and found that test accuracy decreases linearly with $\eta$ [the blue dots in Fig.~\ref{fig_two_eta_speed_accuracy_tradeoff}(b)]. In contrast, the convergence speed increases linearly with $\eta$ [the red dots in Fig.~\ref{fig_two_eta_speed_accuracy_tradeoff}(b), see Appendix \ref{sec:method}: Methods for the numerical estimation of the convergence speed and Fig.~\ref{speed_estimation} for one example]. As a result, test accuracy and convergence speed exhibit a linear tradeoff [the inset green line in Fig.~\ref{fig_two_eta_speed_accuracy_tradeoff}(b)].

To understand why Hebb's rule leads to accurate bilateral alignment, we considered a scenario with small input strength $\gamma$ and large decay rate $\beta$, where Eq.~\eqref{Eq neural dynamics} can be linearized due to the small magnitudes of  $\x$, $\GAB$, and $\GBA$ (See Appendix \ref{Rationale and requirements to linearize the neural dynamics} for details): 
\begin{equation}
\begin{aligned}
\tau\dfrac{d\rbfA}{dt} &= - \rbfA + \WA\x + \GAB\rbfB, \\
\tau\dfrac{d\rbfB}{dt} &= - \rbfB + \WB\x + \GBA\rbfA, \\
\end{aligned}\label{Eq neural dynamics linear}
\end{equation}
which has an exact steady state solution (Appendix \ref{The analytical form of Hebbian solution}). Furthermore, with small $\GAB$ and $\GBA$, the steady-state solutions of $\rbfA $ and $\rbfB$ as well as the ipsi- and contra-lateral responses become (Appendix \ref{The analytical form of Hebbian solution})
\begin{equation}
\begin{aligned}
\rbfA &\approx \WA\x, \ \rbfB \approx  \WB\x,\\
\rbfBip &\approx  \WB\x, \ \rbfBct \approx  \GBA\WA\x.
\end{aligned}
\label{Eq lower order approximations for cortical representations}
\end{equation}
Then Hebb's rule Eq.~\eqref{Eq the sparse Hebbian update rule} becomes
\begin{equation}\label{eq:GBA_SDE}
\Delta\GBA\approx \eta\left(  \WB\x\x^\top \WA^\top\odot \GBAhat  - \beta \GBA \right).
\end{equation}
In the continuum time limit, the above learning rule can be described by the following stochastic differential equation (SDE)
\begin{equation}
\frac{d\GBA}{dt} = \eta\beta\left(  \WB\x\x^\top \WA^\top\odot \GBAhat  -  \GBA \right),
\label{update rule with Delta G}
\end{equation}		
where we have made the scaling transformation $\beta\GBA \rightarrow \GBA$, which does not affect the cosine similarity. In steady state where $\langle d\GBA/dt\rangle=\zero$ in Eq.~\eqref{update rule with Delta G} ($\langle  \rangle$ means average over time or samples), we have the Hebbian solution
\begin{equation}
\begin{aligned}
\Gdagger := \WB \langle\x\x^\top \rangle\WA^\top \odot \GBAhat=\WB \Sigmabf\WA^\top \odot \GBAhat,
\end{aligned}
\label{Eq Hebbian solution}
\end{equation}
indicating that the Hebb's rule learned two feedforward projection matrices and the input correlation matrix to achieve bilateral alignment. The above analytical solution was verified by numerical simulations (Fig.~\ref{figS_hebbian_solution}).

To explain the speed-accuracy tradeoff, we vectorize $\GBA$ and write $\gBA(T):=\vectorform[\GBA(T)]=\gdagger+\delta \g(T)$, with $\gdagger$ the vectorized Hebbian solution and $\delta \g$ the variation around it. For convenience, we drop the subscript ``BA" in $\gBA$, and rewrite the continuum SDE for Hebbian learning [Eq.~\eqref{update rule with Delta G}] as:
\begin{equation}
\frac{d\g}{dt} = \eta\beta\left(   \gdagger -  \g \right)+\eta\beta \boldsymbol{\xi},
\label{UO}
\end{equation}		
which is just the well-known Langevin equation for an Ornstein-Uhlenbeck process~\cite{OU_Process} that can be solved analytically. Here, the term $\boldsymbol{\xi}=\vectorform\left[\WB(\x\x^\top-\Sigmabf)^\top \WA^\top\odot \GBAhat\right]$ represents the white noise that originates from input fluctuations (random training odors). Denote its covariance matrix as $\covbf(\boldsymbol{\xi}):=2\Dbf$, where $\Dbf$ is determined by $\WA,\WB,\GBAhat,\Sigmabf$ but independent of $\eta\beta$. It is clear from Eq.~\eqref{UO} that the relaxation timescale to reach the Hebbian solution $\gdagger$ is $\tau\sim\frac{1}{\eta\beta}$, which explains the linear dependence of the learning or convergence speed $(s=1/\tau\sim\eta\beta) $ as observed in Fig.~\ref{fig_two_eta_speed_accuracy_tradeoff}(b). 

By solving Eq.~\eqref{UO}, we can show that in steady state ($T\rightarrow \infty$), $\g$ has a Gausian distribution with the Hebbian solution $\gdagger$ as its mean and a covariance matrix given by:
\begin{equation}
\begin{aligned}
\lim_{T\rightarrow \infty}\covbf[\delta\g(T)] =  \eta\beta\Dbf.
\end{aligned}
\label{The covariance matrix of the Hebbian SDE solution}
\end{equation}
To determine the test BAL, we Taylor-expanded $a(\g)$ to the second order around its mean $a^\dagger=a(\gdagger)$ at the Hebbian solution $\gdagger$:  
\begin{equation}
\begin{aligned}
a(\g ) &=  a^\dagger  + \left(\frac{\partial a}{\partial \mathbf{g}}\vert_{\gdagger} \right)^{\top} \delta \g +\dfrac{1}{2}\delta \g^{\top} \Hbf^\dagger \delta \g,\\
\end{aligned}
\label{Eq a_hat 2nd order expansion}
\end{equation}
where $\Hbf^\dagger:= \Hbf(\gdagger)$ is the Hessian matrix evaluated at $\gdagger$. Averaging over the Gaussian distribution of $\delta  \g$, the stationary test BAL is:
\begin{equation}
\begin{aligned}
    a(\eta,\beta) = a^\dagger +  \dfrac{1}{2}\langle\delta \g^{\top} \Hdagger \delta \g \rangle_{\delta \g} = a^\dagger + \dfrac{\eta\beta}{2}\tr\left(\Dbf \Hdagger\right).
\end{aligned}
\label{Eq a eta beta of delta g}
\end{equation}
Both numerical and analytical results showed that the Hessian matrix $\Hbf^\dagger$ is negative definite: concretely, in Appendices \ref{Gradient and Hessian matrix of test BAL}-\ref{Visualization of the concave landscape of test BAL}, we showed that $a(\g)$ is concave around $\gdagger$, which is verified numerically in Fig.~\ref{figS_a_hat_is_concave}. Furthermore, as a covariance matrix, $\mathbf{D}$ is positive definite. As a result, $\tr (\mathbf{DH}^\dagger) < 0$ (see Appendix \ref{test BAL decreases with eta H negative definite} for the proof). Therefore, the test BAL decreases with $\eta\beta$ linearly, in agreement with Fig.\ref{fig_two_eta_speed_accuracy_tradeoff}(b).

\subsection{Larger networks require sparser connections to achieve bilateral alignment: an inverse scaling relation} 

In the previous section, we demonstrated that the system can achieve bilateral alignment using relatively sparse inter-hemispheric connections (5\% connectivity) in a relatively small network ($n=500$ neurons). However, in realistic biological systems such as the piriform cortex, there are millions of cortical neurons \cite{srinivasan2018distributed,srinivasan2019scaling}, and the inter-hemispheric projections in mammals might be much sparser than the 5\% connectivity (as suggested by the experimental data \cite{hagiwara2012optophysiological} and the typical $10^3-10^4$ synapses onto one neuron \cite{cragg1967density,schuz1989density}). Therefore, in this section, the critical question we aim to address is whether bilateral alignment can be achieved in a larger network with sparser inter-hemispheric connections (smaller $\rhog$ and larger $n$), or equivalently, how the required sparsity to attain a given alignment performance depends on the network size.
    
To investigate this question, we varied the number of cortical neurons $n$ and studied its relationship with the required inter-hemispheric projection density $\rhogstar$ to achieve a desired level of test BAL. In this section, we used a small learning rate $ \eta = 0.001\ \mathrm{s}^2$ to eliminate the effect of learning rate, i.e., $a\approx a^\dagger$, which only depends on the properties of the network ($\rhog$, $\rhow$, $m$ and $n$). For fixed  $m$ and $\rhow$, we varied $\rhog$ for $n$ ranging from 50 to 2000, and plotted the ensemble average of the test BAL versus $\rhog$ in Fig.~\ref{fig Hebbian theory matches simulations and the inverse scaling}(a) (here the ensemble means different samples of the matrices $\WA,\WB,\GABhat, \GBAhat$, given $m,n,\rhow,\rhog$). We found that $a(\rhog)$ has a power law dependence on $\rhog$, with an exponent $\approx 0.5$ (the left red dashed line) for small $\rhog$ before it saturates to near perfect alignment $a\sim 1$. We define $\rhogstar(n)$ as the inter-hemispheric projection density needed to reach a given (high) alignment level $\astar = 0.5$ (the black horizontal line), i.e., $a(\rhogstar(n))=\astar$. 
Our simulation results revealed an inverse scaling relationship between $\rhogstar$ and $n$ [Fig.~\ref{fig Hebbian theory matches simulations and the inverse scaling}(b)], i.e., $\rhogstar(n)\sim n^{-1}$, which suggests that with more cortical neurons, the inter-hemispheric projections can be even sparser. Remarkably, when we rescaled $\rhog$ by $\rhogstar(n)$, i.e., $\dfrac{\rhog}{\rhogstar(n)}\propto n\rhog$, all data points in Fig.~\ref{fig Hebbian theory matches simulations and the inverse scaling}(a) for different values of $n$ collapsed onto one single curve for the full range of sparsity [Fig.~\ref{fig Hebbian theory matches simulations and the inverse scaling}(c)].

To understand the inverse scaling between $\rhogstar$ and $n$, we evaluate the Weight Alignment Level (WAL) defined as
\begin{equation}
\begin{aligned}
c &:= \cos(\WB,\Gdagger\WA),\\
\end{aligned}
\end{equation}
which describes the alignment between the two vectorized weight matrices $\WB$ and $\Gdagger\WA$ (here, we have omitted the notation of vectorization, a convention adopted throughout this paper). Intuitively, $c$ is highly correlated with the bilateral alignment level $a$, e.g., perfect alignment ($a=1$) is achieved when $c=1$ (or equivalently, $\Gdagger\WA\propto\WB$) since $\rbfBct=\Gdagger\WA\x\propto\WB\x=\rbfBip, \forall \x$. Indeed, both simulations and derivations demonstrated that for IID Gaussian input, $a\approx c$ in general [see Fig.~\ref{figS Theory of Hebbian learning matches simulations}(a) and Appendix \ref{Proof that Weight Alignment Level (WAL) approximates Test Bilateral Alignment Level (BAL)} for details]. Using the Hebbian solution Eq.~\eqref{Eq Hebbian solution} and averaging over independent elements of $\WA,\WB,\GBAhat$ with law of large numbers, we derived the theoretical expression of WAL and test BAL at the Hebbian solution (see Appendix \ref{The theory for Weight Alignment Level} for details): 
\begin{equation}
a^\dagger \approx  c \approx \left(\dfrac{n \rhog}{m +3/\rhow + n\rhog }\right)^{1/2}.
\label{Eq Weight Alignment Level}
\end{equation}
The above approximation of $a^\dagger$ agrees well with the numerical results for a wide range of parameters, as shown in Fig.~\ref{fig Hebbian theory matches simulations and the inverse scaling}(c) [see also Figs.~\ref{figS Theory of Hebbian learning matches simulations}(b)-(d) where we varied $\rhow,\rhog$ for given $m=20,n=50$].

From Eq.~\eqref{Eq Weight Alignment Level}, we can see that the required density $\rhogstar$ to achieve a given level alignment $\astar$ satisfies
\begin{equation}
n  \rhogstar = \frac{a^{\star 2}(m+3/\rhow)}{1-a^{\star 2}} = \text{const}.
\label{rhogstar}
\end{equation}
Therefore, $\rhogstar$ scales inversely with $n$. Note that Eq.~\eqref{Eq Weight Alignment Level} also explains the power law $a\propto \rhog^{0.5}$ in the sparse regime: when $n\rhog \ll m + 3/\rhow$, $a\approx \sqrt{\dfrac{n \rhog}{m +3/\rhow}}\propto \rhog^{0.5}$.  

The biological interpretation of this inverse scaling is interesting to point out. $n \rhog$ in Eq.~\eqref{Eq Weight Alignment Level} represents the average number of projections that one cortical neuron receives from the contralateral cortical neurons. Thus, our results show that a good alignment performance (e.g., $a=0.5$) can be achieved when the average number of connections per neuron is larger than a threshold that is independent of the network size. Therefore, to maintain the same alignment performance, the total number of required inter-hemispheric connections only scales linearly with $n$ instead of $n^2$, much sparser than all-to-all connections. 

As an application of Eq.~\eqref{Eq Weight Alignment Level}, we estimated the required density of the inter-hemispheric projections, using biologically realistic parameters \cite{10.1073/pnas.1007931107,babadi2014sparseness,srinivasan2018distributed,srinivasan2019scaling}: with $m = 3700, n= 5\times 10^5, \rhow= 0.1$ and the required BAL $\astar=0.31$, the corresponding $\rhogstar\approx 7.93\times 10^{-4}$ and $n\rhogstar\approx 397$, meaning $\sim400$ inter-hemispheric projection synapses per neuron, which is reasonable given the $10^3-10^4$ synapses per neuron typically \cite{cragg1967density,schuz1989density}. Note that, the above estimation is insensitive to the exact value of $\rhow$ provided that $m\gg 3/\rhow$. The exact sparsity of interhemispheric connections has not been measured: given the many components of OC such AON and APC, the optogenetic experiments in \cite{hagiwara2012optophysiological} only showed that the inter-hemispheric projections from AON to contralateral AON are sparser than those from AON to ipsilateral aPC. Therefore, this rough estimation $n\rhogstar\approx 400$ may inspire future experiments to get more quantitative measurement. 

Finally, we confirmed that the fine structure of OC does not change our results: with several subregions and diverse neuron types in the olfactory cortex \cite{hagiwara2012optophysiological, bekkers2013neurons}, it is likely that only a subpopulation of cortical neurons project contralaterally. While the exact fraction is unknown, we verified that this fraction does not alter the results. We denoted this fraction as $f$ and assumed that these neurons project to all the contralateral cortical neurons with equal probability, which suggests that each of these neurons needs to project to  $\dfrac{n^2\rhog}{nf}=\dfrac{n\rhog}{f}$ neurons. A natural constraint is that $f\geqslant \rhog$. 

An example of the $f=\rhog$ case is shown in Fig.~\ref{subpopulations projecting}(a): for $(m,n,\rhow,\rhog) = (20,500,0.1,0.05)$, within one olfactory cortex, only $n\times f=25$ neurons will project contralaterally and each of them has to project to all 500 neurons in the other olfactory cortex. However, even in this extreme case, the Hebbian solution $\Gdagger$ matches well with the temporal mean $\langle \GBA\rangle$ [Fig.~\ref{subpopulations projecting}(b) and the pink dot in Fig.~\ref{subpopulations projecting}(c)], and same for the analytical expression of test BAL Eq.~\eqref{Eq Weight Alignment Level} [the pink dot in Fig.~\ref{subpopulations projecting}(d)]. We also studied other values of $f $ from 0.1 to 1 with $\rhog=0.05$, and found that the results do not depend on $f$ [see Fig.~\ref{subpopulations projecting}(c)-(d)], which is expected as long as all elements in $\GBAhat,\WA,\WB$ are chosen independently (see Appendix ~\ref{The theory for Weight Alignment Level} for details). These results indicate that significant alignment can be achieved even if only a small fraction of OC neurons project contralaterally.

\subsection{Global vs local learning rules: SGD modestly outperforms Hebbian rule but obeys the same inverse scaling relation}
While the motivation of using local Hebbian learning rule was to respect the biological constraint, we next asked whether global learning rules such as stochastic gradient descent (SGD), the predominant learning algorithm in artificial neural networks, could achieve better test BAL. To investigate this question, we introduced the following global alignment loss function (other global loss functions such as the cosine similarity itself were tested and did not affect the general results):
\begin{equation}
\cl =  \dfrac{1}{2}\sum_{\mu=1}^{N} ||\rbfB^\mu-\lambda
\Gbf_{\mathrm{BA}}\rbfA^\mu||^2, 
\label{alignment loss function}
\end{equation}
where $\mu$ is the odor index and $\lambda$ is the scaling factor between ipsi- and contra-lateral responses. For each sequentially presented sample $\x^\mu$, the ``learning" is implemented by optimizing $\GBA$ via the online SGD update rule:
\begin{equation}\label{Eq SGD}
\Delta\GBA^{\mu} = \eta( \rbfB^{\mu}\rbfA^{\mu\top} - \lambda \Gbf_{\mathrm{BA}}  \rbfA^{\mu}\rbfA^{\mu\top})\odot \GBAhat.  \\
\end{equation}
This SGD rule is non-local, because in the second term, $(\GBA  \rbfA^{\mu}\rbfA^{\mu\top})_{ij} =\sum_{k}\GBAik  \rbf^{\mu}_{\mathrm{A},k}\rbf^{\mu}_{\mathrm{A},j}$ involves other synapses and neurons. Note that the first term is the same as in Hebbian update rule Eq.~\eqref{Eq the sparse Hebbian update rule}, therefore, to make SGD and Hebbian updates (and their solutions) comparable in size, we adopted $\lambda= \dfrac{\beta}{m\rhow\gamma^2 }$ (see Appendix \ref{Derivations of the SGD scaling factor lambda} for details).

We found that with global information, SGD indeed outperforms Hebbian learning but showed similar dependence on the network properties such as connection sparsity and network size. In particular, using the same parameters as the Hebbian learning simulations shown in Figs.~\ref{fig Hebbian theory matches simulations and the inverse scaling}(a), we varied $\rhog$ for different $n$ and computed the bilateral alignment level $a^{\mathrm{S}}$ for SGD. Quantitatively, we found that $a^{\mathrm{S}}(\rhog) > a^{\mathrm{H}}(\rhog)$ [compare the solid (SGD) and the dashed (Hebbian) purple lines in Fig.~\ref{fig4_SGD_alpha}(a) for $n=2000$]. However, $a^{\mathrm{S}}$ has similar dependence on $\rhog$ and $n$ as in Hebbian learning [Fig.~\ref{fig Hebbian theory matches simulations and the inverse scaling}(a)], e.g. the power law relation between $a$ and $\rhog$ for small $\rhog$ with the same exponent $0.5$ (the left blue dashed line). Furthermore, following the analysis in the previous section, we defined $\rhogstarS(n)$ as the density for SGD to achieve a given level of alignment $\astar$, i.e., $a^{\mathrm{S}}[\rhogstarS(n)]=\astar$, and found the same inverse scaling between $\rhogstarS$ and $n$ [the red line in Fig.~\ref{fig4_SGD_alpha}(b)].

Since Hebbian learning and SGD both exhibit the inverse scaling, the ratio of $\rhogstarS(n)$ to $\rhogstarH(n)$ is a constant. Therefore, to quantitatively compare Hebb's rule and SGD, we defined $\alpha(m,\rhow):=\left\langle \dfrac{\rhogstarS(n)}{\rhogstarH(n)}\right\rangle_n$, the mean ratio of required SGD inter-hemispheric connection density to that of Hebbian learning. A smaller $\alpha$ means that SGD needs fewer inter-hemispheric connections to achieve the same performance as the Hebbian learning. For $m=20, \rhow = 0.1$ used in Fig.~\ref{fig Hebbian theory matches simulations and the inverse scaling}, we found that $\alpha \approx 0.42$ (Fig.~\ref{figS_alpha_one_example}) corresponding to modestly sparser connectivity. By varying $m$ and $\rhow$, we found that $\alpha(m,\rhow)$ increases with $m$ and $\rhow$ [Fig.~\ref{fig4_SGD_alpha}(c)] and approaches $\sim 1$ at large values of $m$ and $\rhow$, meaning that Hebbian larning has almost the same performance as SGD with more OB neurons and denser feedforward projections. 

To explain the similarity between SGD and Hebb's rule for this alignment task, we compared their weight updates upon receiving the same input (training) data at the same time point along the learning trajectory. 
Specifically, the Hebbian and SGD updates at step $T$ are given by: 
\begin{equation}
    \begin{aligned} 
    \Delta\Gbf^\mathrm{H}_T &=  \eta \left( \rbfB\rbfA^\top - \beta\Gbf^\mathrm{H}_T \right)\odot \GBAhat,\\
    \Delta\Gbf^\mathrm{S}_T &=  \eta \left( \rbfB\rbfA^\top - \lambda\Gbf^\mathrm{S}_T\rbfA \rbfA^\top \right)\odot \GBAhat.
    \end{aligned}
\end{equation}
We calculated their cosine similarity, i.e., $\cos(\Delta\g^\mathrm{H}_T,\Delta\g^\mathrm{S}_T)$ where $\Delta \g$ is the vectorized $\Delta\Gbf$. We found significant overlap between the two update vectors during the entire learning process as shown in Fig.~\ref{fig5_gradient_alignment}(a). Furthermore, this update alignment level increases with $m$ and $\rhow$ [see Fig.~\ref{fig5_gradient_alignment}(b), where we averaged over $T$ and $n$ to get $\langle\cos(\Delta\g^\mathrm{H},\Delta\g^\mathrm{S})\rangle$]. We also calculated the solution alignment level, i.e., the cosine similarity between the final SGD solution at step $T=N$ and the Hebbian solution [$\langle\cos(\g^\mathrm{S},\gdagger)\rangle$ in Fig.~\ref{fig5_gradient_alignment}(c)]. We found that the solution alignment level also increases with $m$ and $\rhow$. In fact, both the solution alignment level and $\alpha$ are strongly correlated with the update alignment level [the blue and the red dots in Fig.~\ref{fig5_gradient_alignment}(d)].

To understand the origin of the update alignment and its dependence on $m$ and $\rhow$, we decomposed the weight update vectors $\dbf_1(\x):=\dgHT$ and $\dbf_2(\x):=\dgST$ as 
\begin{equation}
\begin{aligned} 
\dbf_1(\x) &= \h(\x) - \h_1(\x),\\
\dbf_2(\x) &= \h(\x) - \h_2(\x),\\
\end{aligned}
\end{equation}
where $\h(\x)=\vectorform(\eta \rbfB\rbfA^\top\odot \GBAhat)$ is the Hebbian correlation term shared by both update rules, $ \h_1(\x)=\eta \beta \g^{\mathrm{H}}_T$ is the Hebbian-specific update vector, and $\h_2(\x)=\vectorform(\eta\lambda\Gbf^{\mathrm{S}}_T\rbfA \rbfA^\top\odot \GBAhat)$ is the SGD-specific update vector, as illustrated in Fig.~\ref{fig_h_h1_h2}(a). We then evaluated the averaged cosine similarity and amplitude ratio between the shared component $\h(\x)$ and the specific components, $\h_1(\x)$ and $\h_2(\x)$, for Hebbian rule and SGD respectively, which are denoted as
\begin{equation}
    \begin{aligned}
c_1&=\langle\cos(\h(\x),\h_1(\x))\rangle_\x, 
        &&r_1 =\langle|\h(\x)|/|\h_1(\x)|\rangle_\x; \\
        c_2 &=\langle\cos(\h(\x),\h_2(\x))\rangle_\x, 
        && r_2 =\langle|\h(\x)|/|\h_2(\x)|\rangle_\x.
    \end{aligned}
\end{equation}
We found that $r_1\gg 1$ and it increases significantly with $m$ as shown in Fig.~\ref{fig_h_h1_h2}(b). This indicates that $\dbf_1(\x)$ is highly aligned with $\h(\x)$  and the alignment increases with $m$ as illustrated by Fig.~\ref{fig_h_h1_h2}(a) and shown by Fig.~\ref{figS_cos_h_h1_h2}(b) in the SM. For SGD, $r_2$ is larger than $1$ and it increases with $m$ and $\rhow$ as shown in Fig.~\ref{fig_h_h1_h2}(c), and $c_2(<1)$ decreases with $m$ and $\rhow$ [see Fig.~\ref{figS_cos_h_h1_h2}(c) in SM]. Since $r_1\gg 1$, we can approximate $\dbf_1\approx \h$ and show that $\dbf_1 \cdot \dbf_2\approx |\h|^2(1-c_2/r_2)>0$, which means that the Hebbian and SGD update vectors are positively aligned. Furthermore, as $r_2$ increases and $c_2$ decreases with $m$ and $\rhow$,  $\dbf_1(\x)$ and $\dbf_2(\x)$ align better with higher input dimension and denser feedforward projections as shown in Fig.~\ref{fig5_gradient_alignment}(b). Consequently, a higher alignment between the update vectors of the two learning rules results in a higher similarity between the two solutions [Fig.~\ref{fig5_gradient_alignment}(c)] and thus a larger required-density ratio $\alpha$ [Fig.~\ref{fig5_gradient_alignment}(d)], which means that Hebbian learning is closer to SGD in bilateral alignment performance.

\section{Discussion}
We have demonstrated that a simple online Hebbian learning rule is sufficient to align bilateral cortical representations even with sparse inter-hemispheric projections. Through numerical simulations and analytical calculations of a simplified neural network model, we discovered that to achieve a certain level of bilateral alignment, each cortical neuron (on average) must receive a fixed number of inter-hemispheric projections that is independent of the total number of cortical neurons. Consequently, the required projection density scales inversely with the number of cortical neurons. Remarkably, this scaling law also holds for a global learning rule, SGD, that optimizes a similar alignment loss function. Although SGD outperforms the local Hebbian rule in terms of achieving higher bilateral alignment levels, the difference diminishes with higher input dimensions and denser feedforward projections. Our analysis revealed that this reduction is due to a higher overlap between the Hebbian update vector and the SGD update vector.

Our model offers a mechanistic understanding of how the olfactory system achieves bilateral alignment, a mechanism that could extend to other neural systems. Additionally, our model generates several predictions or insights that are potentially testable. Firstly, the inverse scaling relation between the inter-hemispheric connectivity and the cortical neuron number could be be examined through analysis of connectomic data across different species and brain regions. In addition to this inverse scaling, Eq.~\eqref{rhogstar} predicts other quantitative and testable relations, specifically, given $n$ and $\astar$, $\rhogstar$ increases with $m$ and $1/\rhow$. Finally, our model predicts that the fine structure of OC (the fraction of subpopulation projecting contralaterally) does not affect the alignment performance or the inverse scaling.

The update alignment between SGD and Hebb's rule shares the same feature as many recently-developed biologically plausible learning algorithms for training deep neural networks, such as the feedback alignment algorthm and its variants \cite{Random_synaptic_feedback,nokland2016direct}. There, the learning signals generally aligns well with that of the backproporgation algorithm (BP) \cite{refinetti2021align,launay2019principled}. 
In our bilateral alignment task, local and unsupervised Hebb's rule exhibits overlap with the weight update vector of SGD, which optimizes a global alignment loss function. These results suggest that a biologically plausible learning algorithm can perform well if its weight update contains information of the corresponding error gradient. This insight provides a guidance to search for biologically plausible learning algorithms in a much larger admissible solution space. This is because, the algorithm does not have to be an approximation of the BP as long as its learning signal has significant overlap with that in the BP. Future studies in comparing learning dynamics with the Hebbian-like rule and the global SGD rule are necessary for developing brain-inspired efficient local learning algorithms for solving more complex problems.

The current model, aimed at simplicity and analytical tractability, inevitably simplifies biological complexity, which entails certain limitations. We discuss these limitations and possible future directions to address them.   

For the training data used in our model for alignment, we opted for IID Gaussian input to facilitate analytical tractability. However, realistic OB signals may be correlated and sparse \cite{ma2012distributed}. It is thus interesting to study whether and how correlated and sparse inputs change the test BAL formula Eq.~\eqref{Eq Weight Alignment Level} and the inverse scaling. 

Our model for the OB-OC projections is simplified as it only included the ``effective" direct connections $\WA$ and $\WB$. For analytical tractability, elements of $\WA$ and $\WB$ take randomly assigned positive or negative values, which seemingly violates Dale's principle \cite{eccles1976electrical}. However, such simplification may be justified by considering the indirect inhibitory connections through the feed-forward inhibitory neurons in the OC \cite{bekkers2013neurons,franks2011recurrent,stern_transformation_2018}, which can lead to odor-evoked suppressed cortical responses \cite{poo2009odor,Grimaud2021}. In addition, our work adopted random OB-to-OC projections, but a recent work showed that these projections exhibit some structure \cite{chen2022high}.  Studying how local inhibitory circuits and other structural constraints affect distributions of $\WA$ and $\WB$ and the bilateral alignment performance provides another avenue for future exploration.

Last but not least, our current model did not explicitly incorporate the recurrent connections within each olfactory cortex \cite{bekkers2013neurons,franks2011recurrent}. Such recurrent connections have long been postulated to play a significant role in associative memory and pattern completion: with distorted inputs, it can recover the true patterns through internal dynamics \cite{hopfield1982neural,haberly1989olfactory,bolding2020recurrent,pashkovski2020structure}. How such recurrent dynamics contributes to the bilateral alignment is unknown. One hypothesis is that given the required level of bilateral alignment, these recurrent connections can relax the structural requirement of the inter-hemispheric projections (potentially allowing for sparser or more random projections, thus easier implementation during development). Exploring this hypothesis in future models could provide valuable insights on the role of recurrent connections.

Our work may also have implications for other brain areas where representations in the two hemispheres have to be matched or integrated. In sensory regions with topographic representations, a coarse spatial template for organizing cross-hemispheric projections can potentially be genetically specified, with further refinements arising through activity-dependent mechanisms \cite{mizuno2007evidence,suarez2014balanced,lee2019functional}. However, in other brain regions with more scattered, mosaic representations, coordination is likely to occur through more flexible activity-dependent mechanisms. Commissural projections in the medial entorhinal cortex, an area strongly implicated in spatial and episodic memory, play an active role in memory retrieval \cite{caputi2022medial}. Whether this inter-hemispheric interaction requires detailed functional alignment is an interesting question for future work.  In another intriguing example, interhemispheric projections in a brain region called anterior lateral motor cortex, has been shown to play a role in working memory by helping restore population activity when perturbed \cite{li2016robust,chen2021modularity}. Remarkably, this restoration is a key hallmark of attractor networks, which may be formed flexibly during learning, potentially through Hebbian-like mechanisms in interhemispheric connections \cite{inagaki2019discrete,chen2021modularity}. Therefore, our modeling and analytical framework could be extended to these other regions and functions.

\section{Acknowledgement}
We thank Cengiz Pehlevan for helpful discussions, and Julien Grimaud for sharing the illustrations. The work of YT was partially supported by a NIH grant (R35GM131734). This work was supported by Harvard Mind Brain Behavior Graduate Student Award and a grant from NTT Research, and was partially done when Bo Liu was a summer intern at IBM Research. YT would like to thank the Center for Computational Biology at the Flatiron Institute for hospitality while a portion of this work was carried out.  

\bibliography{biblio}

\begin{thebibliography}{67}%
\makeatletter
\providecommand \@ifxundefined [1]{%
 \@ifx{#1\undefined}
}%
\providecommand \@ifnum [1]{%
 \ifnum #1\expandafter \@firstoftwo
 \else \expandafter \@secondoftwo
 \fi
}%
\providecommand \@ifx [1]{%
 \ifx #1\expandafter \@firstoftwo
 \else \expandafter \@secondoftwo
 \fi
}%
\providecommand \natexlab [1]{#1}%
\providecommand \enquote  [1]{``#1''}%
\providecommand \bibnamefont  [1]{#1}%
\providecommand \bibfnamefont [1]{#1}%
\providecommand \citenamefont [1]{#1}%
\providecommand \href@noop [0]{\@secondoftwo}%
\providecommand \href [0]{\begingroup \@sanitize@url \@href}%
\providecommand \@href[1]{\@@startlink{#1}\@@href}%
\providecommand \@@href[1]{\endgroup#1\@@endlink}%
\providecommand \@sanitize@url [0]{\catcode `\\12\catcode `\$12\catcode `\&12\catcode `\#12\catcode `\^12\catcode `\_12\catcode `\%12\relax}%
\providecommand \@@startlink[1]{}%
\providecommand \@@endlink[0]{}%
\providecommand \url  [0]{\begingroup\@sanitize@url \@url }%
\providecommand \@url [1]{\endgroup\@href {#1}{\urlprefix }}%
\providecommand \urlprefix  [0]{URL }%
\providecommand \Eprint [0]{\href }%
\providecommand \doibase [0]{https://doi.org/}%
\providecommand \selectlanguage [0]{\@gobble}%
\providecommand \bibinfo  [0]{\@secondoftwo}%
\providecommand \bibfield  [0]{\@secondoftwo}%
\providecommand \translation [1]{[#1]}%
\providecommand \BibitemOpen [0]{}%
\providecommand \bibitemStop [0]{}%
\providecommand \bibitemNoStop [0]{.\EOS\space}%
\providecommand \EOS [0]{\spacefactor3000\relax}%
\providecommand \BibitemShut  [1]{\csname bibitem#1\endcsname}%
\let\auto@bib@innerbib\@empty
\bibitem [{\citenamefont {Grothe}\ and\ \citenamefont {Pecka}(2014)}]{grothe2014natural}%
  \BibitemOpen
  \bibfield  {author} {\bibinfo {author} {\bibfnamefont {B.}~\bibnamefont {Grothe}}\ and\ \bibinfo {author} {\bibfnamefont {M.}~\bibnamefont {Pecka}},\ }\bibfield  {title} {\bibinfo {title} {The natural history of sound localization in mammals--a story of neuronal inhibition},\ }\href@noop {} {\bibfield  {journal} {\bibinfo  {journal} {Frontiers in neural circuits}\ }\textbf {\bibinfo {volume} {8}},\ \bibinfo {pages} {116} (\bibinfo {year} {2014})}\BibitemShut {NoStop}%
\bibitem [{\citenamefont {McAlpine}(2009)}]{McAlpine2009}%
  \BibitemOpen
  \bibfield  {author} {\bibinfo {author} {\bibfnamefont {D.}~\bibnamefont {McAlpine}},\ }\bibinfo {title} {Binaural pathways and processing},\ in\ \href {https://doi.org/10.1007/978-3-540-29678-2_625} {\emph {\bibinfo {booktitle} {Encyclopedia of Neuroscience}}},\ \bibinfo {editor} {edited by\ \bibinfo {editor} {\bibfnamefont {M.~D.}\ \bibnamefont {Binder}}, \bibinfo {editor} {\bibfnamefont {N.}~\bibnamefont {Hirokawa}},\ and\ \bibinfo {editor} {\bibfnamefont {U.}~\bibnamefont {Windhorst}}}\ (\bibinfo  {publisher} {Springer Berlin Heidelberg},\ \bibinfo {address} {Berlin, Heidelberg},\ \bibinfo {year} {2009})\ pp.\ \bibinfo {pages} {383--388}\BibitemShut {NoStop}%
\bibitem [{\citenamefont {Blake}\ and\ \citenamefont {Fox}(1973)}]{blake1973psychophysical}%
  \BibitemOpen
  \bibfield  {author} {\bibinfo {author} {\bibfnamefont {R.}~\bibnamefont {Blake}}\ and\ \bibinfo {author} {\bibfnamefont {R.}~\bibnamefont {Fox}},\ }\bibfield  {title} {\bibinfo {title} {The psychophysical inquiry into binocular summation},\ }\href@noop {} {\bibfield  {journal} {\bibinfo  {journal} {Perception \& psychophysics}\ }\textbf {\bibinfo {volume} {14}},\ \bibinfo {pages} {161} (\bibinfo {year} {1973})}\BibitemShut {NoStop}%
\bibitem [{\citenamefont {Cumming}\ and\ \citenamefont {DeAngelis}(2001)}]{cumming2001physiology}%
  \BibitemOpen
  \bibfield  {author} {\bibinfo {author} {\bibfnamefont {B.~G.}\ \bibnamefont {Cumming}}\ and\ \bibinfo {author} {\bibfnamefont {G.~C.}\ \bibnamefont {DeAngelis}},\ }\bibfield  {title} {\bibinfo {title} {The physiology of stereopsis},\ }\href@noop {} {\bibfield  {journal} {\bibinfo  {journal} {Annual review of neuroscience}\ }\textbf {\bibinfo {volume} {24}},\ \bibinfo {pages} {203} (\bibinfo {year} {2001})}\BibitemShut {NoStop}%
\bibitem [{\citenamefont {Rajan}\ \emph {et~al.}(2006)\citenamefont {Rajan}, \citenamefont {Clement},\ and\ \citenamefont {Bhalla}}]{rajan2006rats}%
  \BibitemOpen
  \bibfield  {author} {\bibinfo {author} {\bibfnamefont {R.}~\bibnamefont {Rajan}}, \bibinfo {author} {\bibfnamefont {J.~P.}\ \bibnamefont {Clement}},\ and\ \bibinfo {author} {\bibfnamefont {U.~S.}\ \bibnamefont {Bhalla}},\ }\bibfield  {title} {\bibinfo {title} {Rats smell in stereo},\ }\href@noop {} {\bibfield  {journal} {\bibinfo  {journal} {Science}\ }\textbf {\bibinfo {volume} {311}},\ \bibinfo {pages} {666} (\bibinfo {year} {2006})}\BibitemShut {NoStop}%
\bibitem [{\citenamefont {Porter}\ \emph {et~al.}(2005)\citenamefont {Porter}, \citenamefont {Anand}, \citenamefont {Johnson}, \citenamefont {Khan},\ and\ \citenamefont {Sobel}}]{porter2005brain}%
  \BibitemOpen
  \bibfield  {author} {\bibinfo {author} {\bibfnamefont {J.}~\bibnamefont {Porter}}, \bibinfo {author} {\bibfnamefont {T.}~\bibnamefont {Anand}}, \bibinfo {author} {\bibfnamefont {B.}~\bibnamefont {Johnson}}, \bibinfo {author} {\bibfnamefont {R.~M.}\ \bibnamefont {Khan}},\ and\ \bibinfo {author} {\bibfnamefont {N.}~\bibnamefont {Sobel}},\ }\bibfield  {title} {\bibinfo {title} {Brain mechanisms for extracting spatial information from smell},\ }\href@noop {} {\bibfield  {journal} {\bibinfo  {journal} {Neuron}\ }\textbf {\bibinfo {volume} {47}},\ \bibinfo {pages} {581} (\bibinfo {year} {2005})}\BibitemShut {NoStop}%
\bibitem [{\citenamefont {Mainland}\ \emph {et~al.}(2002)\citenamefont {Mainland}, \citenamefont {Bremner}, \citenamefont {Young}, \citenamefont {Johnson}, \citenamefont {Khan}, \citenamefont {Bensafi},\ and\ \citenamefont {Sobel}}]{mainland2002one}%
  \BibitemOpen
  \bibfield  {author} {\bibinfo {author} {\bibfnamefont {J.~D.}\ \bibnamefont {Mainland}}, \bibinfo {author} {\bibfnamefont {E.~A.}\ \bibnamefont {Bremner}}, \bibinfo {author} {\bibfnamefont {N.}~\bibnamefont {Young}}, \bibinfo {author} {\bibfnamefont {B.~N.}\ \bibnamefont {Johnson}}, \bibinfo {author} {\bibfnamefont {R.~M.}\ \bibnamefont {Khan}}, \bibinfo {author} {\bibfnamefont {M.}~\bibnamefont {Bensafi}},\ and\ \bibinfo {author} {\bibfnamefont {N.}~\bibnamefont {Sobel}},\ }\bibfield  {title} {\bibinfo {title} {One nostril knows what the other learns},\ }\href@noop {} {\bibfield  {journal} {\bibinfo  {journal} {Nature}\ }\textbf {\bibinfo {volume} {419}},\ \bibinfo {pages} {802} (\bibinfo {year} {2002})}\BibitemShut {NoStop}%
\bibitem [{\citenamefont {Rabell}\ \emph {et~al.}(2017)\citenamefont {Rabell}, \citenamefont {Mutlu}, \citenamefont {Noutel}, \citenamefont {Del~Olmo},\ and\ \citenamefont {Haesler}}]{rabell2017spontaneous}%
  \BibitemOpen
  \bibfield  {author} {\bibinfo {author} {\bibfnamefont {J.~E.}\ \bibnamefont {Rabell}}, \bibinfo {author} {\bibfnamefont {K.}~\bibnamefont {Mutlu}}, \bibinfo {author} {\bibfnamefont {J.}~\bibnamefont {Noutel}}, \bibinfo {author} {\bibfnamefont {P.~M.}\ \bibnamefont {Del~Olmo}},\ and\ \bibinfo {author} {\bibfnamefont {S.}~\bibnamefont {Haesler}},\ }\bibfield  {title} {\bibinfo {title} {Spontaneous rapid odor source localization behavior requires interhemispheric communication},\ }\href@noop {} {\bibfield  {journal} {\bibinfo  {journal} {Current Biology}\ }\textbf {\bibinfo {volume} {27}},\ \bibinfo {pages} {1542} (\bibinfo {year} {2017})}\BibitemShut {NoStop}%
\bibitem [{\citenamefont {Dalal}\ \emph {et~al.}(2020)\citenamefont {Dalal}, \citenamefont {Gupta},\ and\ \citenamefont {Haddad}}]{dalal2020bilateral}%
  \BibitemOpen
  \bibfield  {author} {\bibinfo {author} {\bibfnamefont {T.}~\bibnamefont {Dalal}}, \bibinfo {author} {\bibfnamefont {N.}~\bibnamefont {Gupta}},\ and\ \bibinfo {author} {\bibfnamefont {R.}~\bibnamefont {Haddad}},\ }\bibfield  {title} {\bibinfo {title} {Bilateral and unilateral odor processing and odor perception},\ }\href@noop {} {\bibfield  {journal} {\bibinfo  {journal} {Communications Biology}\ }\textbf {\bibinfo {volume} {3}},\ \bibinfo {pages} {150} (\bibinfo {year} {2020})}\BibitemShut {NoStop}%
\bibitem [{\citenamefont {Kikuta}\ \emph {et~al.}(2008)\citenamefont {Kikuta}, \citenamefont {Kashiwadani},\ and\ \citenamefont {Mori}}]{kikuta2008compensatory}%
  \BibitemOpen
  \bibfield  {author} {\bibinfo {author} {\bibfnamefont {S.}~\bibnamefont {Kikuta}}, \bibinfo {author} {\bibfnamefont {H.}~\bibnamefont {Kashiwadani}},\ and\ \bibinfo {author} {\bibfnamefont {K.}~\bibnamefont {Mori}},\ }\bibfield  {title} {\bibinfo {title} {Compensatory rapid switching of binasal inputs in the olfactory cortex},\ }\href@noop {} {\bibfield  {journal} {\bibinfo  {journal} {Journal of Neuroscience}\ }\textbf {\bibinfo {volume} {28}},\ \bibinfo {pages} {11989} (\bibinfo {year} {2008})}\BibitemShut {NoStop}%
\bibitem [{\citenamefont {Wilson}\ and\ \citenamefont {Sullivan}(1999)}]{wilson1999respiratory}%
  \BibitemOpen
  \bibfield  {author} {\bibinfo {author} {\bibfnamefont {D.~A.}\ \bibnamefont {Wilson}}\ and\ \bibinfo {author} {\bibfnamefont {R.~M.}\ \bibnamefont {Sullivan}},\ }\bibfield  {title} {\bibinfo {title} {Respiratory airflow pattern at the rat’s snout and an hypothesis regarding its role in olfaction},\ }\href@noop {} {\bibfield  {journal} {\bibinfo  {journal} {Physiology \& behavior}\ }\textbf {\bibinfo {volume} {66}},\ \bibinfo {pages} {41} (\bibinfo {year} {1999})}\BibitemShut {NoStop}%
\bibitem [{\citenamefont {Eccles}(2000)}]{eccles2000nasal}%
  \BibitemOpen
  \bibfield  {author} {\bibinfo {author} {\bibfnamefont {R.}~\bibnamefont {Eccles}},\ }\bibfield  {title} {\bibinfo {title} {Nasal airflow in health and disease},\ }\href@noop {} {\bibfield  {journal} {\bibinfo  {journal} {Acta oto-laryngologica}\ }\textbf {\bibinfo {volume} {120}},\ \bibinfo {pages} {580} (\bibinfo {year} {2000})}\BibitemShut {NoStop}%
\bibitem [{\citenamefont {Catania}(2013)}]{catania2013stereo}%
  \BibitemOpen
  \bibfield  {author} {\bibinfo {author} {\bibfnamefont {K.~C.}\ \bibnamefont {Catania}},\ }\bibfield  {title} {\bibinfo {title} {Stereo and serial sniffing guide navigation to an odour source in a mammal},\ }\href@noop {} {\bibfield  {journal} {\bibinfo  {journal} {Nature Communications}\ }\textbf {\bibinfo {volume} {4}},\ \bibinfo {pages} {1441} (\bibinfo {year} {2013})}\BibitemShut {NoStop}%
\bibitem [{\citenamefont {Gire}\ \emph {et~al.}(2016)\citenamefont {Gire}, \citenamefont {Kapoor}, \citenamefont {Arrighi-Allisan}, \citenamefont {Seminara},\ and\ \citenamefont {Murthy}}]{gire2016mice}%
  \BibitemOpen
  \bibfield  {author} {\bibinfo {author} {\bibfnamefont {D.~H.}\ \bibnamefont {Gire}}, \bibinfo {author} {\bibfnamefont {V.}~\bibnamefont {Kapoor}}, \bibinfo {author} {\bibfnamefont {A.}~\bibnamefont {Arrighi-Allisan}}, \bibinfo {author} {\bibfnamefont {A.}~\bibnamefont {Seminara}},\ and\ \bibinfo {author} {\bibfnamefont {V.~N.}\ \bibnamefont {Murthy}},\ }\bibfield  {title} {\bibinfo {title} {Mice develop efficient strategies for foraging and navigation using complex natural stimuli},\ }\href@noop {} {\bibfield  {journal} {\bibinfo  {journal} {Current Biology}\ }\textbf {\bibinfo {volume} {26}},\ \bibinfo {pages} {1261} (\bibinfo {year} {2016})}\BibitemShut {NoStop}%
\bibitem [{\citenamefont {Bhattacharyya}\ and\ \citenamefont {Bhalla}(2015)}]{bhattacharyya2015robust}%
  \BibitemOpen
  \bibfield  {author} {\bibinfo {author} {\bibfnamefont {U.}~\bibnamefont {Bhattacharyya}}\ and\ \bibinfo {author} {\bibfnamefont {U.~S.}\ \bibnamefont {Bhalla}},\ }\bibfield  {title} {\bibinfo {title} {Robust and rapid air-borne odor tracking without casting},\ }\href@noop {} {\bibfield  {journal} {\bibinfo  {journal} {Eneuro}\ }\textbf {\bibinfo {volume} {2}} (\bibinfo {year} {2015})}\BibitemShut {NoStop}%
\bibitem [{\citenamefont {Findley}\ \emph {et~al.}(2021)\citenamefont {Findley}, \citenamefont {Wyrick}, \citenamefont {Cramer}, \citenamefont {Brown}, \citenamefont {Holcomb}, \citenamefont {Attey}, \citenamefont {Yeh}, \citenamefont {Monasevitch}, \citenamefont {Nouboussi}, \citenamefont {Cullen} \emph {et~al.}}]{findley2021sniff}%
  \BibitemOpen
  \bibfield  {author} {\bibinfo {author} {\bibfnamefont {T.~M.}\ \bibnamefont {Findley}}, \bibinfo {author} {\bibfnamefont {D.~G.}\ \bibnamefont {Wyrick}}, \bibinfo {author} {\bibfnamefont {J.~L.}\ \bibnamefont {Cramer}}, \bibinfo {author} {\bibfnamefont {M.~A.}\ \bibnamefont {Brown}}, \bibinfo {author} {\bibfnamefont {B.}~\bibnamefont {Holcomb}}, \bibinfo {author} {\bibfnamefont {R.}~\bibnamefont {Attey}}, \bibinfo {author} {\bibfnamefont {D.}~\bibnamefont {Yeh}}, \bibinfo {author} {\bibfnamefont {E.}~\bibnamefont {Monasevitch}}, \bibinfo {author} {\bibfnamefont {N.}~\bibnamefont {Nouboussi}}, \bibinfo {author} {\bibfnamefont {I.}~\bibnamefont {Cullen}}, \emph {et~al.},\ }\bibfield  {title} {\bibinfo {title} {Sniff-synchronized, gradient-guided olfactory search by freely moving mice},\ }\href@noop {} {\bibfield  {journal} {\bibinfo  {journal} {Elife}\ }\textbf {\bibinfo {volume} {10}},\ \bibinfo {pages} {e58523} (\bibinfo {year} {2021})}\BibitemShut {NoStop}%
\bibitem [{\citenamefont {Wu}\ \emph {et~al.}(2020)\citenamefont {Wu}, \citenamefont {Chen}, \citenamefont {Ye}, \citenamefont {Zhang},\ and\ \citenamefont {Zhou}}]{wu2020humans}%
  \BibitemOpen
  \bibfield  {author} {\bibinfo {author} {\bibfnamefont {Y.}~\bibnamefont {Wu}}, \bibinfo {author} {\bibfnamefont {K.}~\bibnamefont {Chen}}, \bibinfo {author} {\bibfnamefont {Y.}~\bibnamefont {Ye}}, \bibinfo {author} {\bibfnamefont {T.}~\bibnamefont {Zhang}},\ and\ \bibinfo {author} {\bibfnamefont {W.}~\bibnamefont {Zhou}},\ }\bibfield  {title} {\bibinfo {title} {Humans navigate with stereo olfaction},\ }\href@noop {} {\bibfield  {journal} {\bibinfo  {journal} {Proceedings of the National Academy of Sciences}\ }\textbf {\bibinfo {volume} {117}},\ \bibinfo {pages} {16065} (\bibinfo {year} {2020})}\BibitemShut {NoStop}%
\bibitem [{\citenamefont {Kucharski}\ and\ \citenamefont {Hall}(1987)}]{kucharski1987new}%
  \BibitemOpen
  \bibfield  {author} {\bibinfo {author} {\bibfnamefont {D.}~\bibnamefont {Kucharski}}\ and\ \bibinfo {author} {\bibfnamefont {W.}~\bibnamefont {Hall}},\ }\bibfield  {title} {\bibinfo {title} {New routes to early memories},\ }\href@noop {} {\bibfield  {journal} {\bibinfo  {journal} {Science}\ }\textbf {\bibinfo {volume} {238}},\ \bibinfo {pages} {786} (\bibinfo {year} {1987})}\BibitemShut {NoStop}%
\bibitem [{\citenamefont {Kucharski}\ \emph {et~al.}(1990)\citenamefont {Kucharski}, \citenamefont {Burka},\ and\ \citenamefont {Hall}}]{kucharski1990anterior}%
  \BibitemOpen
  \bibfield  {author} {\bibinfo {author} {\bibfnamefont {D.}~\bibnamefont {Kucharski}}, \bibinfo {author} {\bibfnamefont {N.}~\bibnamefont {Burka}},\ and\ \bibinfo {author} {\bibfnamefont {W.}~\bibnamefont {Hall}},\ }\bibfield  {title} {\bibinfo {title} {The anterior limb of the anterior commissure is an access route to contralateral stored olfactory preference memories},\ }\href@noop {} {\bibfield  {journal} {\bibinfo  {journal} {Psychobiology}\ }\textbf {\bibinfo {volume} {18}},\ \bibinfo {pages} {195} (\bibinfo {year} {1990})}\BibitemShut {NoStop}%
\bibitem [{\citenamefont {Brunjes}\ \emph {et~al.}(2005)\citenamefont {Brunjes}, \citenamefont {Illig},\ and\ \citenamefont {Meyer}}]{brunjes2005field}%
  \BibitemOpen
  \bibfield  {author} {\bibinfo {author} {\bibfnamefont {P.~C.}\ \bibnamefont {Brunjes}}, \bibinfo {author} {\bibfnamefont {K.~R.}\ \bibnamefont {Illig}},\ and\ \bibinfo {author} {\bibfnamefont {E.~A.}\ \bibnamefont {Meyer}},\ }\bibfield  {title} {\bibinfo {title} {A field guide to the anterior olfactory nucleus (cortex)},\ }\href@noop {} {\bibfield  {journal} {\bibinfo  {journal} {Brain research reviews}\ }\textbf {\bibinfo {volume} {50}},\ \bibinfo {pages} {305} (\bibinfo {year} {2005})}\BibitemShut {NoStop}%
\bibitem [{\citenamefont {Kikuta}\ \emph {et~al.}(2010)\citenamefont {Kikuta}, \citenamefont {Sato}, \citenamefont {Kashiwadani}, \citenamefont {Tsunoda}, \citenamefont {Yamasoba},\ and\ \citenamefont {Mori}}]{kikuta2010neurons}%
  \BibitemOpen
  \bibfield  {author} {\bibinfo {author} {\bibfnamefont {S.}~\bibnamefont {Kikuta}}, \bibinfo {author} {\bibfnamefont {K.}~\bibnamefont {Sato}}, \bibinfo {author} {\bibfnamefont {H.}~\bibnamefont {Kashiwadani}}, \bibinfo {author} {\bibfnamefont {K.}~\bibnamefont {Tsunoda}}, \bibinfo {author} {\bibfnamefont {T.}~\bibnamefont {Yamasoba}},\ and\ \bibinfo {author} {\bibfnamefont {K.}~\bibnamefont {Mori}},\ }\bibfield  {title} {\bibinfo {title} {Neurons in the anterior olfactory nucleus pars externa detect right or left localization of odor sources},\ }\href@noop {} {\bibfield  {journal} {\bibinfo  {journal} {Proceedings of the National Academy of Sciences}\ }\textbf {\bibinfo {volume} {107}},\ \bibinfo {pages} {12363} (\bibinfo {year} {2010})}\BibitemShut {NoStop}%
\bibitem [{\citenamefont {Wilson}(1997)}]{wilson1997binaral}%
  \BibitemOpen
  \bibfield  {author} {\bibinfo {author} {\bibfnamefont {D.~A.}\ \bibnamefont {Wilson}},\ }\bibfield  {title} {\bibinfo {title} {Binaral interactions in the rat piriform cortex},\ }\href@noop {} {\bibfield  {journal} {\bibinfo  {journal} {Journal of neurophysiology}\ }\textbf {\bibinfo {volume} {78}},\ \bibinfo {pages} {160} (\bibinfo {year} {1997})}\BibitemShut {NoStop}%
\bibitem [{\citenamefont {Yan}\ \emph {et~al.}(2008)\citenamefont {Yan}, \citenamefont {Tan}, \citenamefont {Qin}, \citenamefont {Lu}, \citenamefont {Ding},\ and\ \citenamefont {Luo}}]{yan2008precise}%
  \BibitemOpen
  \bibfield  {author} {\bibinfo {author} {\bibfnamefont {Z.}~\bibnamefont {Yan}}, \bibinfo {author} {\bibfnamefont {J.}~\bibnamefont {Tan}}, \bibinfo {author} {\bibfnamefont {C.}~\bibnamefont {Qin}}, \bibinfo {author} {\bibfnamefont {Y.}~\bibnamefont {Lu}}, \bibinfo {author} {\bibfnamefont {C.}~\bibnamefont {Ding}},\ and\ \bibinfo {author} {\bibfnamefont {M.}~\bibnamefont {Luo}},\ }\bibfield  {title} {\bibinfo {title} {Precise circuitry links bilaterally symmetric olfactory maps},\ }\href@noop {} {\bibfield  {journal} {\bibinfo  {journal} {Neuron}\ }\textbf {\bibinfo {volume} {58}},\ \bibinfo {pages} {613} (\bibinfo {year} {2008})}\BibitemShut {NoStop}%
\bibitem [{\citenamefont {Hagiwara}\ \emph {et~al.}(2012)\citenamefont {Hagiwara}, \citenamefont {Pal}, \citenamefont {Sato}, \citenamefont {Wienisch},\ and\ \citenamefont {Murthy}}]{hagiwara2012optophysiological}%
  \BibitemOpen
  \bibfield  {author} {\bibinfo {author} {\bibfnamefont {A.}~\bibnamefont {Hagiwara}}, \bibinfo {author} {\bibfnamefont {S.~K.}\ \bibnamefont {Pal}}, \bibinfo {author} {\bibfnamefont {T.~F.}\ \bibnamefont {Sato}}, \bibinfo {author} {\bibfnamefont {M.}~\bibnamefont {Wienisch}},\ and\ \bibinfo {author} {\bibfnamefont {V.~N.}\ \bibnamefont {Murthy}},\ }\bibfield  {title} {\bibinfo {title} {Optophysiological analysis of associational circuits in the olfactory cortex},\ }\href@noop {} {\bibfield  {journal} {\bibinfo  {journal} {Frontiers in neural circuits}\ }\textbf {\bibinfo {volume} {6}},\ \bibinfo {pages} {18} (\bibinfo {year} {2012})}\BibitemShut {NoStop}%
\bibitem [{\citenamefont {Haberly}\ and\ \citenamefont {Price}(1978)}]{haberly1978association}%
  \BibitemOpen
  \bibfield  {author} {\bibinfo {author} {\bibfnamefont {L.~B.}\ \bibnamefont {Haberly}}\ and\ \bibinfo {author} {\bibfnamefont {J.~L.}\ \bibnamefont {Price}},\ }\bibfield  {title} {\bibinfo {title} {Association and commissural fiber systems of the olfactory cortex of the rat. i. systems originating in the piriform cortex and adjacent areas},\ }\href@noop {} {\bibfield  {journal} {\bibinfo  {journal} {Journal of Comparative Neurology}\ }\textbf {\bibinfo {volume} {178}},\ \bibinfo {pages} {711} (\bibinfo {year} {1978})}\BibitemShut {NoStop}%
\bibitem [{\citenamefont {Russo}\ \emph {et~al.}(2020)\citenamefont {Russo}, \citenamefont {Franks}, \citenamefont {Oghaz}, \citenamefont {Axel},\ and\ \citenamefont {Siegelbaum}}]{russo2020synaptic}%
  \BibitemOpen
  \bibfield  {author} {\bibinfo {author} {\bibfnamefont {M.~J.}\ \bibnamefont {Russo}}, \bibinfo {author} {\bibfnamefont {K.~M.}\ \bibnamefont {Franks}}, \bibinfo {author} {\bibfnamefont {R.}~\bibnamefont {Oghaz}}, \bibinfo {author} {\bibfnamefont {R.}~\bibnamefont {Axel}},\ and\ \bibinfo {author} {\bibfnamefont {S.~A.}\ \bibnamefont {Siegelbaum}},\ }\bibfield  {title} {\bibinfo {title} {Synaptic organization of anterior olfactory nucleus inputs to piriform cortex},\ }\href@noop {} {\bibfield  {journal} {\bibinfo  {journal} {Journal of Neuroscience}\ }\textbf {\bibinfo {volume} {40}},\ \bibinfo {pages} {9414} (\bibinfo {year} {2020})}\BibitemShut {NoStop}%
\bibitem [{\citenamefont {Martin-Lopez}\ \emph {et~al.}(2018)\citenamefont {Martin-Lopez}, \citenamefont {Meller},\ and\ \citenamefont {Greer}}]{martin2018development}%
  \BibitemOpen
  \bibfield  {author} {\bibinfo {author} {\bibfnamefont {E.}~\bibnamefont {Martin-Lopez}}, \bibinfo {author} {\bibfnamefont {S.~J.}\ \bibnamefont {Meller}},\ and\ \bibinfo {author} {\bibfnamefont {C.~A.}\ \bibnamefont {Greer}},\ }\bibfield  {title} {\bibinfo {title} {Development of piriform cortex interhemispheric connections via the anterior commissure: progressive and regressive strategies},\ }\href@noop {} {\bibfield  {journal} {\bibinfo  {journal} {Brain Structure and Function}\ }\textbf {\bibinfo {volume} {223}},\ \bibinfo {pages} {4067} (\bibinfo {year} {2018})}\BibitemShut {NoStop}%
\bibitem [{\citenamefont {Dike{\c{c}}ligil}\ \emph {et~al.}(2023)\citenamefont {Dike{\c{c}}ligil}, \citenamefont {Yang}, \citenamefont {Sanghani}, \citenamefont {Lucas}, \citenamefont {Chen}, \citenamefont {Davis},\ and\ \citenamefont {Gottfried}}]{dikeccligil2023odor}%
  \BibitemOpen
  \bibfield  {author} {\bibinfo {author} {\bibfnamefont {G.~N.}\ \bibnamefont {Dike{\c{c}}ligil}}, \bibinfo {author} {\bibfnamefont {A.~I.}\ \bibnamefont {Yang}}, \bibinfo {author} {\bibfnamefont {N.}~\bibnamefont {Sanghani}}, \bibinfo {author} {\bibfnamefont {T.}~\bibnamefont {Lucas}}, \bibinfo {author} {\bibfnamefont {H.~I.}\ \bibnamefont {Chen}}, \bibinfo {author} {\bibfnamefont {K.~A.}\ \bibnamefont {Davis}},\ and\ \bibinfo {author} {\bibfnamefont {J.~A.}\ \bibnamefont {Gottfried}},\ }\bibfield  {title} {\bibinfo {title} {Odor representations from the two nostrils are temporally segregated in human piriform cortex},\ }\href@noop {} {\bibfield  {journal} {\bibinfo  {journal} {Current Biology}\ } (\bibinfo {year} {2023})}\BibitemShut {NoStop}%
\bibitem [{\citenamefont {Grimaud}\ \emph {et~al.}(2021)\citenamefont {Grimaud}, \citenamefont {Dorrell}, \citenamefont {Pehlevan},\ and\ \citenamefont {Murthy}}]{Grimaud2021}%
  \BibitemOpen
  \bibfield  {author} {\bibinfo {author} {\bibfnamefont {J.}~\bibnamefont {Grimaud}}, \bibinfo {author} {\bibfnamefont {W.}~\bibnamefont {Dorrell}}, \bibinfo {author} {\bibfnamefont {C.}~\bibnamefont {Pehlevan}},\ and\ \bibinfo {author} {\bibfnamefont {V.}~\bibnamefont {Murthy}},\ }\bibfield  {title} {\bibinfo {title} {Bilateral alignment of receptive fields in the olfactory cortex},\ }\href@noop {} {\bibfield  {journal} {\bibinfo  {journal} {bioRxiv}\ } (\bibinfo {year} {2021})}\BibitemShut {NoStop}%
\bibitem [{\citenamefont {Haberly}(2001)}]{haberly2001parallel}%
  \BibitemOpen
  \bibfield  {author} {\bibinfo {author} {\bibfnamefont {L.~B.}\ \bibnamefont {Haberly}},\ }\bibfield  {title} {\bibinfo {title} {Parallel-distributed processing in olfactory cortex: new insights from morphological and physiological analysis of neuronal circuitry},\ }\href@noop {} {\bibfield  {journal} {\bibinfo  {journal} {Chemical senses}\ }\textbf {\bibinfo {volume} {26}},\ \bibinfo {pages} {551} (\bibinfo {year} {2001})}\BibitemShut {NoStop}%
\bibitem [{\citenamefont {Sosulski}\ \emph {et~al.}(2011)\citenamefont {Sosulski}, \citenamefont {Bloom}, \citenamefont {Cutforth}, \citenamefont {Axel},\ and\ \citenamefont {Datta}}]{sosulski2011distinct}%
  \BibitemOpen
  \bibfield  {author} {\bibinfo {author} {\bibfnamefont {D.~L.}\ \bibnamefont {Sosulski}}, \bibinfo {author} {\bibfnamefont {M.~L.}\ \bibnamefont {Bloom}}, \bibinfo {author} {\bibfnamefont {T.}~\bibnamefont {Cutforth}}, \bibinfo {author} {\bibfnamefont {R.}~\bibnamefont {Axel}},\ and\ \bibinfo {author} {\bibfnamefont {S.~R.}\ \bibnamefont {Datta}},\ }\bibfield  {title} {\bibinfo {title} {Distinct representations of olfactory information in different cortical centres},\ }\href@noop {} {\bibfield  {journal} {\bibinfo  {journal} {Nature}\ }\textbf {\bibinfo {volume} {472}},\ \bibinfo {pages} {213} (\bibinfo {year} {2011})}\BibitemShut {NoStop}%
\bibitem [{\citenamefont {Miyamichi}\ \emph {et~al.}(2011)\citenamefont {Miyamichi}, \citenamefont {Amat}, \citenamefont {Moussavi}, \citenamefont {Wang}, \citenamefont {Wickersham}, \citenamefont {Wall}, \citenamefont {Taniguchi}, \citenamefont {Tasic}, \citenamefont {Huang}, \citenamefont {He} \emph {et~al.}}]{miyamichi2011cortical}%
  \BibitemOpen
  \bibfield  {author} {\bibinfo {author} {\bibfnamefont {K.}~\bibnamefont {Miyamichi}}, \bibinfo {author} {\bibfnamefont {F.}~\bibnamefont {Amat}}, \bibinfo {author} {\bibfnamefont {F.}~\bibnamefont {Moussavi}}, \bibinfo {author} {\bibfnamefont {C.}~\bibnamefont {Wang}}, \bibinfo {author} {\bibfnamefont {I.}~\bibnamefont {Wickersham}}, \bibinfo {author} {\bibfnamefont {N.~R.}\ \bibnamefont {Wall}}, \bibinfo {author} {\bibfnamefont {H.}~\bibnamefont {Taniguchi}}, \bibinfo {author} {\bibfnamefont {B.}~\bibnamefont {Tasic}}, \bibinfo {author} {\bibfnamefont {Z.~J.}\ \bibnamefont {Huang}}, \bibinfo {author} {\bibfnamefont {Z.}~\bibnamefont {He}}, \emph {et~al.},\ }\bibfield  {title} {\bibinfo {title} {Cortical representations of olfactory input by trans-synaptic tracing},\ }\href@noop {} {\bibfield  {journal} {\bibinfo  {journal} {Nature}\ }\textbf {\bibinfo {volume} {472}},\ \bibinfo {pages} {191} (\bibinfo {year} {2011})}\BibitemShut {NoStop}%
\bibitem [{\citenamefont {Ghosh}\ \emph {et~al.}(2011)\citenamefont {Ghosh}, \citenamefont {Larson}, \citenamefont {Hefzi}, \citenamefont {Marnoy}, \citenamefont {Cutforth}, \citenamefont {Dokka},\ and\ \citenamefont {Baldwin}}]{ghosh2011sensory}%
  \BibitemOpen
  \bibfield  {author} {\bibinfo {author} {\bibfnamefont {S.}~\bibnamefont {Ghosh}}, \bibinfo {author} {\bibfnamefont {S.~D.}\ \bibnamefont {Larson}}, \bibinfo {author} {\bibfnamefont {H.}~\bibnamefont {Hefzi}}, \bibinfo {author} {\bibfnamefont {Z.}~\bibnamefont {Marnoy}}, \bibinfo {author} {\bibfnamefont {T.}~\bibnamefont {Cutforth}}, \bibinfo {author} {\bibfnamefont {K.}~\bibnamefont {Dokka}},\ and\ \bibinfo {author} {\bibfnamefont {K.~K.}\ \bibnamefont {Baldwin}},\ }\bibfield  {title} {\bibinfo {title} {Sensory maps in the olfactory cortex defined by long-range viral tracing of single neurons},\ }\href@noop {} {\bibfield  {journal} {\bibinfo  {journal} {Nature}\ }\textbf {\bibinfo {volume} {472}},\ \bibinfo {pages} {217} (\bibinfo {year} {2011})}\BibitemShut {NoStop}%
\bibitem [{\citenamefont {Schaffer}\ \emph {et~al.}(2018)\citenamefont {Schaffer}, \citenamefont {Stettler}, \citenamefont {Kato}, \citenamefont {Choi}, \citenamefont {Axel},\ and\ \citenamefont {Abbott}}]{schaffer2018odor}%
  \BibitemOpen
  \bibfield  {author} {\bibinfo {author} {\bibfnamefont {E.~S.}\ \bibnamefont {Schaffer}}, \bibinfo {author} {\bibfnamefont {D.~D.}\ \bibnamefont {Stettler}}, \bibinfo {author} {\bibfnamefont {D.}~\bibnamefont {Kato}}, \bibinfo {author} {\bibfnamefont {G.~B.}\ \bibnamefont {Choi}}, \bibinfo {author} {\bibfnamefont {R.}~\bibnamefont {Axel}},\ and\ \bibinfo {author} {\bibfnamefont {L.}~\bibnamefont {Abbott}},\ }\bibfield  {title} {\bibinfo {title} {Odor perception on the two sides of the brain: consistency despite randomness},\ }\href@noop {} {\bibfield  {journal} {\bibinfo  {journal} {Neuron}\ }\textbf {\bibinfo {volume} {98}} (\bibinfo {year} {2018})}\BibitemShut {NoStop}%
\bibitem [{\citenamefont {Srinivasan}\ and\ \citenamefont {Stevens}(2019)}]{srinivasan2019scaling}%
  \BibitemOpen
  \bibfield  {author} {\bibinfo {author} {\bibfnamefont {S.}~\bibnamefont {Srinivasan}}\ and\ \bibinfo {author} {\bibfnamefont {C.~F.}\ \bibnamefont {Stevens}},\ }\bibfield  {title} {\bibinfo {title} {Scaling principles of distributed circuits},\ }\href@noop {} {\bibfield  {journal} {\bibinfo  {journal} {Current Biology}\ }\textbf {\bibinfo {volume} {29}} (\bibinfo {year} {2019})}\BibitemShut {NoStop}%
\bibitem [{\citenamefont {Cragg}(1967)}]{cragg1967density}%
  \BibitemOpen
  \bibfield  {author} {\bibinfo {author} {\bibfnamefont {B.}~\bibnamefont {Cragg}},\ }\bibfield  {title} {\bibinfo {title} {The density of synapses and neurones in the motor and visual areas of the cerebral cortex.},\ }\href@noop {} {\bibfield  {journal} {\bibinfo  {journal} {Journal of Anatomy}\ }\textbf {\bibinfo {volume} {101}},\ \bibinfo {pages} {639} (\bibinfo {year} {1967})}\BibitemShut {NoStop}%
\bibitem [{\citenamefont {Sch{\"u}z}\ and\ \citenamefont {Palm}(1989)}]{schuz1989density}%
  \BibitemOpen
  \bibfield  {author} {\bibinfo {author} {\bibfnamefont {A.}~\bibnamefont {Sch{\"u}z}}\ and\ \bibinfo {author} {\bibfnamefont {G.}~\bibnamefont {Palm}},\ }\bibfield  {title} {\bibinfo {title} {Density of neurons and synapses in the cerebral cortex of the mouse},\ }\href@noop {} {\bibfield  {journal} {\bibinfo  {journal} {Journal of Comparative Neurology}\ }\textbf {\bibinfo {volume} {286}},\ \bibinfo {pages} {442} (\bibinfo {year} {1989})}\BibitemShut {NoStop}%
\bibitem [{\citenamefont {Wang}\ \emph {et~al.}(2020)\citenamefont {Wang}, \citenamefont {Zhang}, \citenamefont {Chen}, \citenamefont {Manyande}, \citenamefont {Haddad}, \citenamefont {Liu},\ and\ \citenamefont {Xu}}]{wang2020cell}%
  \BibitemOpen
  \bibfield  {author} {\bibinfo {author} {\bibfnamefont {L.}~\bibnamefont {Wang}}, \bibinfo {author} {\bibfnamefont {Z.}~\bibnamefont {Zhang}}, \bibinfo {author} {\bibfnamefont {J.}~\bibnamefont {Chen}}, \bibinfo {author} {\bibfnamefont {A.}~\bibnamefont {Manyande}}, \bibinfo {author} {\bibfnamefont {R.}~\bibnamefont {Haddad}}, \bibinfo {author} {\bibfnamefont {Q.}~\bibnamefont {Liu}},\ and\ \bibinfo {author} {\bibfnamefont {F.}~\bibnamefont {Xu}},\ }\bibfield  {title} {\bibinfo {title} {Cell-type-specific whole-brain direct inputs to the anterior and posterior piriform cortex},\ }\href@noop {} {\bibfield  {journal} {\bibinfo  {journal} {Frontiers in neural circuits}\ }\textbf {\bibinfo {volume} {14}},\ \bibinfo {pages} {4} (\bibinfo {year} {2020})}\BibitemShut {NoStop}%
\bibitem [{\citenamefont {Murthy}(2011)}]{murthy2011olfactory}%
  \BibitemOpen
  \bibfield  {author} {\bibinfo {author} {\bibfnamefont {V.~N.}\ \bibnamefont {Murthy}},\ }\bibfield  {title} {\bibinfo {title} {Olfactory maps in the brain},\ }\href@noop {} {\bibfield  {journal} {\bibinfo  {journal} {Annual review of neuroscience}\ }\textbf {\bibinfo {volume} {34}},\ \bibinfo {pages} {233} (\bibinfo {year} {2011})}\BibitemShut {NoStop}%
\bibitem [{\citenamefont {Uchida}\ \emph {et~al.}(2014)\citenamefont {Uchida}, \citenamefont {Poo},\ and\ \citenamefont {Haddad}}]{uchida2014coding}%
  \BibitemOpen
  \bibfield  {author} {\bibinfo {author} {\bibfnamefont {N.}~\bibnamefont {Uchida}}, \bibinfo {author} {\bibfnamefont {C.}~\bibnamefont {Poo}},\ and\ \bibinfo {author} {\bibfnamefont {R.}~\bibnamefont {Haddad}},\ }\bibfield  {title} {\bibinfo {title} {Coding and transformations in the olfactory system},\ }\href@noop {} {\bibfield  {journal} {\bibinfo  {journal} {Annual review of neuroscience}\ }\textbf {\bibinfo {volume} {37}},\ \bibinfo {pages} {363} (\bibinfo {year} {2014})}\BibitemShut {NoStop}%
\bibitem [{\citenamefont {Babadi}\ and\ \citenamefont {Sompolinsky}(2014)}]{babadi2014sparseness}%
  \BibitemOpen
  \bibfield  {author} {\bibinfo {author} {\bibfnamefont {B.}~\bibnamefont {Babadi}}\ and\ \bibinfo {author} {\bibfnamefont {H.}~\bibnamefont {Sompolinsky}},\ }\bibfield  {title} {\bibinfo {title} {Sparseness and expansion in sensory representations},\ }\href@noop {} {\bibfield  {journal} {\bibinfo  {journal} {Neuron}\ }\textbf {\bibinfo {volume} {83}},\ \bibinfo {pages} {1213} (\bibinfo {year} {2014})}\BibitemShut {NoStop}%
\bibitem [{\citenamefont {Uhlenbeck}\ and\ \citenamefont {Ornstein}(1930)}]{OU_Process}%
  \BibitemOpen
  \bibfield  {author} {\bibinfo {author} {\bibfnamefont {G.~E.}\ \bibnamefont {Uhlenbeck}}\ and\ \bibinfo {author} {\bibfnamefont {L.~S.}\ \bibnamefont {Ornstein}},\ }\bibfield  {title} {\bibinfo {title} {On the theory of the brownian motion},\ }\href {https://doi.org/10.1103/PhysRev.36.823} {\bibfield  {journal} {\bibinfo  {journal} {Phys. Rev.}\ }\textbf {\bibinfo {volume} {36}},\ \bibinfo {pages} {823} (\bibinfo {year} {1930})}\BibitemShut {NoStop}%
\bibitem [{\citenamefont {Srinivasan}\ and\ \citenamefont {Stevens}(2018)}]{srinivasan2018distributed}%
  \BibitemOpen
  \bibfield  {author} {\bibinfo {author} {\bibfnamefont {S.}~\bibnamefont {Srinivasan}}\ and\ \bibinfo {author} {\bibfnamefont {C.~F.}\ \bibnamefont {Stevens}},\ }\bibfield  {title} {\bibinfo {title} {The distributed circuit within the piriform cortex makes odor discrimination robust},\ }\href@noop {} {\bibfield  {journal} {\bibinfo  {journal} {Journal of Comparative Neurology}\ }\textbf {\bibinfo {volume} {526}},\ \bibinfo {pages} {2725} (\bibinfo {year} {2018})}\BibitemShut {NoStop}%
\bibitem [{\citenamefont {Richard}\ \emph {et~al.}(2010)\citenamefont {Richard}, \citenamefont {Taylor},\ and\ \citenamefont {Greer}}]{10.1073/pnas.1007931107}%
  \BibitemOpen
  \bibfield  {author} {\bibinfo {author} {\bibfnamefont {M.~B.}\ \bibnamefont {Richard}}, \bibinfo {author} {\bibfnamefont {S.~R.}\ \bibnamefont {Taylor}},\ and\ \bibinfo {author} {\bibfnamefont {C.~A.}\ \bibnamefont {Greer}},\ }\bibfield  {title} {\bibinfo {title} {{Age-induced disruption of selective olfactory bulb synaptic circuits}},\ }\href {https://doi.org/10.1073/pnas.1007931107} {\bibfield  {journal} {\bibinfo  {journal} {Proceedings of the National Academy of Sciences}\ }\textbf {\bibinfo {volume} {107}},\ \bibinfo {pages} {15613} (\bibinfo {year} {2010})}\BibitemShut {NoStop}%
\bibitem [{\citenamefont {Bekkers}\ and\ \citenamefont {Suzuki}(2013)}]{bekkers2013neurons}%
  \BibitemOpen
  \bibfield  {author} {\bibinfo {author} {\bibfnamefont {J.~M.}\ \bibnamefont {Bekkers}}\ and\ \bibinfo {author} {\bibfnamefont {N.}~\bibnamefont {Suzuki}},\ }\bibfield  {title} {\bibinfo {title} {Neurons and circuits for odor processing in the piriform cortex},\ }\href@noop {} {\bibfield  {journal} {\bibinfo  {journal} {Trends in neurosciences}\ }\textbf {\bibinfo {volume} {36}},\ \bibinfo {pages} {429} (\bibinfo {year} {2013})}\BibitemShut {NoStop}%
\bibitem [{\citenamefont {Lillicrap}\ \emph {et~al.}(2016)\citenamefont {Lillicrap}, \citenamefont {Cownden}, \citenamefont {Tweed},\ and\ \citenamefont {Akerman}}]{Random_synaptic_feedback}%
  \BibitemOpen
  \bibfield  {author} {\bibinfo {author} {\bibfnamefont {T.~P.}\ \bibnamefont {Lillicrap}}, \bibinfo {author} {\bibfnamefont {D.}~\bibnamefont {Cownden}}, \bibinfo {author} {\bibfnamefont {D.~B.}\ \bibnamefont {Tweed}},\ and\ \bibinfo {author} {\bibfnamefont {C.~J.}\ \bibnamefont {Akerman}},\ }\bibfield  {title} {\bibinfo {title} {{Random synaptic feedback weights support error backpropagation for deep learning}},\ }\href {https://doi.org/10.1038/ncomms13276} {\bibfield  {journal} {\bibinfo  {journal} {Nature Communications}\ }\textbf {\bibinfo {volume} {7}},\ \bibinfo {pages} {13276} (\bibinfo {year} {2016})}\BibitemShut {NoStop}%
\bibitem [{\citenamefont {N{\o}kland}(2016)}]{nokland2016direct}%
  \BibitemOpen
  \bibfield  {author} {\bibinfo {author} {\bibfnamefont {A.}~\bibnamefont {N{\o}kland}},\ }\bibfield  {title} {\bibinfo {title} {Direct feedback alignment provides learning in deep neural networks},\ }\href@noop {} {\bibfield  {journal} {\bibinfo  {journal} {Advances in neural information processing systems}\ }\textbf {\bibinfo {volume} {29}} (\bibinfo {year} {2016})}\BibitemShut {NoStop}%
\bibitem [{\citenamefont {Refinetti}\ \emph {et~al.}(2021)\citenamefont {Refinetti}, \citenamefont {d’Ascoli}, \citenamefont {Ohana},\ and\ \citenamefont {Goldt}}]{refinetti2021align}%
  \BibitemOpen
  \bibfield  {author} {\bibinfo {author} {\bibfnamefont {M.}~\bibnamefont {Refinetti}}, \bibinfo {author} {\bibfnamefont {S.}~\bibnamefont {d’Ascoli}}, \bibinfo {author} {\bibfnamefont {R.}~\bibnamefont {Ohana}},\ and\ \bibinfo {author} {\bibfnamefont {S.}~\bibnamefont {Goldt}},\ }\bibfield  {title} {\bibinfo {title} {Align, then memorise: the dynamics of learning with feedback alignment},\ }in\ \href@noop {} {\emph {\bibinfo {booktitle} {International Conference on Machine Learning}}}\ (\bibinfo {organization} {PMLR},\ \bibinfo {year} {2021})\ pp.\ \bibinfo {pages} {8925--8935}\BibitemShut {NoStop}%
\bibitem [{\citenamefont {Launay}\ \emph {et~al.}(2019)\citenamefont {Launay}, \citenamefont {Poli},\ and\ \citenamefont {Krzakala}}]{launay2019principled}%
  \BibitemOpen
  \bibfield  {author} {\bibinfo {author} {\bibfnamefont {J.}~\bibnamefont {Launay}}, \bibinfo {author} {\bibfnamefont {I.}~\bibnamefont {Poli}},\ and\ \bibinfo {author} {\bibfnamefont {F.}~\bibnamefont {Krzakala}},\ }\bibfield  {title} {\bibinfo {title} {Principled training of neural networks with direct feedback alignment},\ }\href@noop {} {\bibfield  {journal} {\bibinfo  {journal} {arXiv preprint arXiv:1906.04554}\ } (\bibinfo {year} {2019})}\BibitemShut {NoStop}%
\bibitem [{\citenamefont {Ma}\ \emph {et~al.}(2012)\citenamefont {Ma}, \citenamefont {Qiu}, \citenamefont {Gradwohl}, \citenamefont {Scott}, \citenamefont {Yu}, \citenamefont {Alexander}, \citenamefont {Wiegraebe},\ and\ \citenamefont {Yu}}]{ma2012distributed}%
  \BibitemOpen
  \bibfield  {author} {\bibinfo {author} {\bibfnamefont {L.}~\bibnamefont {Ma}}, \bibinfo {author} {\bibfnamefont {Q.}~\bibnamefont {Qiu}}, \bibinfo {author} {\bibfnamefont {S.}~\bibnamefont {Gradwohl}}, \bibinfo {author} {\bibfnamefont {A.}~\bibnamefont {Scott}}, \bibinfo {author} {\bibfnamefont {E.~Q.}\ \bibnamefont {Yu}}, \bibinfo {author} {\bibfnamefont {R.}~\bibnamefont {Alexander}}, \bibinfo {author} {\bibfnamefont {W.}~\bibnamefont {Wiegraebe}},\ and\ \bibinfo {author} {\bibfnamefont {C.~R.}\ \bibnamefont {Yu}},\ }\bibfield  {title} {\bibinfo {title} {Distributed representation of chemical features and tunotopic organization of glomeruli in the mouse olfactory bulb},\ }\href@noop {} {\bibfield  {journal} {\bibinfo  {journal} {Proceedings of the National Academy of Sciences}\ }\textbf {\bibinfo {volume} {109}},\ \bibinfo {pages} {5481} (\bibinfo {year} {2012})}\BibitemShut {NoStop}%
\bibitem [{\citenamefont {Eccles}(1976)}]{eccles1976electrical}%
  \BibitemOpen
  \bibfield  {author} {\bibinfo {author} {\bibfnamefont {J.~C.}\ \bibnamefont {Eccles}},\ }\bibfield  {title} {\bibinfo {title} {From electrical to chemical transmission in the central nervous system: the closing address of the sir henry dale centennial symposium cambridge, 19 september 1975},\ }\href@noop {} {\bibfield  {journal} {\bibinfo  {journal} {Notes and records of the Royal Society of London}\ }\textbf {\bibinfo {volume} {30}},\ \bibinfo {pages} {219} (\bibinfo {year} {1976})}\BibitemShut {NoStop}%
\bibitem [{\citenamefont {Franks}\ \emph {et~al.}(2011)\citenamefont {Franks}, \citenamefont {Russo}, \citenamefont {Sosulski}, \citenamefont {Mulligan}, \citenamefont {Siegelbaum},\ and\ \citenamefont {Axel}}]{franks2011recurrent}%
  \BibitemOpen
  \bibfield  {author} {\bibinfo {author} {\bibfnamefont {K.~M.}\ \bibnamefont {Franks}}, \bibinfo {author} {\bibfnamefont {M.~J.}\ \bibnamefont {Russo}}, \bibinfo {author} {\bibfnamefont {D.~L.}\ \bibnamefont {Sosulski}}, \bibinfo {author} {\bibfnamefont {A.~A.}\ \bibnamefont {Mulligan}}, \bibinfo {author} {\bibfnamefont {S.~A.}\ \bibnamefont {Siegelbaum}},\ and\ \bibinfo {author} {\bibfnamefont {R.}~\bibnamefont {Axel}},\ }\bibfield  {title} {\bibinfo {title} {Recurrent circuitry dynamically shapes the activation of piriform cortex},\ }\href@noop {} {\bibfield  {journal} {\bibinfo  {journal} {Neuron}\ }\textbf {\bibinfo {volume} {72}},\ \bibinfo {pages} {49} (\bibinfo {year} {2011})}\BibitemShut {NoStop}%
\bibitem [{\citenamefont {Stern}\ \emph {et~al.}(2018)\citenamefont {Stern}, \citenamefont {Bolding}, \citenamefont {Abbott},\ and\ \citenamefont {Franks}}]{stern_transformation_2018}%
  \BibitemOpen
  \bibfield  {author} {\bibinfo {author} {\bibfnamefont {M.}~\bibnamefont {Stern}}, \bibinfo {author} {\bibfnamefont {K.~A.}\ \bibnamefont {Bolding}}, \bibinfo {author} {\bibfnamefont {L.}~\bibnamefont {Abbott}},\ and\ \bibinfo {author} {\bibfnamefont {K.~M.}\ \bibnamefont {Franks}},\ }\bibfield  {title} {\bibinfo {title} {A transformation from temporal to ensemble coding in a model of piriform cortex},\ }\href {https://doi.org/10.7554/eLife.34831} {\bibfield  {journal} {\bibinfo  {journal} {eLife}\ }\textbf {\bibinfo {volume} {7}},\ \bibinfo {pages} {e34831} (\bibinfo {year} {2018})},\ \bibinfo {note} {publisher: eLife Sciences Publications, Ltd}\BibitemShut {NoStop}%
\bibitem [{\citenamefont {Poo}\ and\ \citenamefont {Isaacson}(2009)}]{poo2009odor}%
  \BibitemOpen
  \bibfield  {author} {\bibinfo {author} {\bibfnamefont {C.}~\bibnamefont {Poo}}\ and\ \bibinfo {author} {\bibfnamefont {J.~S.}\ \bibnamefont {Isaacson}},\ }\bibfield  {title} {\bibinfo {title} {Odor representations in olfactory cortex:“sparse” coding, global inhibition, and oscillations},\ }\href@noop {} {\bibfield  {journal} {\bibinfo  {journal} {Neuron}\ }\textbf {\bibinfo {volume} {62}},\ \bibinfo {pages} {850} (\bibinfo {year} {2009})}\BibitemShut {NoStop}%
\bibitem [{\citenamefont {Chen}\ \emph {et~al.}(2022)\citenamefont {Chen}, \citenamefont {Chen}, \citenamefont {Baserdem}, \citenamefont {Zhan}, \citenamefont {Li}, \citenamefont {Davis}, \citenamefont {Kebschull}, \citenamefont {Zador}, \citenamefont {Koulakov},\ and\ \citenamefont {Albeanu}}]{chen2022high}%
  \BibitemOpen
  \bibfield  {author} {\bibinfo {author} {\bibfnamefont {Y.}~\bibnamefont {Chen}}, \bibinfo {author} {\bibfnamefont {X.}~\bibnamefont {Chen}}, \bibinfo {author} {\bibfnamefont {B.}~\bibnamefont {Baserdem}}, \bibinfo {author} {\bibfnamefont {H.}~\bibnamefont {Zhan}}, \bibinfo {author} {\bibfnamefont {Y.}~\bibnamefont {Li}}, \bibinfo {author} {\bibfnamefont {M.~B.}\ \bibnamefont {Davis}}, \bibinfo {author} {\bibfnamefont {J.~M.}\ \bibnamefont {Kebschull}}, \bibinfo {author} {\bibfnamefont {A.~M.}\ \bibnamefont {Zador}}, \bibinfo {author} {\bibfnamefont {A.~A.}\ \bibnamefont {Koulakov}},\ and\ \bibinfo {author} {\bibfnamefont {D.~F.}\ \bibnamefont {Albeanu}},\ }\bibfield  {title} {\bibinfo {title} {High-throughput sequencing of single neuron projections reveals spatial organization in the olfactory cortex},\ }\href@noop {} {\bibfield  {journal} {\bibinfo  {journal} {Cell}\ }\textbf {\bibinfo {volume} {185}} (\bibinfo {year} {2022})}\BibitemShut {NoStop}%
\bibitem [{\citenamefont {Hopfield}(1982)}]{hopfield1982neural}%
  \BibitemOpen
  \bibfield  {author} {\bibinfo {author} {\bibfnamefont {J.~J.}\ \bibnamefont {Hopfield}},\ }\bibfield  {title} {\bibinfo {title} {Neural networks and physical systems with emergent collective computational abilities.},\ }\href@noop {} {\bibfield  {journal} {\bibinfo  {journal} {Proceedings of the national academy of sciences}\ }\textbf {\bibinfo {volume} {79}},\ \bibinfo {pages} {2554} (\bibinfo {year} {1982})}\BibitemShut {NoStop}%
\bibitem [{\citenamefont {Haberly}\ and\ \citenamefont {Bower}(1989)}]{haberly1989olfactory}%
  \BibitemOpen
  \bibfield  {author} {\bibinfo {author} {\bibfnamefont {L.~B.}\ \bibnamefont {Haberly}}\ and\ \bibinfo {author} {\bibfnamefont {J.~M.}\ \bibnamefont {Bower}},\ }\bibfield  {title} {\bibinfo {title} {Olfactory cortex: model circuit for study of associative memory?},\ }\href@noop {} {\bibfield  {journal} {\bibinfo  {journal} {Trends in neurosciences}\ }\textbf {\bibinfo {volume} {12}},\ \bibinfo {pages} {258} (\bibinfo {year} {1989})}\BibitemShut {NoStop}%
\bibitem [{\citenamefont {Bolding}\ \emph {et~al.}(2020)\citenamefont {Bolding}, \citenamefont {Nagappan}, \citenamefont {Han}, \citenamefont {Wang},\ and\ \citenamefont {Franks}}]{bolding2020recurrent}%
  \BibitemOpen
  \bibfield  {author} {\bibinfo {author} {\bibfnamefont {K.~A.}\ \bibnamefont {Bolding}}, \bibinfo {author} {\bibfnamefont {S.}~\bibnamefont {Nagappan}}, \bibinfo {author} {\bibfnamefont {B.-X.}\ \bibnamefont {Han}}, \bibinfo {author} {\bibfnamefont {F.}~\bibnamefont {Wang}},\ and\ \bibinfo {author} {\bibfnamefont {K.~M.}\ \bibnamefont {Franks}},\ }\bibfield  {title} {\bibinfo {title} {Recurrent circuitry is required to stabilize piriform cortex odor representations across brain states},\ }\href@noop {} {\bibfield  {journal} {\bibinfo  {journal} {Elife}\ }\textbf {\bibinfo {volume} {9}},\ \bibinfo {pages} {e53125} (\bibinfo {year} {2020})}\BibitemShut {NoStop}%
\bibitem [{\citenamefont {Pashkovski}\ \emph {et~al.}(2020)\citenamefont {Pashkovski}, \citenamefont {Iurilli}, \citenamefont {Brann}, \citenamefont {Chicharro}, \citenamefont {Drummey}, \citenamefont {Franks}, \citenamefont {Panzeri},\ and\ \citenamefont {Datta}}]{pashkovski2020structure}%
  \BibitemOpen
  \bibfield  {author} {\bibinfo {author} {\bibfnamefont {S.~L.}\ \bibnamefont {Pashkovski}}, \bibinfo {author} {\bibfnamefont {G.}~\bibnamefont {Iurilli}}, \bibinfo {author} {\bibfnamefont {D.}~\bibnamefont {Brann}}, \bibinfo {author} {\bibfnamefont {D.}~\bibnamefont {Chicharro}}, \bibinfo {author} {\bibfnamefont {K.}~\bibnamefont {Drummey}}, \bibinfo {author} {\bibfnamefont {K.~M.}\ \bibnamefont {Franks}}, \bibinfo {author} {\bibfnamefont {S.}~\bibnamefont {Panzeri}},\ and\ \bibinfo {author} {\bibfnamefont {S.~R.}\ \bibnamefont {Datta}},\ }\bibfield  {title} {\bibinfo {title} {Structure and flexibility in cortical representations of odour space},\ }\href@noop {} {\bibfield  {journal} {\bibinfo  {journal} {Nature}\ }\textbf {\bibinfo {volume} {583}},\ \bibinfo {pages} {253} (\bibinfo {year} {2020})}\BibitemShut {NoStop}%
\bibitem [{\citenamefont {Mizuno}\ \emph {et~al.}(2007)\citenamefont {Mizuno}, \citenamefont {Hirano},\ and\ \citenamefont {Tagawa}}]{mizuno2007evidence}%
  \BibitemOpen
  \bibfield  {author} {\bibinfo {author} {\bibfnamefont {H.}~\bibnamefont {Mizuno}}, \bibinfo {author} {\bibfnamefont {T.}~\bibnamefont {Hirano}},\ and\ \bibinfo {author} {\bibfnamefont {Y.}~\bibnamefont {Tagawa}},\ }\bibfield  {title} {\bibinfo {title} {Evidence for activity-dependent cortical wiring: formation of interhemispheric connections in neonatal mouse visual cortex requires projection neuron activity},\ }\href@noop {} {\bibfield  {journal} {\bibinfo  {journal} {Journal of Neuroscience}\ }\textbf {\bibinfo {volume} {27}},\ \bibinfo {pages} {6760} (\bibinfo {year} {2007})}\BibitemShut {NoStop}%
\bibitem [{\citenamefont {Suarez}\ \emph {et~al.}(2014)\citenamefont {Suarez}, \citenamefont {Fenlon}, \citenamefont {Marek}, \citenamefont {Avitan}, \citenamefont {Sah}, \citenamefont {Goodhill},\ and\ \citenamefont {Richards}}]{suarez2014balanced}%
  \BibitemOpen
  \bibfield  {author} {\bibinfo {author} {\bibfnamefont {R.}~\bibnamefont {Suarez}}, \bibinfo {author} {\bibfnamefont {L.~R.}\ \bibnamefont {Fenlon}}, \bibinfo {author} {\bibfnamefont {R.}~\bibnamefont {Marek}}, \bibinfo {author} {\bibfnamefont {L.}~\bibnamefont {Avitan}}, \bibinfo {author} {\bibfnamefont {P.}~\bibnamefont {Sah}}, \bibinfo {author} {\bibfnamefont {G.~J.}\ \bibnamefont {Goodhill}},\ and\ \bibinfo {author} {\bibfnamefont {L.~J.}\ \bibnamefont {Richards}},\ }\bibfield  {title} {\bibinfo {title} {Balanced interhemispheric cortical activity is required for correct targeting of the corpus callosum},\ }\href@noop {} {\bibfield  {journal} {\bibinfo  {journal} {Neuron}\ }\textbf {\bibinfo {volume} {82}},\ \bibinfo {pages} {1289} (\bibinfo {year} {2014})}\BibitemShut {NoStop}%
\bibitem [{\citenamefont {Lee}\ \emph {et~al.}(2019)\citenamefont {Lee}, \citenamefont {Vandemark}, \citenamefont {Mezey}, \citenamefont {Shultz},\ and\ \citenamefont {Fitzpatrick}}]{lee2019functional}%
  \BibitemOpen
  \bibfield  {author} {\bibinfo {author} {\bibfnamefont {K.-S.}\ \bibnamefont {Lee}}, \bibinfo {author} {\bibfnamefont {K.}~\bibnamefont {Vandemark}}, \bibinfo {author} {\bibfnamefont {D.}~\bibnamefont {Mezey}}, \bibinfo {author} {\bibfnamefont {N.}~\bibnamefont {Shultz}},\ and\ \bibinfo {author} {\bibfnamefont {D.}~\bibnamefont {Fitzpatrick}},\ }\bibfield  {title} {\bibinfo {title} {Functional synaptic architecture of callosal inputs in mouse primary visual cortex},\ }\href@noop {} {\bibfield  {journal} {\bibinfo  {journal} {Neuron}\ }\textbf {\bibinfo {volume} {101}},\ \bibinfo {pages} {421} (\bibinfo {year} {2019})}\BibitemShut {NoStop}%
\bibitem [{\citenamefont {Caputi}\ \emph {et~al.}(2022)\citenamefont {Caputi}, \citenamefont {Liu}, \citenamefont {Fuchs}, \citenamefont {Liu},\ and\ \citenamefont {Monyer}}]{caputi2022medial}%
  \BibitemOpen
  \bibfield  {author} {\bibinfo {author} {\bibfnamefont {A.}~\bibnamefont {Caputi}}, \bibinfo {author} {\bibfnamefont {X.}~\bibnamefont {Liu}}, \bibinfo {author} {\bibfnamefont {E.~C.}\ \bibnamefont {Fuchs}}, \bibinfo {author} {\bibfnamefont {Y.-C.}\ \bibnamefont {Liu}},\ and\ \bibinfo {author} {\bibfnamefont {H.}~\bibnamefont {Monyer}},\ }\bibfield  {title} {\bibinfo {title} {Medial entorhinal cortex commissural input regulates the activity of spatially and object-tuned cells contributing to episodic memory},\ }\href@noop {} {\bibfield  {journal} {\bibinfo  {journal} {Neuron}\ }\textbf {\bibinfo {volume} {110}},\ \bibinfo {pages} {3389} (\bibinfo {year} {2022})}\BibitemShut {NoStop}%
\bibitem [{\citenamefont {Li}\ \emph {et~al.}(2016)\citenamefont {Li}, \citenamefont {Daie}, \citenamefont {Svoboda},\ and\ \citenamefont {Druckmann}}]{li2016robust}%
  \BibitemOpen
  \bibfield  {author} {\bibinfo {author} {\bibfnamefont {N.}~\bibnamefont {Li}}, \bibinfo {author} {\bibfnamefont {K.}~\bibnamefont {Daie}}, \bibinfo {author} {\bibfnamefont {K.}~\bibnamefont {Svoboda}},\ and\ \bibinfo {author} {\bibfnamefont {S.}~\bibnamefont {Druckmann}},\ }\bibfield  {title} {\bibinfo {title} {Robust neuronal dynamics in premotor cortex during motor planning},\ }\href@noop {} {\bibfield  {journal} {\bibinfo  {journal} {Nature}\ }\textbf {\bibinfo {volume} {532}},\ \bibinfo {pages} {459} (\bibinfo {year} {2016})}\BibitemShut {NoStop}%
\bibitem [{\citenamefont {Chen}\ \emph {et~al.}(2021)\citenamefont {Chen}, \citenamefont {Kang}, \citenamefont {Lindsey}, \citenamefont {Druckmann},\ and\ \citenamefont {Li}}]{chen2021modularity}%
  \BibitemOpen
  \bibfield  {author} {\bibinfo {author} {\bibfnamefont {G.}~\bibnamefont {Chen}}, \bibinfo {author} {\bibfnamefont {B.}~\bibnamefont {Kang}}, \bibinfo {author} {\bibfnamefont {J.}~\bibnamefont {Lindsey}}, \bibinfo {author} {\bibfnamefont {S.}~\bibnamefont {Druckmann}},\ and\ \bibinfo {author} {\bibfnamefont {N.}~\bibnamefont {Li}},\ }\bibfield  {title} {\bibinfo {title} {Modularity and robustness of frontal cortical networks},\ }\href@noop {} {\bibfield  {journal} {\bibinfo  {journal} {Cell}\ }\textbf {\bibinfo {volume} {184}},\ \bibinfo {pages} {3717} (\bibinfo {year} {2021})}\BibitemShut {NoStop}%
\bibitem [{\citenamefont {Inagaki}\ \emph {et~al.}(2019)\citenamefont {Inagaki}, \citenamefont {Fontolan}, \citenamefont {Romani},\ and\ \citenamefont {Svoboda}}]{inagaki2019discrete}%
  \BibitemOpen
  \bibfield  {author} {\bibinfo {author} {\bibfnamefont {H.~K.}\ \bibnamefont {Inagaki}}, \bibinfo {author} {\bibfnamefont {L.}~\bibnamefont {Fontolan}}, \bibinfo {author} {\bibfnamefont {S.}~\bibnamefont {Romani}},\ and\ \bibinfo {author} {\bibfnamefont {K.}~\bibnamefont {Svoboda}},\ }\bibfield  {title} {\bibinfo {title} {Discrete attractor dynamics underlies persistent activity in the frontal cortex},\ }\href@noop {} {\bibfield  {journal} {\bibinfo  {journal} {Nature}\ }\textbf {\bibinfo {volume} {566}},\ \bibinfo {pages} {212} (\bibinfo {year} {2019})}\BibitemShut {NoStop}%
\bibitem [{\citenamefont {Feng}\ and\ \citenamefont {Tu}(2021)}]{feng2021inverse}%
  \BibitemOpen
  \bibfield  {author} {\bibinfo {author} {\bibfnamefont {Y.}~\bibnamefont {Feng}}\ and\ \bibinfo {author} {\bibfnamefont {Y.}~\bibnamefont {Tu}},\ }\bibfield  {title} {\bibinfo {title} {The inverse variance--flatness relation in stochastic gradient descent is critical for finding flat minima},\ }\href@noop {} {\bibfield  {journal} {\bibinfo  {journal} {Proceedings of the National Academy of Sciences}\ }\textbf {\bibinfo {volume} {118}} (\bibinfo {year} {2021})}\BibitemShut {NoStop}%
\end{thebibliography}%


\onecolumngrid

\newpage
\section{Figures}

\captionsetup[figure]{font=small}
\begin{figure}[H]%
\centering  
\includegraphics[width=0.95\textwidth]{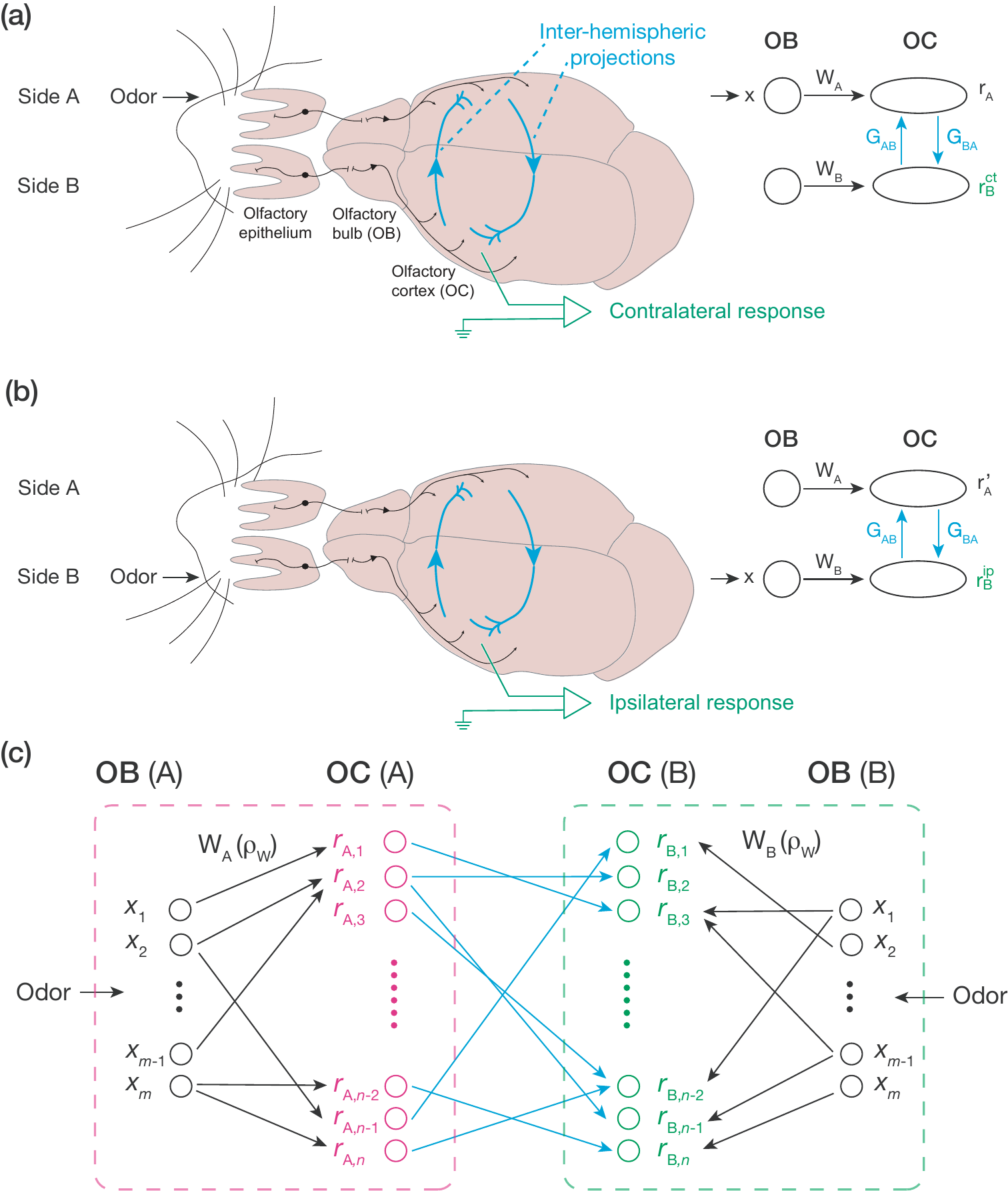}
\caption{\textbf{Background and model setup.}   \textbf{(a)} A sketch of the early mouse olfactory system and the experimental setup to measure the contralateral response (left), and the corresponding neural circuit (right). Odor is delivered to the nostril on Side A, and the contralateral response is measured by the electrode placed in the Side-B olfactory cortex. Adapted from \cite{Grimaud2021}.
}

\label{fig Background and model setup}
\end{figure}

\newpage
\noindent  \textbf{(b)} A sketch of the experimental setup to measure the ipsilateral response (left), and the corresponding neural circuit (right).
\textbf{(c)} Neural network of the bilateral alignment circuit during training. The pink box represents Side-A hemisphere, and the green one is for Side-B. Training odors are delivered to both nostrils. Due to the two random and different OB-to-OC projections ($\WA$ and $\WB$), the cortical representations $\rbfA, \rbfB$ are also different. The blue arrows ($\GBA$) illustrate the inter-hemispheric projections that Side-B OC receives from Side-A OC, with density $\rhog$. $\GAB$ is not shown in this network.

\newpage
\begin{figure}[H]%
\centering
\includegraphics[width=0.8\textwidth]{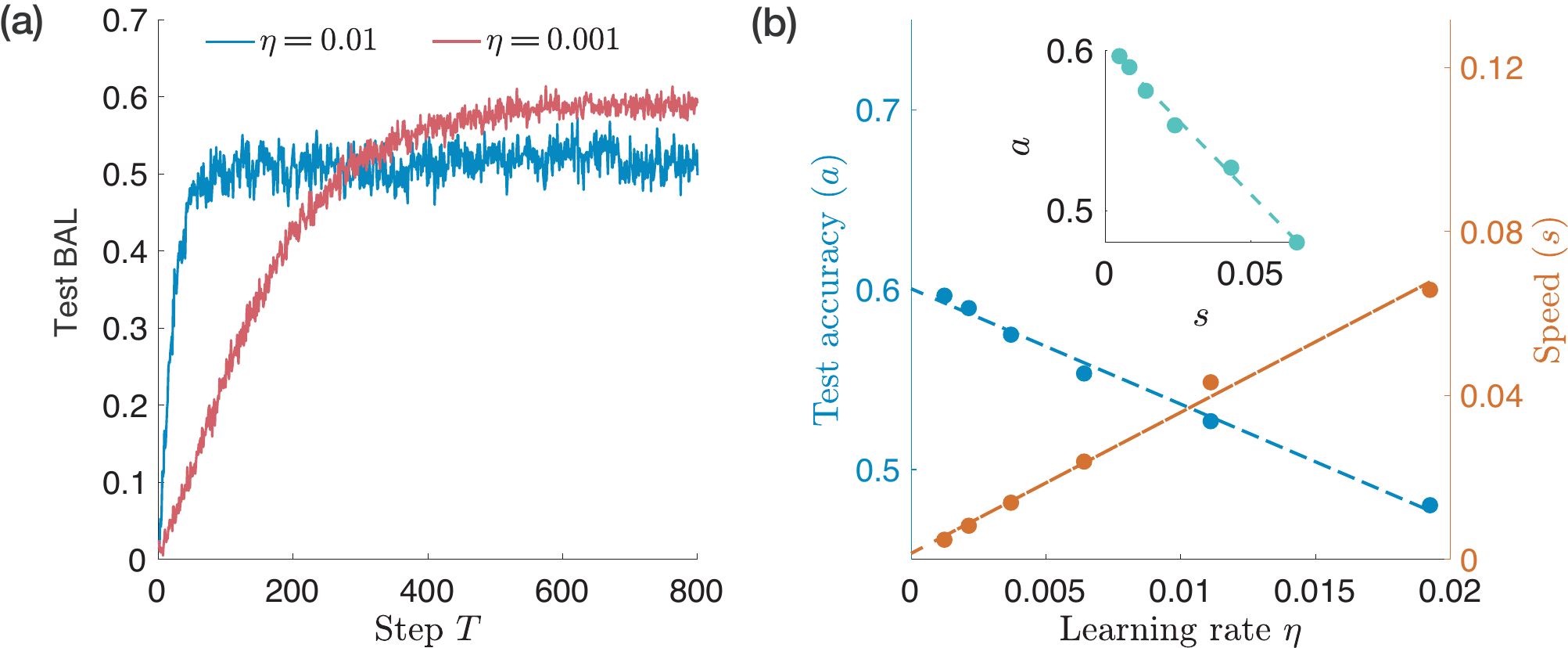}
\caption{\textbf{Hebbian learning can align bilateral responses using sparse connections, with learning rate $\eta$ controlling the linear speed-accuracy tradeoff.} \textbf{(a)} Two learning curves with $\eta= 0.01\ \mathrm{s}^2$ (blue) and $\eta= 0.001\ \mathrm{s}^2$ (red), given $(m,n,\rhow,\rhog)=(20,500,0.1,0.05)$.  \textbf{(b)} Test accuracy ($a$, left $y$-axis) and convergence speed ($s$, right $y$-axis) both depend on $\eta$ linearly but with opposite signs, resulting in a linear tradeoff (the green line of the inset). Here each dot is from numerical results, and the dashed lines are from linear fitting. } 
\label{fig_two_eta_speed_accuracy_tradeoff}
\end{figure}

\newpage
\begin{figure}[H]%
\centering
\includegraphics[width=\textwidth]{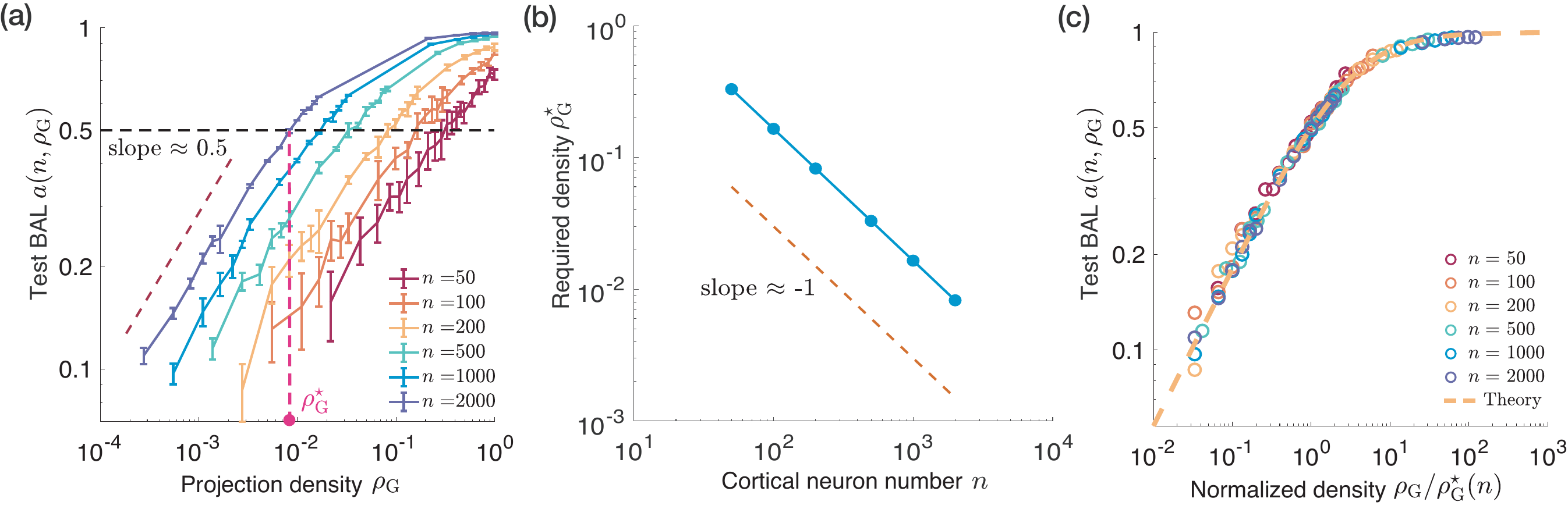}
\caption{\textbf{Numerical and theoretical results of test BAL, and the inverse scaling between $\rhogstar$ and $n$.}
\textbf{(a)} Test BAL increases with $\rhog$ in a power law fashion with an exponent around 0.5 for small $\rhog$. The required density $\rhogstar$ to achieve an alignment of 0.5 are determined by the black dashed line, e.g., the purple $\rhogstar\approx 8.3\permil$ for $n=2000$. Colors indicate different values of $n$ with fixed $m=20, \rhow= 0.1$. \textbf{(b)}  $\rhogstar$ scale inversely with $n $. \textbf{(c)} All the data points in Panel A collapse onto a single curve with normalized $\rhog$, and match well with theoretical prediction (the orange line).
}
\label{fig Hebbian theory matches simulations and the inverse scaling}
\end{figure}

\newpage
\begin{figure}[H]%
\centering
\includegraphics[width=\textwidth]{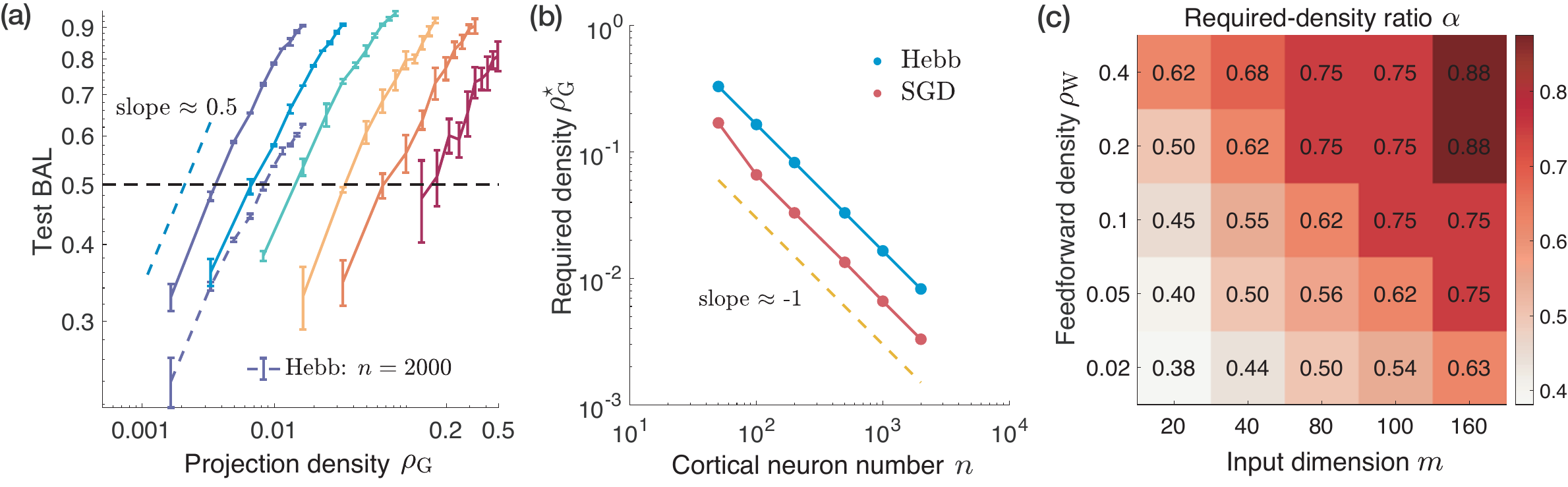}
\caption{\textbf{Global SGD learning rule outperforms local Hebb's rule but has the same inverse scaling, and two learning rules get closer with larger $m$ and $\rhow$.} 
\textbf{(a)} Test BAL under SGD learning shows similar power-law scaling with $\rhog$ as that in the Hebbian learning. Colors indicate different $n$ as in Fig.~\ref{fig Hebbian theory matches simulations and the inverse scaling}(c). \textbf{(b)} $\rhogstar$ in both Hebbian and SGD learning rules scale inversely with $n$. \textbf{(c)} Larger $m$ and $\rhow$ lead to better alignment between Hebbian and SGD learning rules, as shown by the heatmap of $\alpha(m,\rhow)$.
}
\label{fig4_SGD_alpha}
\end{figure}

\newpage
\begin{figure}[H]%
\centering
\includegraphics[width=0.8\textwidth]{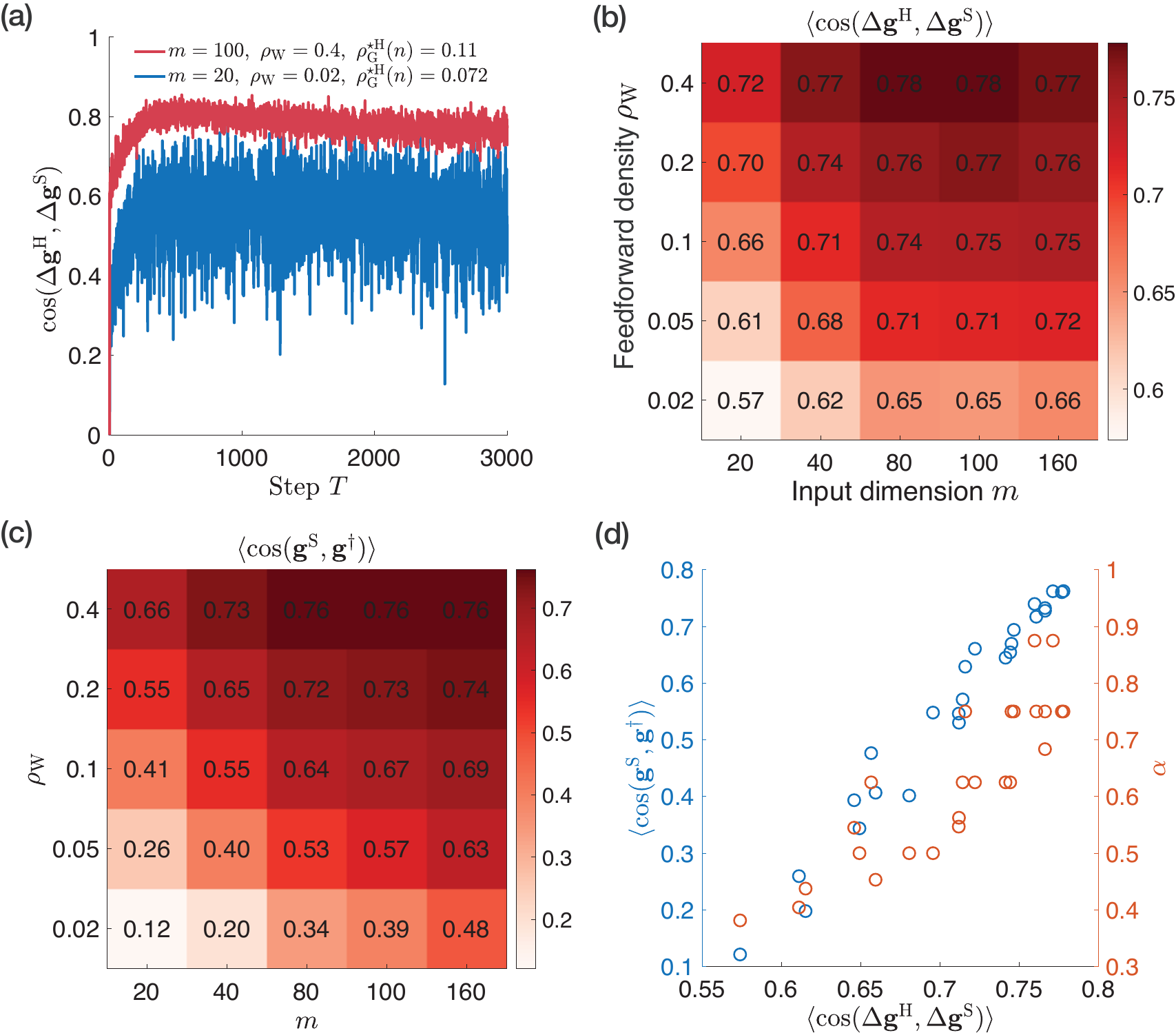}
\caption{\textbf{The weight update vector of Hebbian learning aligns better with the corresponding SGD vector for larger $m$ and $\rhow$, leading to higher solution alignment level and larger $\alpha$.} \textbf{(a)} The temporal dynamics of update alignment $\cos(\Delta\mathbf{g}^\mathrm{H}, \Delta\mathbf{g}^\mathrm{S} )$ for two pairs of parameters [red for $m=100,\rhow=0.4, \rhogstarH(n=500)=0.11$, and blue for $m=40,\rhow=0.02, \rhogstarH(n=500)=0.072$]. \textbf{(b)} The update alignment level increases with $m$ and $\rho_W$. \textbf{(c)} Similar trend is observed for the final SGD solution and the theoretical Hebbian solution, i.e., $\cos(\g^\mathrm{S},\gdagger)$. \textbf{(d)} Both the solution alignment level $\cos(\g^\mathrm{S},\gdagger)$ (blue dots for the left $y$-axis) and $\alpha$ (red dots for the right $y$-axis) strongly correlate with update alignment level $\cos(\Delta\g^\mathrm{H},\Delta\g^\mathrm{S})$. Here each dot corresponds to one pair of $(m,\rhow)$ in Panels (b)-(c).
}
\label{fig5_gradient_alignment}
\end{figure}

\newpage
\begin{figure}[H]%
\centering
\includegraphics[width=\textwidth]{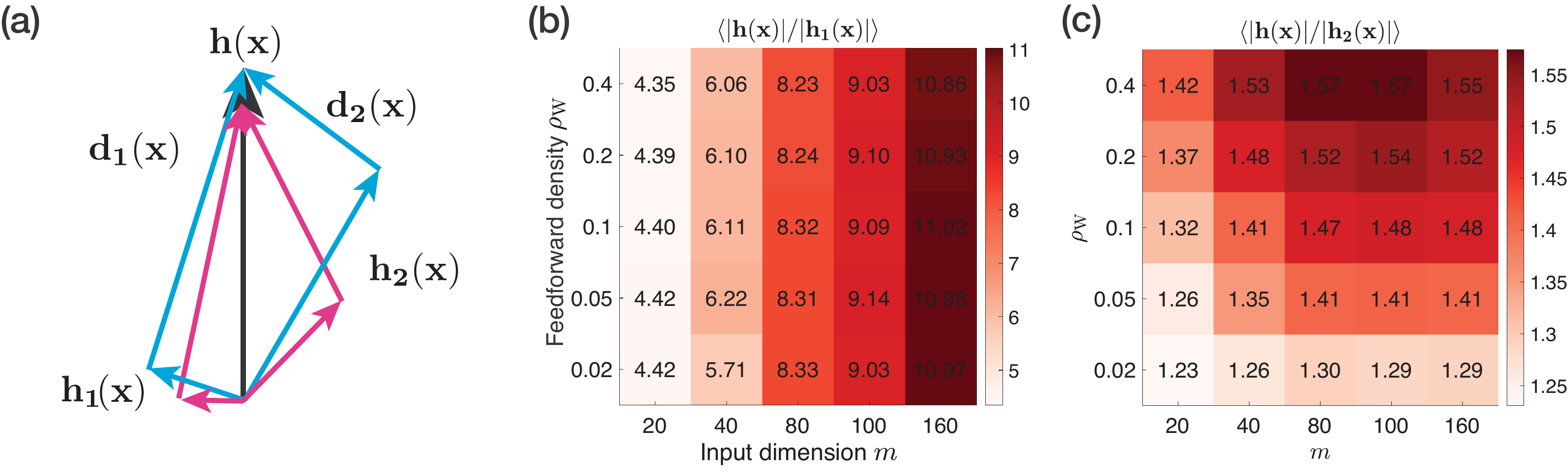}
\caption{\textbf{Both Hebbian and SGD weight update vectors overlap more with the common Hebbian term with larger $m$ and $\rhow$.} \textbf{(a)} The geometric representation of how $\h_1(\x), \h_2(\x), \dbf_1(\x), \dbf_2(\x)$ change with $(m,\rhow)$, compared to $\h(\x)$ (the central arrow). Blue arrows: small $m$ and $\rhow$. Pink arrows: large $m$ and $\rhow$. \textbf{(b)} $|\h(\x)|/|\h_1(\x)|\gg 1$ and increases with $m$. Here, for each pair of $(m,\rhow)$, the mean is obtained by averaging over $\x$ and $n$. \textbf{(c)} In general, $\h(\x)$ is more dominant over $\h_2(\x)$ in magnitude for larger $m$ and $\rhow$. } 
\label{fig_h_h1_h2}
\end{figure}

\newpage
\appendix

\section*{Supplementary materials}
\section{Analysis of the neural dynamics}
\subsection{Equations of ipsi- and contra-lateral responses}
\label{Equations of ipsi- and contra-lateral responses}
Based on Eq.~\eqref{Eq neural dynamics} in the main text (MT, for short) and the diagram in Fig.~\ref{fig Background and model setup}(b), the stationary ipsilateral response $\rbfBip$ is determined by the fixed point of the following ordinary differential equations (ODEs)
\begin{equation}
\begin{aligned}
   \tau\dfrac{d\rbfA'}{dt} &= - \rbfA' +\tanh( \GAB\rbfBip), \\
   \tau\dfrac{d\rbfBip}{dt} &= - \rbfBip + \tanh(\WB\x + \GBA\rbfA' ). \\
\end{aligned}\label{ipsi-lateral response}
\end{equation}
Similarly, corresponding to the diagram in Fig.~\ref{fig Background and model setup}(a), the stationary contralateral response $\rbfBct$ is determined by the fixed point of
\begin{equation}
\begin{aligned}
\tau\dfrac{d\rbfA}{dt} &= - \rbfA +\tanh(\WA\x + \GAB\rbfBct), \\
\tau\dfrac{d\rbfBct}{dt} &= - \rbfBct + \tanh( \GBA\rbfA ). \\
\end{aligned}\label{contra-lateral response}
\end{equation}

\subsection{The analytical form of the steady-state cortical responses}
\label{The analytical form of Hebbian solution}

The linearized neural dynamics MT Eq.~\eqref{Eq neural dynamics linear} has an exact fixed point solution
\begin{equation}
\begin{aligned}
    \begin{bmatrix}
        \rbfA \\
        \rbfB
    \end{bmatrix} &= \left[ \one_{2n} - \begin{pmatrix}
        0 & \GAB \\
        \GBA & 0
    \end{pmatrix} \right]^{-1} \begin{bmatrix}
        \WA \\
        \WB
    \end{bmatrix} \x,
\end{aligned}
\label{Exact solutions to the original equations}
\end{equation}
where $ \one_{2n}$ is an identity matrix of dimension $2n\times 2n$. Using the fact that the entries of $\Gbf_{\mathrm{AB},ij}\ll 1, \Gbf_{\mathrm{BA},ij} \ll 1$, we can take the Taylor expansion of the above inverse matrix with respect to $\GAB$ and $\GBA$ and keep the first order:
\begin{equation}
\begin{aligned}
    \begin{bmatrix}
        \rbfA \\
        \rbfB
    \end{bmatrix} 
     =
    \left[ \one_{2n} 
    + 
    \begin{pmatrix}
        0 & \GAB \\
        \GBA & 0
    \end{pmatrix} 
    + \cdots 
    \right]
    \begin{bmatrix}
        \WA \\
        \WB
    \end{bmatrix} \x  
    \approx \begin{bmatrix}
        \WA \x + \GAB\WB\x \\
        \WB\x + \GBA\WA\x
    \end{bmatrix},
\end{aligned}
\label{Eqs Its Taylor expansion}
\end{equation}
which gives us the approximate steady-state solutions [MT Eq.~\eqref{Eq lower order approximations for cortical representations}].

\section{Derivations of Hebbian test BAL}
\label{SM: Derivations of Hebbian test BAL}
\subsection{Weight Alignment Level (WAL) approximates test Bilateral Alignment Level (BAL)}
\label{Proof that Weight Alignment Level (WAL) approximates Test Bilateral Alignment Level (BAL)}
In this section, we show that the WAL $c$ and BAL $a^\dagger$ are approximately equal. Start from the stationary ipsilateral and contralateral responses  $\rbfBip\approx\WB \x, \rbfBct\approx \Gdagger \WA\x$,
\begin{equation}
        \begin{aligned}
            \langle\cos (\rbfBip, \rbfBct)\rangle^2_{\x} & := \left\langle\dfrac{\rbfBip \cdot \rbfBct }{|\rbfBip| \cdot |\rbfBct| }\right\rangle_{\x}^2 \approx  \left(\dfrac{\langle\rbfBip \cdot \rbfBct \rangle_\x}{\langle|\rbfBip|\rangle _{\x} \langle |\rbfBct|  \rangle_{\x}  }\right)^2 \approx\dfrac{\langle\rbfBip \cdot \rbfBct \rangle_\x^2}{\langle|\rbfBip|^2\rangle_{\x} \langle |\rbfBct|^2  \rangle_{\x}  }.
        \end{aligned}
        \label{cosine similarity two approximations}
    \end{equation}
In the above approximation, we swapped the average over $\x$ and the fraction. The legitimacy is verified numerically (see Fig.~\ref{figS Theory of Hebbian learning matches simulations}(a)). For $\x\sim \cn(\zero, \gamma^2 \one)$, we have
\begin{align}
    \langle\rbfBip \cdot \rbfBct \rangle_\x &= \langle\rm{Tr}\left((\WB\x)^\top \Gdagger \WA\x\right)\rangle_\x = \gamma^2\rm{Tr}(\Gdagger \WA\WB^\top),\\
    \langle|\rbfBip|^2\rangle_\x &= \langle\rm{Tr}(\WB\x)^\top\WB\x\rangle_\x =\gamma^2 \rm{Tr}(\WB^\top\WB) = \gamma^2\Vert \WB\Vert_F^2,\\
    \langle|\rbfBct|^2\rangle_\x &= \langle\rm{Tr}(\Gdagger\WA\x)^\top\Gdagger\WA\x\rangle_\x =\gamma^2 \Vert \Gdagger\WA\Vert_F^2.
    \label{WAL derivations}
\end{align}
Plug the above equations in \eqref{cosine similarity two approximations}, we have
\begin{equation}\label{EqS Weight Alignment Level}
     a^{\dagger2} \approx \langle\cos (\rbfBip, \rbfBct)\rangle^2_{\x}  = \dfrac{ \tr^2(\WB^\top \Gdagger\WA) }{||\WB||_\mathrm{F}^2\cdot  ||\Gdagger\WA||_\mathrm{F}^2 } = \rm{cos}^2(\Gdagger\WA, \WB) = c^2.
\end{equation}

\subsection{Calculate the Weight Alignment Level}
\label{The theory for Weight Alignment Level}
In this section, we calculate WAL $c$, which is also the analytical expression of test BAL, i.e., MT Eq.~\eqref{Eq Weight Alignment Level}. With the Hebbian solution $\Gdagger = \beta^{-1} \WB \Sigmabf\WA^\top \odot \GBAhat =\beta^{-1} \gamma^2\WB\WA^\top \odot \GBAhat  $ in our study,
\begin{equation}
\begin{aligned}
    \tr(\WB^\top \Gdagger\WA)&\approx  \beta^{-1} \gamma^2 \tr \left[ \WB^\top (\WB \WA^\top \odot \GBAhat) \WA \right]\\
    &= \beta^{-1}\gamma^2 \left[\sum_{il} \hat{G}_{BA,il} \sum_{jk} W_{A,lk}  W_{A,lj}  W_{B,ij} W_{B,ik}  \right] \\
    &= \beta^{-1}\gamma^2 \left[\sum_{il} \hat{G}_{BA,il} \sum_{j} \left( W_{A,lj}^2  W_{B,ij}^2 + \sum_{k\neq j} W_{A,lk}  W_{A,lj}  W_{B,ij} W_{B,ik} \right) \right] \\
    (\textbf{Due to LLN}) \quad & \approx \beta^{-1}\gamma^2 \left[\sum_{il} \hat{G}_{BA,il} \sum_{j}  \left( W_{A,lj}^2  W_{B,ij}^2   \right)  \right] \\
    (\textbf{Due to LLN})  \quad  & \approx \beta^{-1}\gamma^2  mn^2 \rhow^2\rhog,
\end{aligned}
\label{tr WB GBAH WA}
\end{equation} 
where we utilized the law of large numbers (``LLN") in the last two lines, as well as the normal distribution of $ W_{A,lj}$ and $ W_{B,ij}$. Here, $n^2\rhog$ enters the penultimate line via the summation of $\hat{G}_{BA,il}$, therefore, is the total number of inter-hemispheric projections. As we will show soon, $n\rhog$ remains when taking the fraction. Thus, intuitively, we can say that $n\rhog$ in MT Eq.~\eqref{Eq Weight Alignment Level} represents the number of inter-hemispheric projections received per cortical neuron. 

And $||\Gdagger\WA||_\mathrm{F}^2$ in the denominator of Eq.~\eqref{Eq Weight Alignment Level} is
\begin{equation}
\begin{aligned}
    ||\Gdagger\WA||_\mathrm{F}^2&\approx  \beta^{-2} \gamma^4 \left\{\sum_{ij}  \left[(\WB \WA^\top \odot \GBAhat) \WA\right]_{ij}^2 \right\}:=  \beta^{-2}\gamma^4  \sum_{ij} C_{ij}^2, \\            
\end{aligned}
\end{equation} 
where $ C_{ij}= \sum_{kl} W_{B,ik} W_{A,lk} \hat{G}_{BA,il}W_{A,lj}$. Therefore,
\begin{equation}
\begin{aligned}
    C_{ij}^2 
    &= \sum_{kk'll'} \hat{G}_{BA,il}\hat{G}_{BA,il'} W_{B,ik} W_{B,ik'} W_{A,lk}W_{A,l'k'}  W_{A,lj}W_{A,l'j}.
\end{aligned}
\end{equation} 
In order to have nonzero summation after applying LLN, it is required that $\WA$'s and $\WB$'s elements must have even powers, such as $W_{A,lj}^2 $ and $ W_{B,ij}^2 $ in Eq.~\eqref{tr WB GBAH WA}. This includes three scenarios: \\
1. $l'\neq l, \ k'=k=j$, \\
2. $l'= l, \ k'=k\neq j$, \\
3. $l'= l, \ k'=k= j$, \\
which gives 
\begin{equation}
\begin{aligned}
C_{ij}^2= &\sum_{l,l'\neq l} \hat{G}_{BA,il}\hat{G}_{BA,il'} W_{B,ij}^2 W_{A,lj}^2W_{A,l'j}^2 \\
  &+ \sum_{l}\sum_{k\neq j}  \hat{G}_{BA,il}^2 W_{B,ik}^2 W_{A,lk}^2W_{A,lj}^2\\
  &+ \sum_{l}  \hat{G}_{BA,il}^2 W_{B,ij}^2 W_{A,lj}^4\\
  (\textbf{Due to LLN}) \quad  \approx & n^2 \rhog^2 \rhow^3 + mn \rhog\rhow^3 + n\rhog\rhow^2\cdot 1 \cdot 3\\
  = & n\rhog\rhow^2 (n\rhog\rhow + m\rhow + 3),\\
  \therefore  ||\Gdagger\WA||_\mathrm{F}^2\approx &  \beta^{-2}\gamma^4 mn^2\rhog\rhow^2 (n\rhog\rhow + m\rhow + 3).
 \label{GBAHWA norm}
\end{aligned}
\end{equation} 

For the other term in the denominator of Eq.~\eqref{cosine similarity two approximations}, we have
\begin{equation}
\begin{aligned}
||\WB||_\mathrm{F}^2&=\sum_{ij}  W_{B,ij}^2    \approx mn\rhow .
\end{aligned}
\label{WB norm}
\end{equation} 

Combining Eqs.~\eqref{tr WB GBAH WA}, \eqref{GBAHWA norm}, and \eqref{WB norm}, we get MT Eq.~\eqref{Eq Weight Alignment Level}: 
\begin{equation}
\begin{aligned}
    (\adagger)^2&=\langle\cos (\rbfBip, \rbfBct)\rangle^2_\x & \approx \dfrac{ \tr^2(\WB^\top \Gdagger\WA) }{||\WB||_\mathrm{F}^2\cdot  ||\Gdagger\WA||_\mathrm{F}^2 } = \dfrac{n \rhog}{m +3/\rhow +n \rhog }.
\end{aligned}
\label{perfect alignment}
\end{equation} 

Finally, we pointed out that as long as all elements in $\GBAhat,\WA,\WB$ are chosen independently, the LLN still applies in Eqs.~\eqref{tr WB GBAH WA}-\eqref{WB norm}, which explains the result that the fraction of subpopulations projecting contralaterally does not affect bilateral alignment level (Fig.~\ref{subpopulations projecting}).

\subsection{Gradient and Hessian matrix of test BAL $a$, and its 2nd-order Taylor expansion}
\label{Gradient and Hessian matrix of test BAL}

From Eq.~\eqref{Eq Weight Alignment Level}, in general, we can derive the gradient of test BAL $a$ regarding any $\GBA$ under the sparsity constraint $\GBAhat$
\begin{equation}
\begin{aligned} 
\dfrac{\partial a}{\partial\GBA}\odot \GBAhat &= \dfrac{a}{u} \bigg[  \WB\WA^\top-  \dfrac{u}{v^2} \GBA\WA\WA^\top\bigg]\odot \GBAhat,\\
\end{aligned}
\end{equation}
where $v= ||\GBA\WA||_{\mathrm{F}} , \ u = \tr(\WB^\top\GBA\WA)$. To simplify the notation, we denote $\Gbf\equiv\GBA$, $\Gbfhat\equiv\GBAhat $, and $\E\equiv\WA\WA^\top$. 

To derive the Hessian matrix, we firstly assume that $v(\Gbf)$ and $u(\Gbf)$ are slow variables such that they roughly remain constant with small $\Gbf$ fluctuation. Therefore, if we denote $\B:=\dfrac{\partial a}{\partial\Gbf}\odot \Gbfhat$, we have 
\begin{equation}
\begin{aligned} 
\dfrac{\partial B_{ij}}{\partial G_{kl}} &\approx -\dfrac{a}{v^2} \dfrac{\partial } {\partial G_{kl}} (\sum_m G_{im}E_{mj} \Ghat_{ij})  =-\dfrac{a}{v^2} \sum_m E_{mj}\Ghat_{ij}\delta_{ik}\delta_{lm}.
\end{aligned}
\end{equation}
As a result, the second-order term of the Taylor expansion of $a(\Gbf+\delta\Gbf)$ is
\begin{equation}
\begin{aligned} 
    \delta^{(2)}a &:= \dfrac{1}{2}\sum_{ijkl} \delta G_{ij} \cdot \dfrac{\partial B_{ij}}{\partial G_{kl}}\cdot \delta G_{kl}\\
    &= -\dfrac{a}{2v^2}\sum_{ijkl} \delta G_{ij} \cdot\sum_m E_{mj}\Ghat_{ij}\delta_{ik}\delta_{lm}\cdot \delta G_{kl}\\
    &= -\dfrac{a}{2v^2}\sum_{ijl} \delta G_{ij} \cdot E_{lj}\Ghat_{ij}\cdot \delta G_{il}\\
    &= -\dfrac{a}{2v^2}\Tr{ \left[ (\delta\Gbf)\E(\delta\Gbf\odot\Gbfhat)^\top \right] }.
\end{aligned}
    \label{EqS second order Taylor term of hat_a}
\end{equation}
Note that $\E=\WA\WA^\top\approx m\rhow\one_{n}$. Thus, Eq.~\eqref{EqS second order Taylor term of hat_a} becomes
\begin{equation}
    \delta^{(2)}a \approx  -\dfrac{am\rhow}{2v^2}\Tr{ \left[ (\delta\Gbf)(\delta\Gbf\odot\Gbfhat)^\top  \right] }
    =  -\dfrac{am\rhow}{2v^2} ||\delta\Gbf\odot\Gbfhat||^2_{\mathrm{F}}<0,
\end{equation}
indicating that test BAL $a(\Gbf) $ is concave, or equivalently, the Hessian matrix $\Hdagger$ in Eq.~\eqref{Eq a_hat 2nd order expansion} is negative definite.

The 2nd-order Taylor expansion of $a(\Gbf+\delta\Gbf)$ now becomes
\begin{equation}
\begin{aligned} {}
    a(\Gbf+\delta\Gbf) = a(\Gbf) + \Tr(\B^\top \delta\Gbf) - \dfrac{a}{2v^2}\Tr{ \left[ (\delta\Gbf)\E(\delta\Gbf\odot\Gbfhat)^\top \right] }. 
\end{aligned}
    \label{EqS 2nd-order Taylor expansion of a_hat}
\end{equation}

\subsection{Visualization of the concave landscape of test BAL}
\label{Visualization of the concave landscape of test BAL}
To verify the validity of the expansion Eq.~\eqref{EqS 2nd-order Taylor expansion of a_hat} around the Hebbian solution, we adopted the methods in \cite{feng2021inverse} to visualize the landscape of $a(\Gbf)$ or  $a(\g)$. Firstly, we performed principal component analysis (PCA) on the temporal matrix of $\g(T)$:
\begin{equation}
\begin{aligned} 
\begin{bmatrix}
    \g(T=1), \g(T=2), \cdots, \g(T=N) \end{bmatrix},
\end{aligned}
\end{equation}
from which we got the variances at each principal component (PC) direction. Denote the unit vector of $i$-th PC direction as $\p_i$ and the variance of $\g(T)$ at this direction as $\sigma_i^2$. Then we set the perturbation $\g' = \gdagger+\theta_i \p_i$, where $\theta_i$ is the projection onto the $i$-th PC direction. 

From the analytical expression Eq.~\eqref{cosine similarity two approximations}, we calculated the test BAL for $\g'$ as
\begin{equation}
\begin{aligned} 
    \ahat_1(\g') = \dfrac{ \tr(\WB^\top \Gbf'\WA) }{||\WB||_\mathrm{F}\cdot  ||\Gbf'\WA||_\mathrm{F} },
\end{aligned}
    \label{EqS analytical expression of a_hat with PCA}
\end{equation}
where $\Gbf'$ is the matrix form of $\g'$. On the other hand, we could use the expansion Eq.~\eqref{EqS 2nd-order Taylor expansion of a_hat} to approximate the test BAL as
\begin{equation}
  \begin{aligned} 
        \ahat_2(\g') = \adagger + \theta_i\Tr(\B^\top \Pbf_i) - \dfrac{\adagger\theta_i^2}{2v^2}\Tr{ \left[ \Pbf_i\WA\WA^\top(\Pbf_i\odot\Gbfhat)^\top  \right] },
  \end{aligned}
        \label{EqS 2nd-order Taylor expansion of a_hat with PCA}
\end{equation}
where $\Pbf_i$ is the matrix form of $\p_i$.

To reflect the fluctuation of $\g(T)$ during learning, we varied $\theta_i$ from $-3\sigma_i$ to $3\sigma_i$, then computed and compared corresponding $\ahat_1$ and $\ahat_2$. As shown in Fig.~\ref{figS_a_hat_is_concave}(a), for $(m, n, \rho_W, \rhog, \xi) = (20,500,0.1,0.05, 0.01)$, the landscape of $a_1$ along the first PC direction is concave (the blue curve, calculated with Eq.~\eqref{EqS analytical expression of a_hat with PCA}), which is well approximated by $a_2$ from Eq.~\eqref{EqS 2nd-order Taylor expansion of a_hat with PCA}. The concavity and the matching between $\ahat_1$ and $\ahat_2$ are also verified in other PC directions [see Figs.\ref{figS_a_hat_is_concave}(b)-(c) for $i=5$ and $i=10$], as well as with other parameters [see Figs.\ref{figS_a_hat_is_concave}(d)-(e) for a smaller learning rate $\xi=0.001$ and Figs.\ref{figS_a_hat_is_concave}(g)-(i) for $(m, n, \rho_W, \rhog, \xi) = (20,50,0.5,0.5, 0.001)$].

\subsection{The test BAL decreases with $\eta \beta$}
\label{test BAL decreases with eta H negative definite}

In the main text, we asserted that the test BAL decreases with $\eta\beta$. This is built on the fact that $\tr(\Dbf\Hdagger)< 0$. We now show this more explicitly.

Since $\Dbf$ is positive definite, a real matrix $\Qbf$ exists such that $\Dbf = \Qbf\Qbf^\top$. Then
\begin{equation}
    \begin{aligned}
            \tr\left(\Dbf \Hbf\right) = \tr\left(\Qbf\Qbf^\top \Hbf\right) = \tr\left(\Qbf^\top \Hbf\Qbf\right) = \sum_i \qbf_i^\top \Hbf\qbf_i < 0 , \\
    \end{aligned}
\end{equation}
where $\qbf_i$ is the $i$-th column of $\Qbf$.

\section{Requirements of the linear neural dynamics and the Hebbian solution}
\label{Rationale and requirements to linearize the neural dynamics}

In deriving the Hebbian solution Eq.\eqref{Eq Hebbian solution}, we have linearized the neural dynamics Eq.~\eqref{Eq neural dynamics linear}. In this section, we examine the requirements of $\gamma$ and $\beta$ for this linear approximation.

Comparing MT Eqs.~\eqref{Eq neural dynamics} and~\eqref{Eq neural dynamics linear}, to make the latter (linearized dynamics) valid, we require that
\begin{equation}
\begin{aligned}
-\onevec_{n}\ll\WA\x&\ll \onevec_{n}, \  &-\onevec_{n}\ll\GAB\rbfB\ll \onevec_{n},\\
 -\onevec_{n}\ll\WB\x&\ll \onevec_{n}, \  &-\onevec_{n}\ll\GBA\rbfA\ll \onevec_{n},\\
\end{aligned}
\end{equation}
where $\onevec_{n} $ is a $n$-dimensional vector with all elements being 1 and the ``much less than" symbol applies elementwise. Focusing on the second row, with MT Eq.~\eqref{Eq lower order approximations for cortical representations}, the requirement becomes
\begin{equation}
    \begin{aligned}
        -\onevec_{n}\ll\WB\x&\ll \onevec_{n}, \  &-\onevec_{n}\ll\GBA\rbfA\ll \onevec_{n},\\
    \end{aligned}
\end{equation}
Furthermore, to perform zeroth-order approximation in Eq.~\eqref{Eqs Its Taylor expansion} and obtain the expressions of $\rbfA$ and $\rbfB$ in MT Eq.~\eqref{Eq lower order approximations for cortical representations}, we ask for a stronger condition
\begin{equation}
    \begin{aligned}
        -\onevec_{n}\ll \GBA\WA\x\ll \WB\x&\ll \onevec_{n}.\\
    \end{aligned}
    \label{EqS Linearization condition stronger}
\end{equation}

For the middle inequality in Eq.~\eqref{EqS Linearization condition stronger}, an approximate condition is $\dfrac{||\Gdagger\WA||_\mathrm{F}^2}{||\WB||_\mathrm{F}^2}\ll 1$, which, according to Eqs.~\eqref{GBAHWA norm}-\ref{WB norm}, becomes
\begin{equation}
\begin{aligned}
\beta^{-2}\gamma^4 n\rhog\rhow (n\rhog\rhow + m\rhow + 3)  \ll 1.
\end{aligned}
\label{EqS Linearization condition stronger 1}
\end{equation}
And for the first and the last inequalities in Eq.~\eqref{EqS Linearization condition stronger}, note that $\forall i$, the mean and the variance of $(\WA\x)_i$ are $\langle(\WA\x)_i\rangle_{\x} = \langle \sum_j W_{A,ij}x_j \rangle_{\x} = 0$. We have
\begin{equation}
\var[(\WA\x)_i] = \sum_{jk} W_{A,ij} W_{A,ik}\langle x_jx_k \rangle_\x = \sum_{jk}  W_{A,ij} W_{A,ik} \delta_{jk}  \gamma^2 
= \sum_{j} W_{A,ij}^2 \gamma^2 \approx m\rhow \gamma^2,
\label{EqS Linearization condition stronger 2}
\end{equation}
where we utilized the assumption that $\x$ is an IID Gaussian variable, and implicitly, the law of large numbers in the last approximation. Therefore, an approximate condition is 
\begin{equation}
\begin{aligned}
m\rhow \gamma^2\ll 1,
\end{aligned}
\label{EqS Linearization condition stronger 2 approximate}
\end{equation}
which, combining with Eq.~\eqref{EqS Linearization condition stronger 1}, gives the condition for linearizing the neural dynamics:
\begin{equation}
\begin{aligned}
\gamma &\ll (\dfrac{1}{ m\rhow })^{1/2}, \  \beta\gg \gamma^2[ n\rhog\rhow (n\rhog\rhow + m\rhow + 3)]^{1/2} \\
\end{aligned}
\label{EqS Linearization condition final}
\end{equation}
These conditions are satisfied for the parameters used in Fig.~\ref{fig_two_eta_speed_accuracy_tradeoff}(a), where $(\rhow,\rhog)^m_n=(0.1,0.05)^{20}_{500}, \beta = 3\ \mathrm{Hz}^2, \gamma=\frac{1}{30}  \mathrm{Hz}$ (Table~\ref{Table Parameter values and the interpretations}).

\section{Derivation of the SGD scaling factor $\lambda$}
\label{Derivations of the SGD scaling factor lambda}
Note that the average of the second term $-\lambda \GBA  \rbfA\rbfA^{\top}$ in the SGD learning rule Eq.~\eqref{Eq SGD} is
\begin{equation}
    \left\langle -\lambda\GBA  \rbfA\rbfA^{\top}\right\rangle_{\x}=-\lambda\GBA  \left\langle \rbfA\rbfA^{\top}\right\rangle_{\x} \approx -m\rhow\gamma^2\lambda\GBA.
\end{equation}

Thus to make SGD rule \eqref{Eq SGD} and Hebbian updates \eqref{Eq the sparse Hebbian update rule} comparable in magnitude, we require
\begin{equation}
    \left\langle \lambda\GBA  \rbfA\rbfA^{\top}\right\rangle_{\x} \sim \beta \GBA,
\end{equation}
which leads to
\begin{equation}\label{EqS SGD lambda requirement}
    \lambda = \dfrac{\beta}{m\rhow\gamma^2}
\end{equation}

With the above condition Eq.~\eqref{EqS SGD lambda requirement}, the SGD and Hebbian solutions are also comparable in magnitude. Since the two learning rules share the first term, their second terms are equal at stationary state:
\begin{equation}
    \beta \Gdagger \approx \lambda \GBAS \langle \rbfA\rbfA^\top\rangle_\x \approx m\rho_W\gamma^2\GBAS = \beta \GBAS.
\end{equation}
Thus, $\GBAS$ and $\Gdagger$ should have similar magnitudes.

\section{Methods}
\label{sec:method}

\textbf{Bisection method to identify $\rhogstar(n)$ more accurately.} For each $n$ in Fig.~\ref{fig Hebbian theory matches simulations and the inverse scaling}(a), we adopted ten values of $\rhog$ then identified $\rhogstar(n)$. To find a more accurate $\rhogstar(n)$, we require a relative error $\epsilon$: at the beginning, if $|\frac{a(\rhogstar) - a^\star}{a^\star}|<\epsilon$, the algorithm ends and outputs $\rhogstar$ directly. Otherwise, denote the two adjacent values of $\rhogstar(n)$ as $\rho_1$ and $\rho_2$, where $\rho_1<\rhogstar(n)<\rho_2$. We took the interval $[\rho_1,\rho_2]$, ran simulations to calculate the test BAL for the midpoint $\rho_3 =\dfrac{\rho_1+\rho_2}{2}$, then compared $a(\rho_3)$ with $a^\star$ and $a(\rhogstar)$ to find a new interval. This bisection method was repeated until the relative error or the maximum of searching loops was reached, and the new $\rhogstar(n)$ was output. In this work, we adopted $\epsilon=5\%$, which is sufficient to generate Fig.~\ref{fig Hebbian theory matches simulations and the inverse scaling}(b).

\textbf{Numerical estimation of the
convergence speed for Hebbian learning.} Given the Hebbian solution $\gdagger$, we calculated the cosine similarity $\cos(\g_T,\gdagger)$ at each step. Then we used the exponential function $q\exp(-vT)$ to fit the difference $\Delta\cos(\g_T,\gdagger):=1-\cos(\g_T,\gdagger)$ and determine the two parameters $q,v$. Here, $v$ represents the convergence speed, which is used to generate Fig.~\ref{fig_two_eta_speed_accuracy_tradeoff}(b). See Fig.~\ref{speed_estimation} for the comparison between numerical and fitting results of $\Delta\cos(\g_T,\gdagger)$.

\section{The parameters}

\begin{table}[htbp]
\centering
\begin{tabular}{|m{0.2\textwidth}|m{0.8\textwidth}|}
\hline
\hfil $\tau=10$ ms   & The single-neuron integration time constant. \\
\hline
\hfil $\beta = 3\ \mathrm{Hz}^2$ & The decay constant of the synapses. \\
\hline
\hfil $\gamma = 1/30$ Hz & The input strength. \\
\hline  
\hfil $K=20$  & The number of test odors. \\
\hline
\end{tabular}
\centering \caption{Parameter values and the interpretations.}
\label{Table Parameter values and the interpretations}
\end{table}

\newpage

\newpage
\section*{Supplementary Figures}

\renewcommand{\thefigure}{S\arabic{figure}}
\setcounter{figure}{0}  

\onecolumngrid

\begin{figure}[H]%
\includegraphics[width=\textwidth]{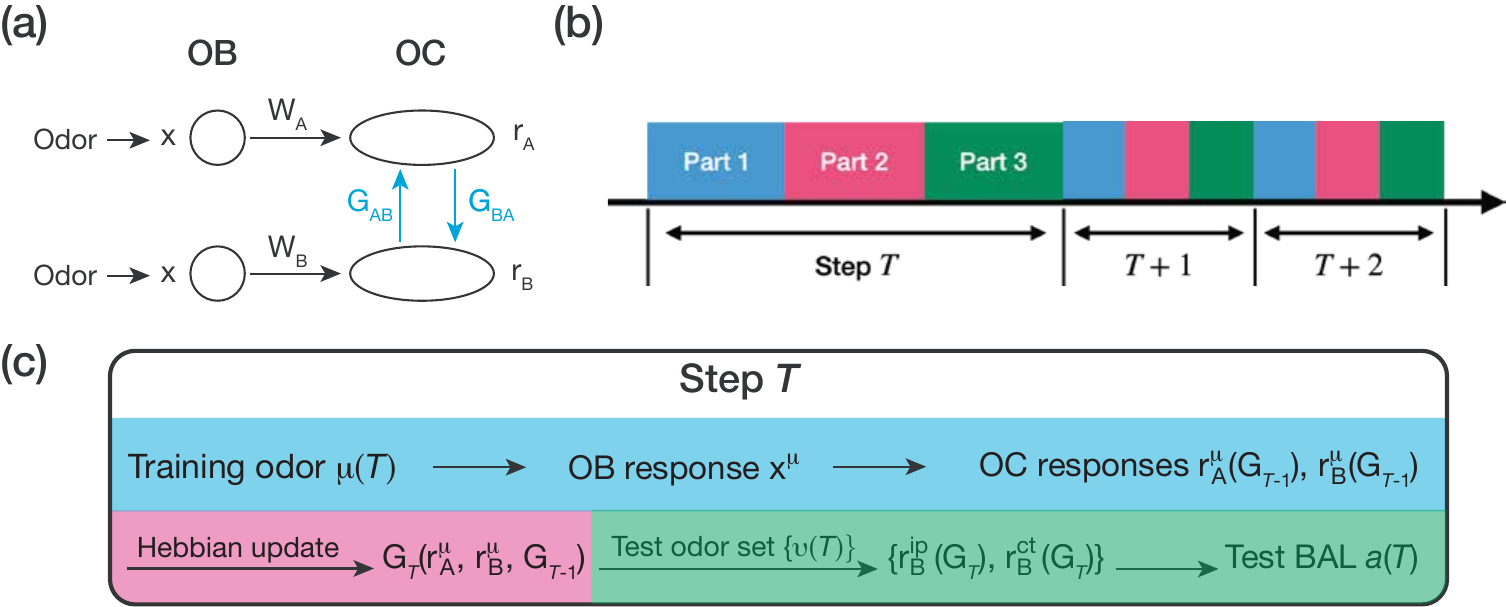}
\caption{\textbf{Model setup and training process.}  \textbf{(a)} Neural circuit to generate the cortical representations $\rbfA$ and $\rbfB$, with the single training odor delivered to both nostrils, which corresponds to the network in Fig.~\ref{fig Background and model setup}(d). \textbf{(b)} During learning, each step contains three sequential parts that are elaborated in panel (c). \textbf{(c)} The three parts for step $T$. Part I (the blue box): a single training odor $\mu(T)$ is ``delivered" to both nostrils, which has the OB response $\x^{\mu}$; the OC responses $\rbfA^{\mu}$ and $\rbfB^{\mu}$ are simulated based on Eq.~\eqref{Eq neural dynamics}, given the inter-hemispheric projection matrices $\GAB$ and $\GBA$ at step $T-1$ (shortened as $\Gbf_{T-1}$ in the bracket). Part II (the pink box): $\GAB$ and $\GBA$ will be updated with Hebb's rule Eq.~\eqref{Eq the sparse Hebbian update rule}, to produce $\GAB$ and $\GBA$ at step $T$ shortened as $\Gbf_{T}$, which depends on $\rbfA^{\mu}, \rbfB^{\mu}$ and $\Gbf_{T-1}$.  Part III (the green box): we randomly generate a test odor set $\{\nu(T)\}$ (each odor is distinguished by its OB response $\x^{\nu}$), and simulate their  ipsi- and contra-lateral responses $\{\rbfB^{\mathrm{ip}}(\Gbf_{T}),\rbfB^{\mathrm{ct}}(\Gbf_{T})\}$; then we calculate the cosine similarity between $\rbfB^{\mathrm{ip,}\nu}$ and $\rbfB^{\mathrm{ct,}\nu}$ for each test odor $\nu(T)$ and average over the test odor set to get the test BAL $a(T)$. }
\label{figS Model setup and training process}
\end{figure}

\newpage

\begin{figure}[H]%
\centering
\includegraphics[width=0.5\textwidth]{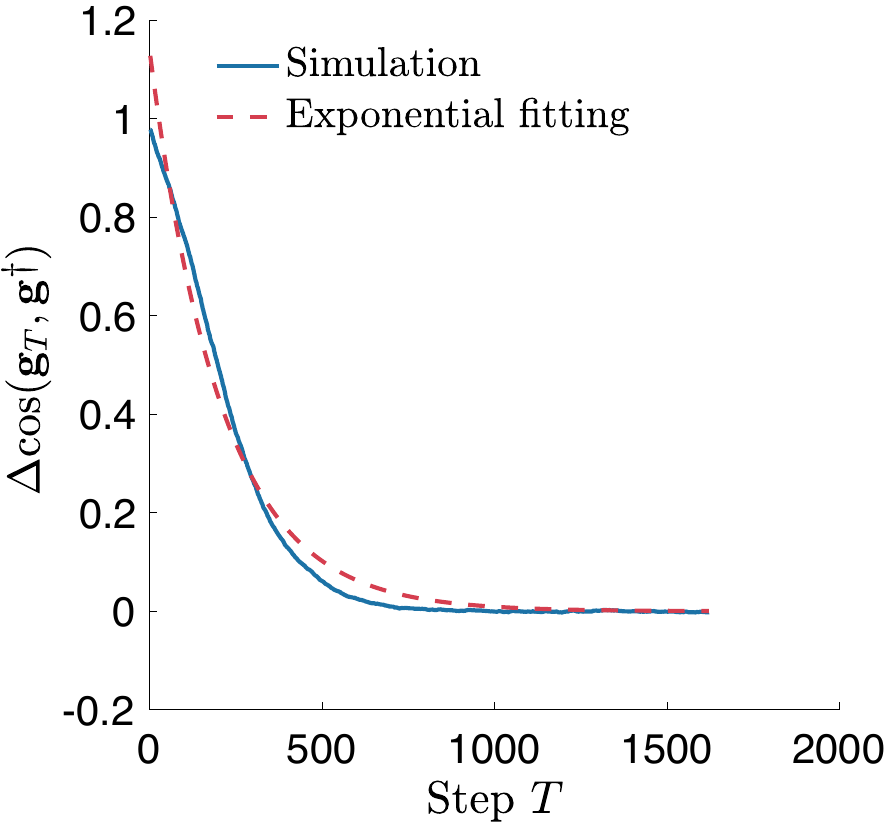}
\caption{\textbf{Numerical estimation of the
convergence speed, for $(m, n, \rhow, \rhog,\xi) = (20, 500, 0.1, 0.05, 0.001)$ in Fig.~\ref{fig_two_eta_speed_accuracy_tradeoff}(a). } Blue line: numerical curve for $\Delta\cos(\g_T,\gdagger)$. Red dashed line: the $q\exp(-vT)$ curve after the exponential fitting. }
\label{speed_estimation}
\end{figure}

\newpage
\begin{figure}[H]%
\centering
\includegraphics[width=0.8\textwidth]{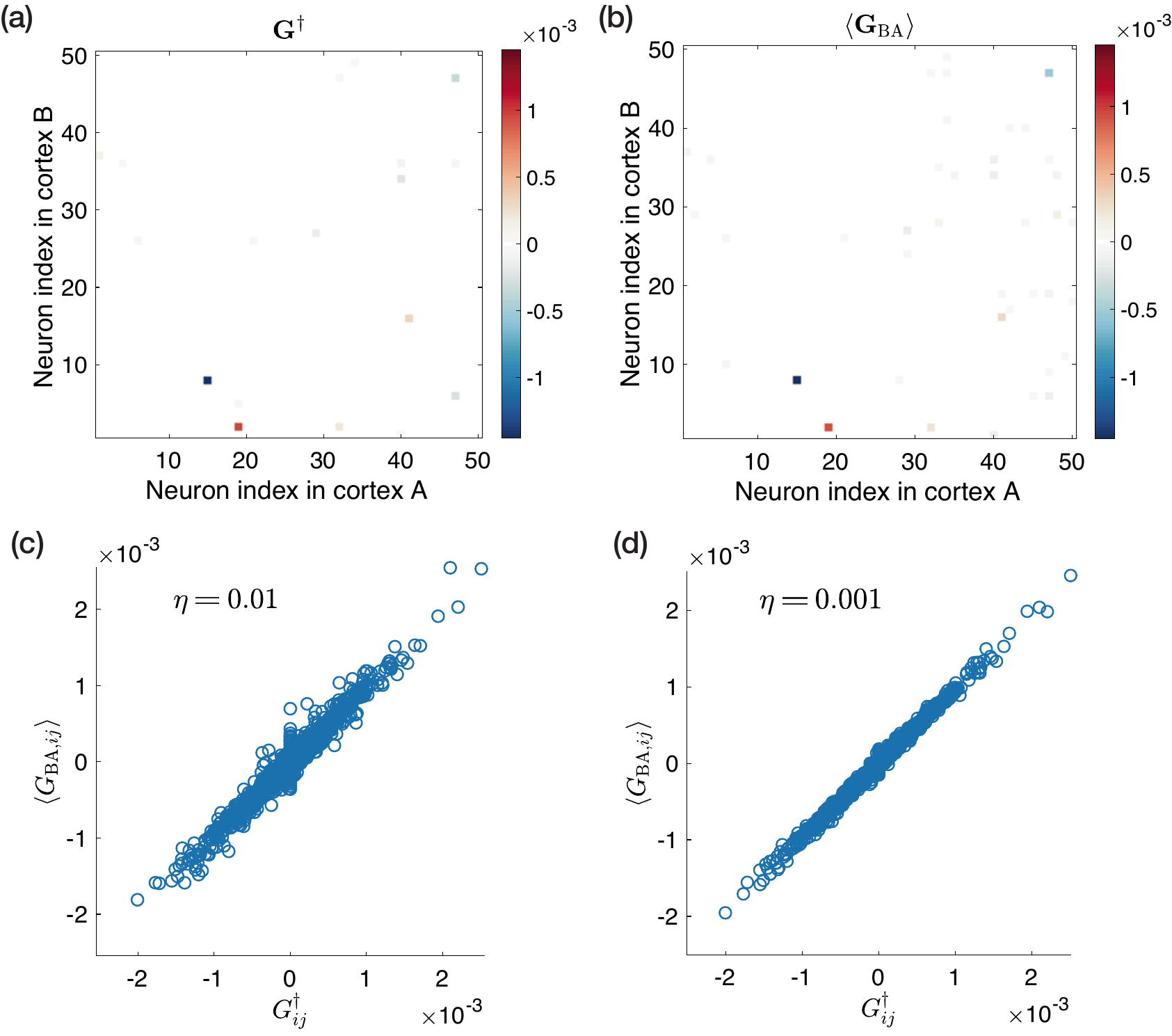}
    \caption{\textbf{Hebbian solution matches numerical results for simulations in Fig.~\ref{fig_two_eta_speed_accuracy_tradeoff}.} \textbf{(a)} The heatmap of the Hebbian solution $\Gdagger$ for $\eta= 0.01\ \mathrm{s}^2$,  only the first 50 neurons in cortex A and those in cortex B are shown. Here, only a handful of the $50^2\cdot\rhog=125$ nonzero elements are clearly visible, as some are significantly smaller than others. \textbf{(b)} Similar to Panel A, but for the temporal mean inter-hemispheric projection matrix $\langle\GBA\rangle$. It is visually almost identical to $\Gdagger$, and their elementwise comparison is in Panel \textbf{(c)} with a cosine similarity of 0.96. \textbf{(d)} Similar to Panel C, but for $\eta= 0.001\ \mathrm{s}^2$ which has the same Hebbian solution. And the cosine similarity between two matrices is 0.99.} 
\label{figS_hebbian_solution}
\end{figure}

\newpage
\begin{figure}[H]%
\centering
\includegraphics[width=0.93\textwidth]{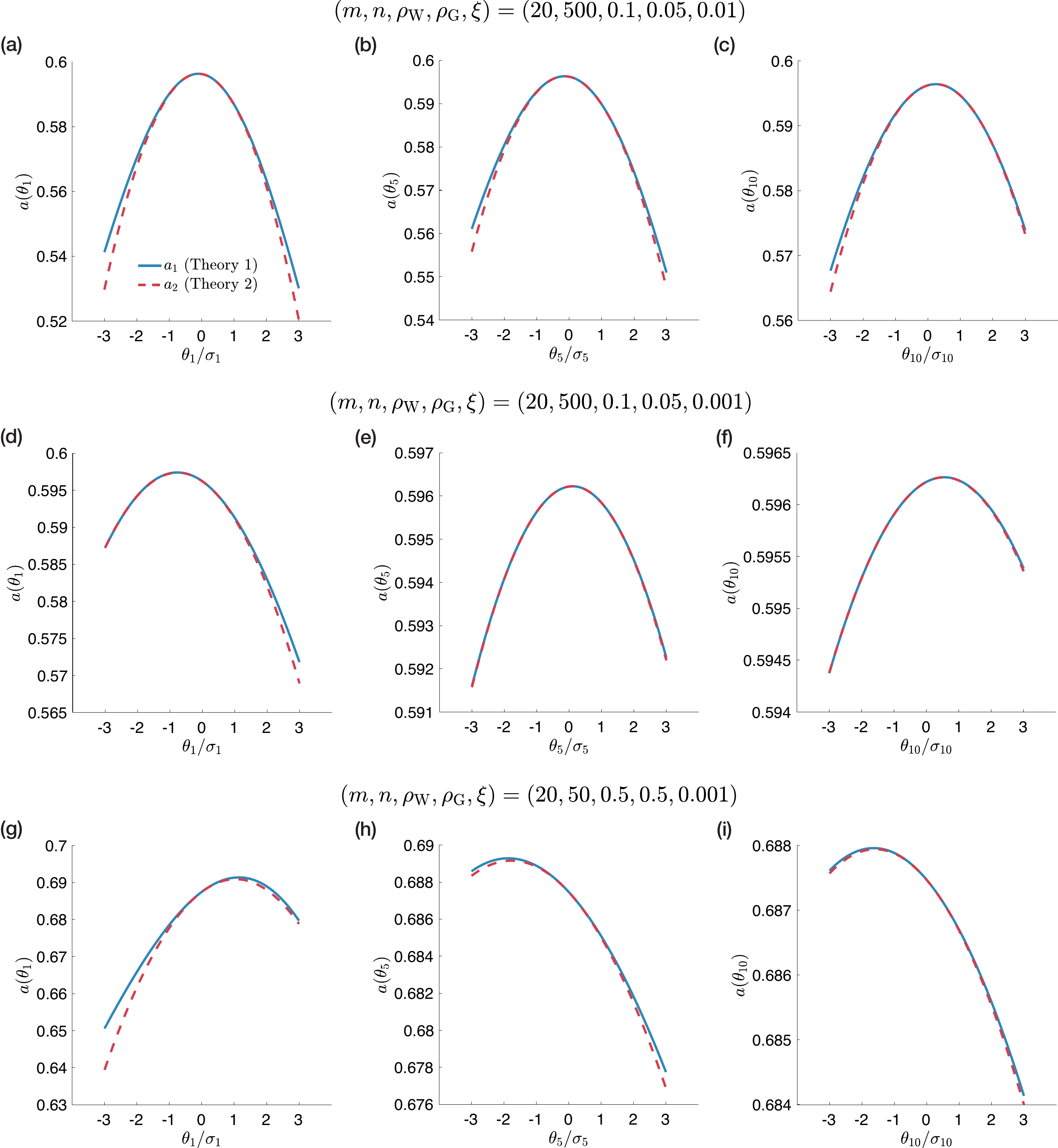}
\caption{\textbf{The concave landscape of test BAL $\ahat$.} \textbf{(a)-(c)} For $(m, n, \rho_W, \rhog, \xi) = (20,500,0.1,0.05, 0.01)$, the landscape of $\ahat$ along the PC directions $i=1,5,10$. \textbf{(a)} The landscape along the first PC direction. $\theta_1$ (the projection onto the first PC) ranges from $-3\sigma_1$ to $3\sigma_1$ ($x$-axis). The blue curve ($\ahat_1$) is the analytical result calculated with Eq.~\eqref{EqS analytical expression of a_hat with PCA}. The red curve ($\ahat_2$) is the second-order approximation using Eq.~\eqref{EqS 2nd-order Taylor expansion of a_hat with PCA}. \textbf{(b-c)} Similar to Panel (a), but for the 5-th and 10-th PC directions. Here, the $y$-axes in (b) and (c) have smaller range, due to smaller fluctuations in these directions. \textbf{(d)-(f)} Similar to Panels (a)-(c) but for a smaller learning rate $ \xi= 0.001$. \textbf{(g)-(i)} Similar to Panels (d)-(f), but for a smaller system with denser projections $(m, n, \rho_W, \rhog) = (20,50,0.5,0.5)$. } 
\label{figS_a_hat_is_concave}
\end{figure}

\newpage
\begin{figure}[H]%
\centering
\includegraphics[width=0.8\textwidth]{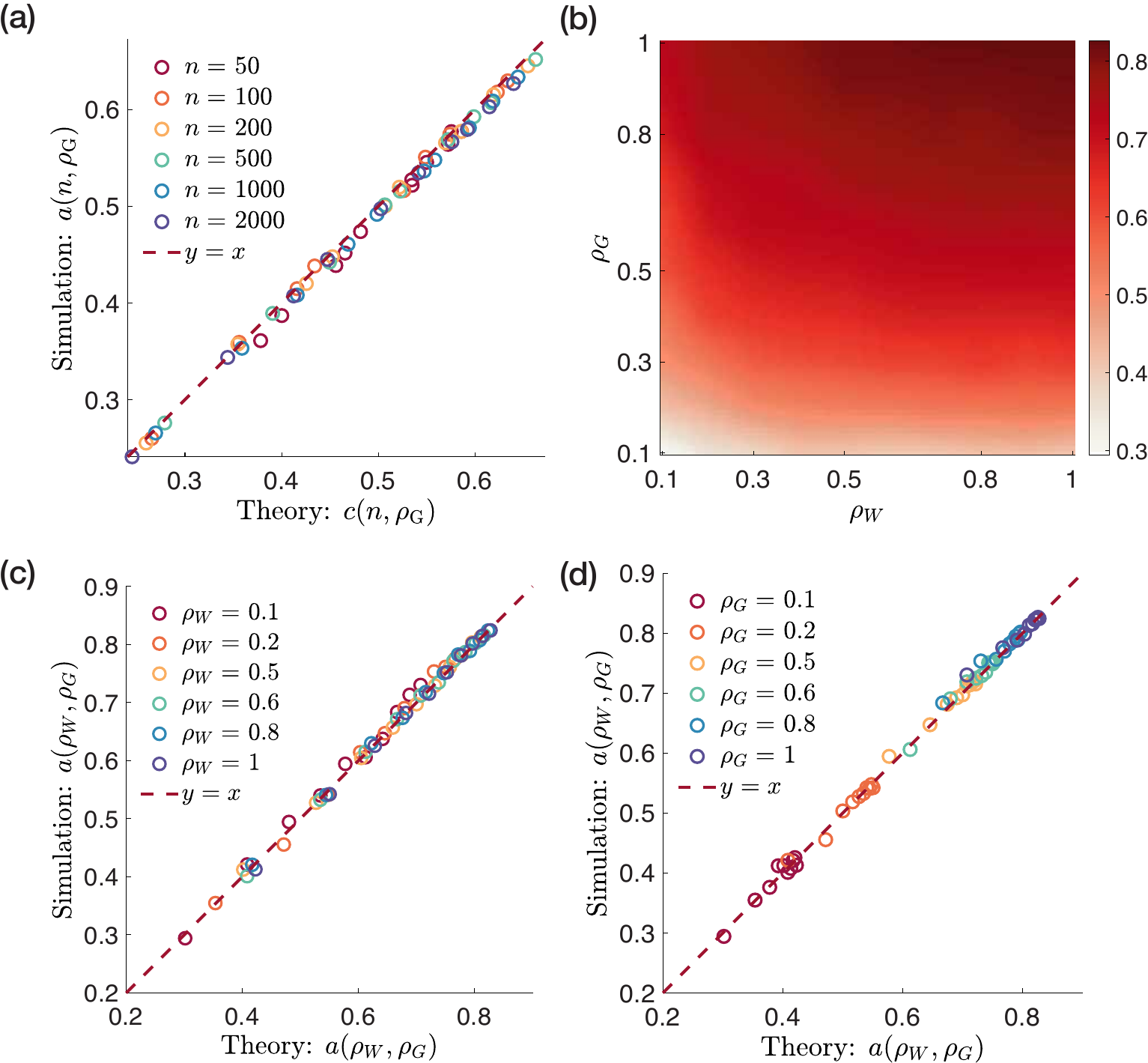}
\caption{\textbf{Theory of Hebbian learning matches simulations.} 
\textbf{(a)} test BAL $a(n,\rhog)$ matches Weight Alignment Level $c(n,\rhog)$ very well, even if $c< 1$. Here, the data for each pair of $(n,\rhog)$ is from Fig.~\ref{fig Hebbian theory matches simulations and the inverse scaling}(a). \textbf{(b)} The heatmap of $a(\rhow,\rhog)$, given $m=20, n=50$. The range of $\rhow$ and $\rhog$ is 0.1 to 1. Consistent with Eq.~\eqref{perfect alignment}, $a$ increases with $\rhow$ and $\rhog$. \textbf{(c)} Pairwise comparison between numerical test BAL and its theoretical values, for chosen data in Panel (b). Here we only showed the data corresponding to $\rhow=0.1,0.2,0.5,0.6,0.8,1$ (different colors), with all values of $\rhog$. The red line: $y=x$. \textbf{(d)} Similar to (c), but for the data corresponding to $\rhog=0.1,0.2,0.5,0.6,0.8,1$ and all values of $\rhow$. Here, the data points are more separate across various $\rhog$ than Panel (c), because $\rhog$ has a larger impact on test BAL than $\rhow$, as predicted by Eq.~\eqref{perfect alignment}. }
\label{figS Theory of Hebbian learning matches simulations}
\end{figure}

\newpage
\begin{figure}[H]%
\centering
\includegraphics[width=0.8\textwidth]{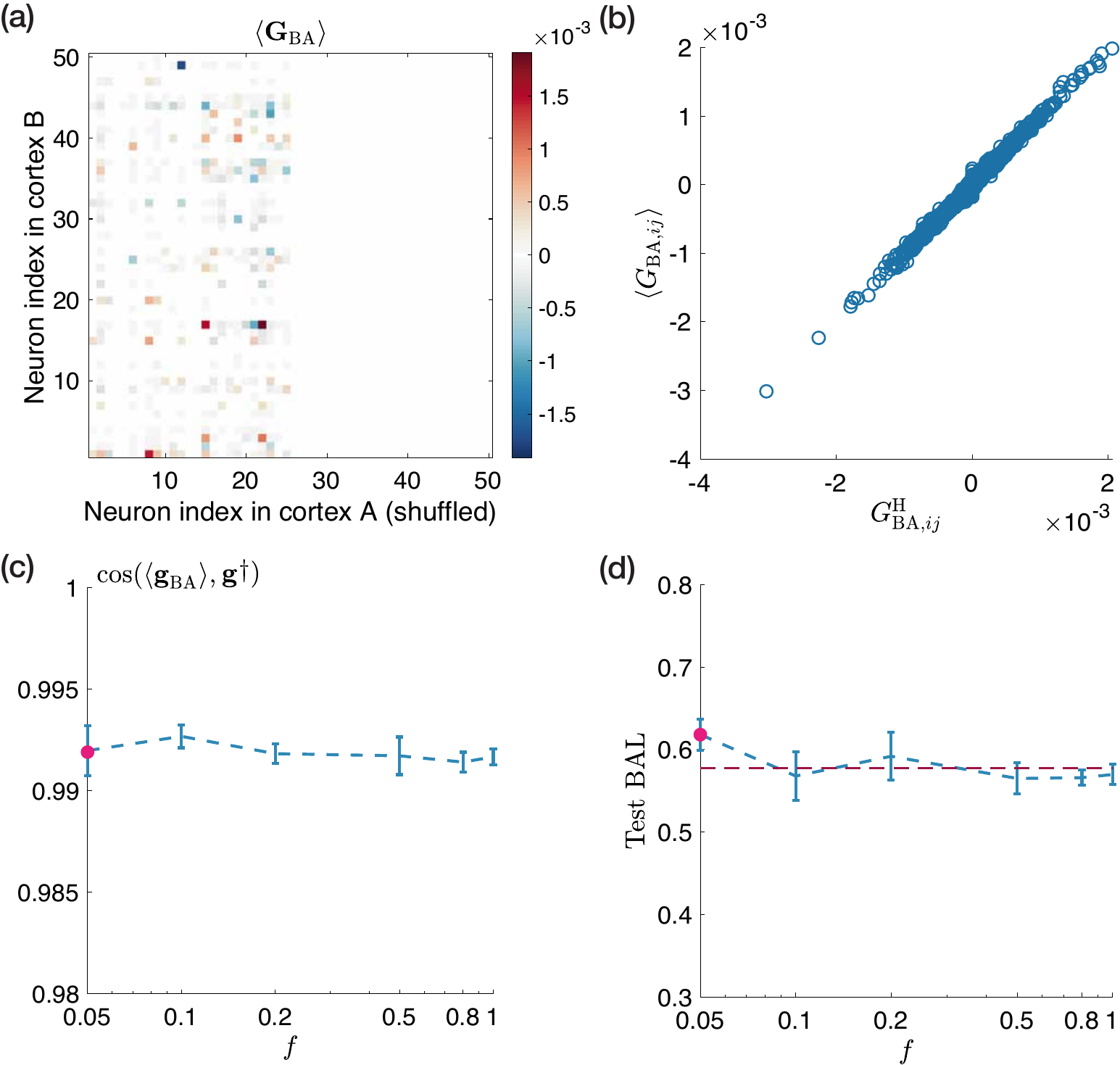}
\caption{\textbf{The fraction of subpopulations projecting contralaterally does not affect bilateral alignment level.} \textbf{(a)} Heatmap for the temporal average of the inter-hemispheric projection matrix $\langle \GBA\rangle$. Here, only 25 neurons from olfactory cortex A project to cortex B (shuffled as columns 1-25 here), with all other elements in  $\langle \GBA\rangle$ being 0). These 25 neurons project to all neurons in cortex B. Parameters:  $(\rhow,\rhog)^m_n = (0.1,0.05)^{20}_{500}, \ \eta= 0.001\ \mathrm{s}^2,$ and $f=\rhog=0.05$. \textbf{(b)} For simulation in Panel (a), the element-wise comparison between the Hebbian solution $\GBAH$ and $\langle \GBA\rangle$. The cosine similarity between two vectorized matrices $\cos( \langle \gBA\rangle,\gdagger)=0.992$. \textbf{(c)} $\cos( \langle \gBA\rangle,\gdagger)$ vs $f=0.1,0.2,0.5,0.6,0.8,1$. The errorbars are over 3 samples. The pink dot: the mean $\cos( \langle \gBA\rangle,\gdagger)=0.992$ for $f=\rhog=0.05$. \textbf{(d)} Test BAL vs $f$. The red dashed line indicates a theoretical value of 0.577, according to Eq.~\eqref{perfect alignment}. The pink dot: the mean numerical test BAL is $0.618$ for $f=\rhog=0.05$.
}
\label{subpopulations projecting}
\end{figure}

\newpage
\begin{figure}[H]%
\centering
\includegraphics[width=0.4\textwidth]{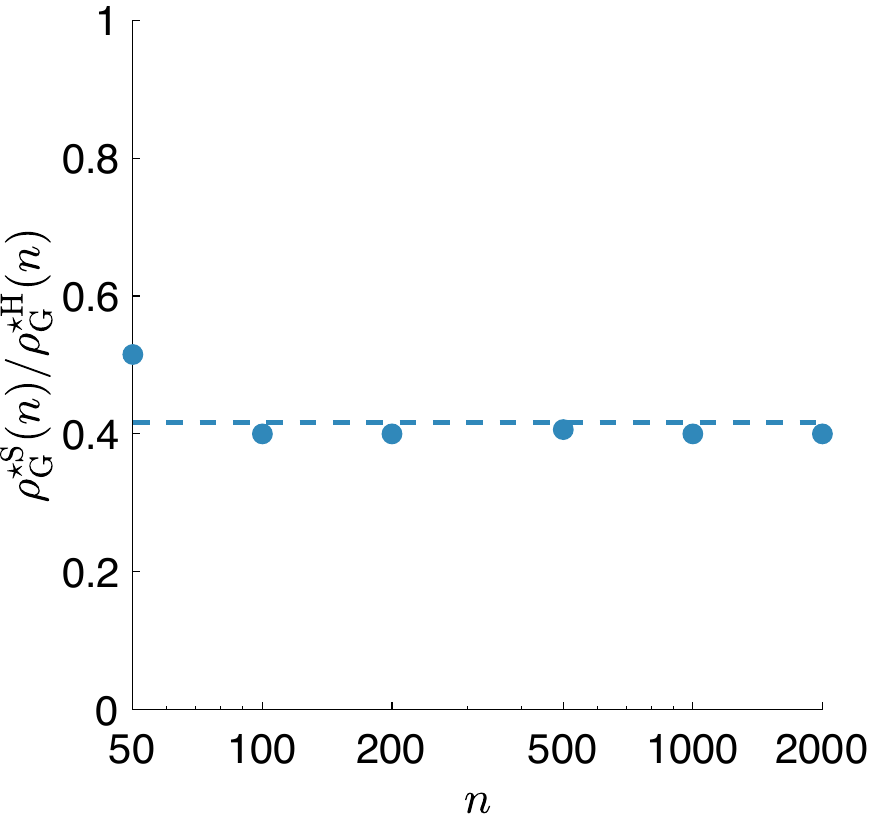}
\caption{\textbf{ The calculation of $\alpha$ for $(m,\rhow) = (20,0.1)$.} With the identified values of $\rhogstarH(n)$ and $\rhogstarS(n)$ in Fig.~\ref{fig4_SGD_alpha}(b), $\dfrac{\rhogstarS(n)}{\rhogstarH (n)}$ was calculated and plotted for each $n$. The blue line corresponds to $\alpha:=\langle \dfrac{\rhogstarS(n)}{\rhogstarH (n)}\rangle_n\approx0.42$. }
\label{figS_alpha_one_example}
\end{figure}

\newpage
\begin{figure}[H]%
\centering
\includegraphics[width=0.8\textwidth]{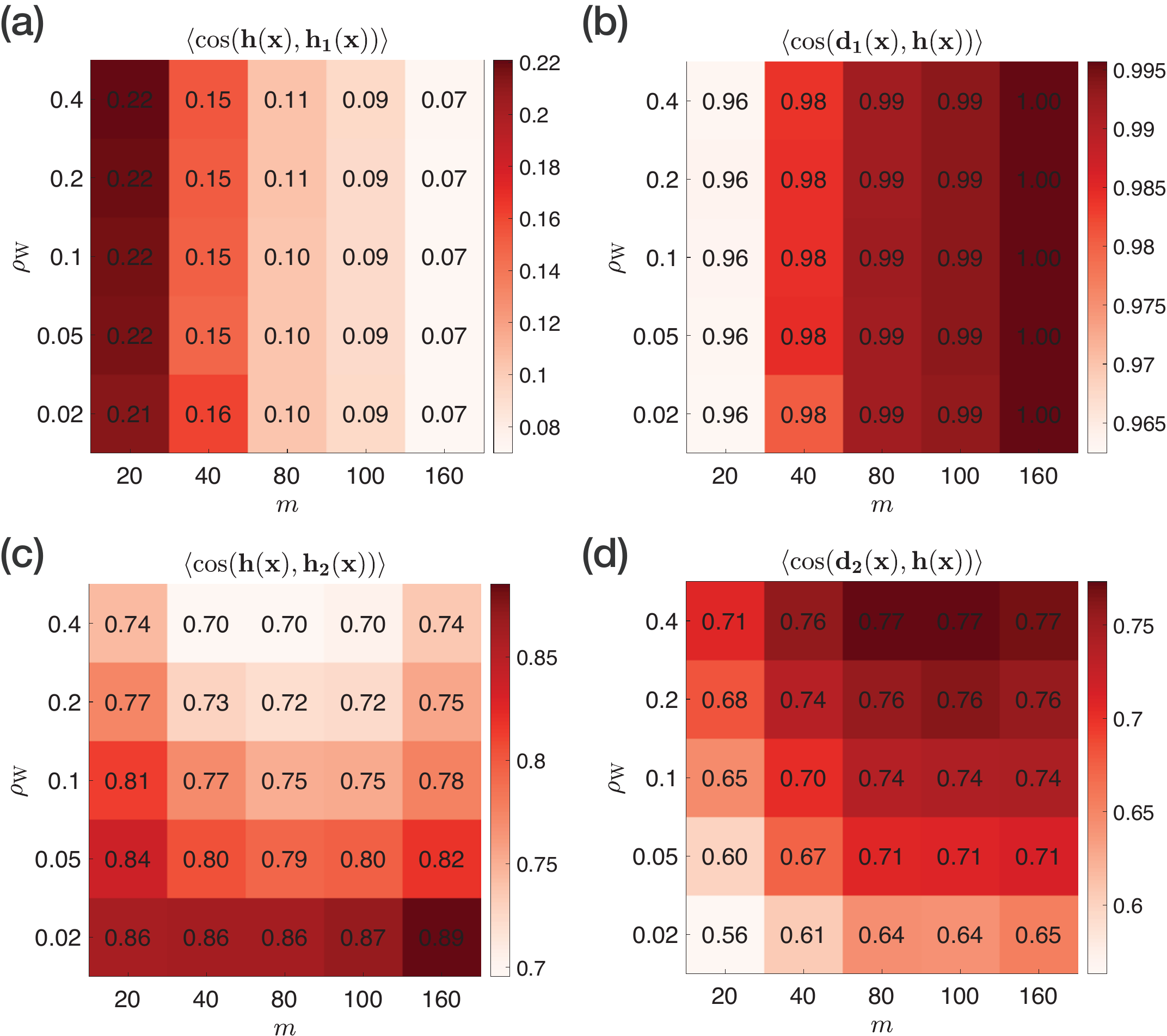}    \caption{\textbf{Numerical results of $\cos(\h(\x),\h_1(\x))$, $\cos(\dbf_1(\x),\h(\x))$, $\cos(\h(\x),\h_2(\x))$, and $\cos(\dbf_1(\x),\dbf_2(\x))$.} \textbf{(a)} $\cos(\h(\x),\h_1(\x))$ decreases with $m$. \textbf{(b)} $\dbf_1$ highly aligns with $\h$, with the alignment level slightly higher for larger $m$. \textbf{(c)} $\cos(\h(\x),\h_2(\x))$ decreases with $\rhow$. \textbf{(d)} In general, $\cos(\dbf_2(\x),\h(\x))$ increases with $m$ and $\rhow$. }
    \label{figS_cos_h_h1_h2}
\end{figure}

\end{document}